\documentclass[%
 reprint,
 amsmath,amssymb,
 aps,
prb,
]{revtex4-2}

\usepackage{graphicx}
\usepackage{dcolumn}
\usepackage{bm}
\usepackage{hyperref}

\usepackage[dvipsnames]{xcolor} 
\usepackage{enumitem}
\hypersetup{colorlinks=true,allcolors=Black}
\usepackage{braket}  
\usepackage{amsmath,amssymb}
\usepackage{graphicx}
\usepackage{bbm}
\usepackage{comment}

\usepackage{cases}

\begin{document}

\preprint{APS/123-QED}

\noindent

\title{ {\textbf{Validity of 
 DFT+U band gaps in all  its known 
functional forms}}}

\author{Andrew C. Burgess$^{1,2}$}
\author{David D. O'Regan$^{2}$}
\email{david.o.regan@tcd.ie}
\affiliation{$^1$School of Physics, University College Dublin, Dublin 4, Ireland \\
 $^2$School of Physics, CRANN Institute and AMBER Research Centre, Trinity College Dublin, The University of Dublin, Ireland}

\date{\today}

\begin{abstract}
 The Density Functional Theory plus Hubbard $U$ (DFT+$U$) technique is one of the most widely used tools by condensed matter physicists  and solid state chemists for the simulation of transition-metal and lanthanide bearing crystals, and increasingly of much more diverse chemistries. 
 Although often synonymous with the corrective functionals of Dudarev et al.~\cite{dudarevElectronenergylossSpectraStructural1998} and Liechtenstein et al.~\cite{liechtensteinDensityfunctionalTheoryStrong1995}, there exists a wide variety of DFT+$U$-type functionals ready to be utilized, and no
 doubt yet to be developed.
 Since the earliest days, the gap in the
 DFT+$U$ single-particle eigenspectrum
 has been associated with the fundamental band gap, and the method has typically found more success for spectra than for total-energy derived properties. 
 There has been some doubt, however, as to the conceptual validity of this association. 
 Here, extending findings from recent years regarding \textcolor{black}{local, semi-local, and hybrid functionals
 within the generalized Kohn-Sham framework, we prove and numerically demonstrate} that the DFT+$U$ eigenspectrum gap is 
 \textcolor{black}{conceptually} 
 valid, in the 
 \textcolor{black}{specific} 
 sense that it matches its own fundamental gap  calculated using total-energy differences.
 \textcolor{black}{We emphasize that this does
 not  imply its agreement with 
 experimental values, 
 and indeed our argument is independent 
 of the Hubbard $U$ parameter.} 
 \textcolor{black}{The  result holds} for pristine periodic systems with converged $k$-point sampling but not, however,
 for defective ones, 
 \textcolor{black}{isolated systems, or
 systems in which added charges exhibit
 spontaneous localization.}
 We show that bandgap validity for 
 \textcolor{black}{pristine} 
 solids holds in the presence of pseudopotentials
 and PAW potentials, when using hybrid functionals, and in  
 DFT+$U$(+$J$) 
 irrespective of the level of subspace projection onto the band-edge states.
 We survey 
 every \textcolor{black}{collinear-spin} DFT+$U$-type functional known to have been published to date,  
 within a unified notation.
 We verify analytically under which conditions the eigenvalue gap equals its
 fundamental gap, for each functional.
 \textcolor{black}{Returning to the
 related but different question of 
 band-gap correction efficiency,  
 we offer fresh analysis of  DFT+$U$ bandgap  projection dependence, 
 and each functional's effect on  energies and gaps for the hydrogen lattice in the Mott-Hubbard limit.}
\end{abstract}

\maketitle

\section{Background}
Kohn-Sham Density Functional Theory (DFT)~\cite{hohenbergInhomogeneousElectronGas1964,kohnSelfConsistentEquationsIncluding1965,m.tealeDFTExchangeSharing2022,perdewFourteenEasyLessons2010} is and has been, for several decades, the most popular tool for predicting the electronic structure of molecules, materials, nano-structures, surfaces and interfaces. It enjoys widespread use across the fields of physics, chemistry and material science~\cite{vannoordenTop100Papers2014,haunschildComprehensiveAnalysisHistory2019}. Central to every Kohn Sham DFT calculation is the selection of an appropriate exchange correlation approximation. Following many years of sustained research on developing explicit approximate functionals for the total energies
of condensed matter systems,
no consensus has yet emerged  within
the electronic structure theory community
as to the level of theoretical complexity that is
 most efficient for use in calculating such energies, 
 their differences and derivatives.
 Even at a given level of theory, there are
 significant differences between different implementations~\cite{Lejaeghereaad3000}. The situation is particularly challenging for 
materials that exhibit significant correlation effects~\cite{Dagotto257}.
While qualitative agreement with experimental  trends can often be sufficient for progress 
in solid state physics~\cite{Curtarolo2013}, 
a very high precision is needed, e.g., to calculate chemical reaction rates~\cite{Norskov2009}, 

At present, there remains insufficient theoretical understanding available for this 
otherwise mature, impactful~\cite{goldbeck2012economic,goldbeck2015industry,RevModPhys.87.897} field to assess whether it is  best to  focus on developing improved  functionals of the electronic density alone, since DFT is exact in principle, or to focus attention on another specific quantity such as the kinetic energy density, the potential, the 
occupied, or even unoccupied Kohn-Sham orbitals~\cite{doi:10.1063/1.2831900}, in pursuit
of accuracy. Beyond DFT, there now exist sophisticated but
typically  very costly functionals of the one-body density matrix~\cite{PhysRevB.78.201103}, the pair-density~\cite{ZIESCHE1994213}, the density response function~\cite{C3CS60456J}, the Green's function~\cite{PhysRev.139.A796,RevModPhys.74.601}, and the many-body wave-function (either in its entirety with appropriate restrictions~\cite{Taft2017}, or for pre-defined 
 subspaces within a  mean-field embedding~\cite{RevModPhys.78.865}). 
 The diversity of this inter-disciplinary field is a strength, 
and it reflects the differing
requirements in its key areas of application.  %
In the present work, we will focus 
on functionals of the occupied Kohn-Sham
orbitals, giving rise to non-local
(implicit)\footnote{\textcolor{black}{The
term `implicit' here is used to mean
that while the potential operator is orbital 
index independent (there is only a single 
potential), its action upon each orbital
may be, and typically will be,  different.
Functionals in this restricted 
class are functionals of the GKS 
density-matrix sufficiently, i.e.,
they don't use the full variational freedom
made available by the orbitals.}}
orbital-dependent 
potentials~\cite{seidl1996generalized,garrick2020exact}.
These fall on functional 
`Rung 4', i.e., the hybrid-functionals, albeit that some
that we will \textcolor{black}{review} have  features
not usually  found in hybrids, 
such as explicit electron count
dependence. 
 
 The situation, overall, is more optimistic
 if one looks  at 
 theoretical spectroscopy.
 It is now possible to predict the 
 spectral functions of many materials, 
 in excellent agreement with photoemission measurements, 
 using first-principles  theoretical methodologies which 
 require no significant external parameters~\cite{PhysRevB.93.085124}.
 Often, progress has been made  by means of 
 the successful combination of electronic band-structure theory,
 e.g., periodic Kohn-Sham DFT and more recently 
 many-body perturbation theory within the $GW$
 approximation~\cite{0034-4885-61-3-002,PhysRevB.76.165106},  
 with algorithms for solving explicitly-correlated
 discrete many-body Hamiltonians~\cite{RevModPhys.78.865}.
 The latter are restricted to the  selected subspaces deemed
 to be  correlation-prone.
 The combination is usually made by means of self-consistent
 Green's function embedding techniques such as
 dynamical mean-field theory, yielding DFT+DMFT~\cite{BIERMANN201617,PhysRevLett.96.226402,Vollhardt2017,
RevModPhys.78.865,PSSB:PSSB200642053,PhysRevLett.110.086401,
doi:10.1146/annurev-conmatphys-020911-125045} and
$GW$+DMFT~\cite{PhysRevLett.90.086402,0295-5075-100-6-67001,PhysRevB.88.165119}, 
 and more recently increasingly 
 on the basis of  
 density-matrix rather than Green's function embedding~\cite{PhysRevLett.109.186404}.

\section{DFT+$U$ in context}
\begin{figure}[b]
\centering
\includegraphics[scale=0.2]{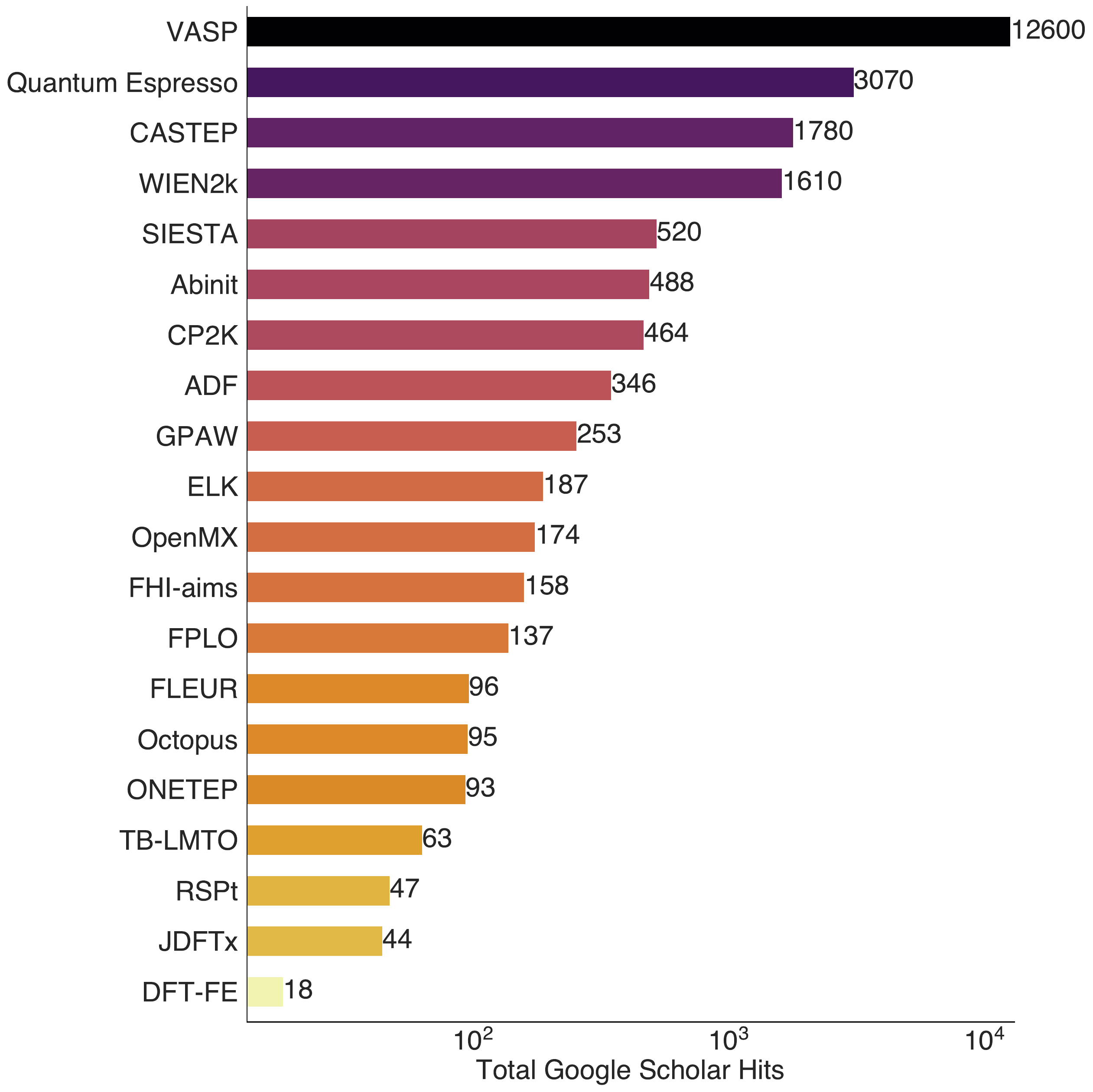}
\caption{The number of Google Scholar hits~\cite{GoogleScholar},
on a logarithmic scale,
for a range of DFT  packages that implement it~\cite{kresseEfficientIterativeSchemes1996,clarkFirstPrinciplesMethods2005,giannozziAdvancedCapabilitiesMaterials2017,kuhneCP2KElectronicStructure2020,teveldeChemistryADF2001,blaha2001wien2k,artachoSIESTAMethodDevelopments2008,gonzeABINITFirstprinciplesApproach2009,elk,mortensenGPAWOpenPython2024,nakataLargeScaleLinear2020,blumInitioMolecularSimulations2009,tancogne-dejeanOctopusComputationalFramework2020,fleurWeb,koepernikFullpotentialNonorthogonalLocalorbital1999,prenticeONETEPLinearscalingDensity2020,geudtnerDeMon2k2012,willsFullPotentialElectronicStructure2010,bigdft,sundararamanJDFTxSoftwareJoint2017,briddonAccurateKohnSham2011,motamarriDFTFEMassivelyParallel2020,DFTKjcon,onoTOMBOAllelectronMixedbasis2015,SEQUEST,jepsen2000stuttgart,ozakiVariationallyOptimizedAtomic2003}. The search was conducted by searching the name of the DFT package in conjunction with DFT+U.}
\label{fig:code_popularity}
\end{figure}

An alternative approach to employing such 
relatively costly methods is to supplement a standard, low cost local or semi-local exchange correlation (XC) functional~\cite{voskoAccurateSpindependentElectron1980,barthLocalExchangecorrelationPotential1972,perdewAccurateSimpleAnalytic1992,perdewGeneralizedGradientApproximation1996,leeDevelopmentColleSalvettiCorrelationenergy1988,perdewRestoringDensityGradientExpansion2008,beckeDensityfunctionalExchangeenergyApproximation1988} with a corrective functional such as Koopmans' Compliant Functionals~\cite{daboKoopmansConditionDensityfunctional2010,borghiKoopmanscompliantFunctionalsTheir2014,colonnaKoopmansSpectralFunctionals2022,linscottKoopmansOpenSourcePackage2023}, self-interaction error (SIE) correction functionals~\cite{perdewSelfinteractionCorrectionDensityfunctional1981,pedersonCommunicationSelfinteractionCorrection2014} and van der Waals functionals~\cite{grimmeSemiempiricalGGAtypeDensity2006,grimmeConsistentAccurateInitio2010,tkatchenkoAccurateMolecularVan2009,kimUMBDMaterialsReadyDispersion2020}. In particular, the DFT plus Hubbard $U$ (DFT$+U$) method~\cite{himmetogluHubbardcorrectedDFTEnergy2014,anisimovFirstprinciplesCalculationsElectronic1997,kirchner-hallExtensiveBenchmarkingDFT2021} has been devised to treat systems with localized electronic states such as 3d-transition metal oxides~\cite{saigautamEvaluatingTransitionMetal2018,artrithDatadrivenApproachParameterize2022,crespoElectronicMagneticProperties2013,florisVibrationalPropertiesMnO2011,franchiniGroundstatePropertiesMultivalent2007,huChoiceDFT+UCalculations2011,kiejnaSurfacePropertiesClean2012,limImprovedPseudopotentialTransferability2016,longEvaluatingOptimal$U$2020,padilha$mathrmDFT+U$Simulation$mathrmTi_4mathrmO_7textensuremathmathrmTimathrmO_2$2015,rodlQuasiparticleBandStructures2009,rollmannFirstprinciplesCalculationStructure2004,tancogne-dejeanUltrafastModificationHubbard2018,tompsettImportanceAnisotropicCoulomb2012,zhouFirstprinciplesPredictionRedox2004,zhouInitioElectronPhononInteractions2021}, 4f-lanthanide oxides~\cite{carchiniUnderstandingTuningIntrinsic2016,castletonTuningLDA+UElectron2007,colonsantanaEffectGadoliniumDoping2012,hamadaNewConstraintDFT2018,huangCOAdsorptionOxidation2008,liAssessmentPBE+UHSE062020,reshakEffectElectronicProperties2009,yangEffectEnvironmentReaction2010}, and transition metal complexes~\cite{scherlisSimulationHemeUsing2007,panchmatiaHalideLigatedIron2010,panchmatiaGGA+UModeling2008,kulikFirstPrinciplesStudyNonheme2009,kulikDensityFunctionalTheory2006,brumboiuInfluenceElectronCorrelation2016,caoStronglyCorrelatedElectrons2008}. Although originally developed for application with the Local Spin Density Approximation (LSDA)~\cite{anisimovBandTheoryMott1991}, it is now common practice to also apply this corrective functional with Generalized Gradient Approximations (GGAs)~\cite{hunnestad3DOxygenVacancy2024,koksalChernZ2Topological2019,wangDualisticInsulatorStates2024,floresStructuralElectronicProperties2020,loschenFirstprinciples$mathrmLDA+mathrmU$$mathrmGGA+mathrmU$2007,maCorrelationEffectsLattice2013,morganLithiumIntercalationTiO$_2$B2012,nguyenWaterAdsorptionDissociation2013,stoefflerInitioElectronicStructure2006,yangEffectsSelfconsistentExtended2024} and more recent studies have investigated its application in conjunction with meta-GGAs~\cite{longEvaluatingOptimal$U$2020,saigautamEvaluatingTransitionMetal2018,zengInvestigationDoublePerovskites2022,kimLatticeElectronicProperties2021}. One of the strengths of the DFT$+U$ method is its ease of implementation as part of a broader DFT code package. Indeed, the method has been implemented in a wide variety of DFT codes, many of which have garnered a large, active community of users. The total number of Google Scholar~\cite{GoogleScholar} `hits' associated with the DFT$+U$ 
method for a range of DFT  packages~\cite{kresseEfficientIterativeSchemes1996,clarkFirstPrinciplesMethods2005,giannozziAdvancedCapabilitiesMaterials2017,kuhneCP2KElectronicStructure2020,teveldeChemistryADF2001,blaha2001wien2k,artachoSIESTAMethodDevelopments2008,gonzeABINITFirstprinciplesApproach2009,elk,mortensenGPAWOpenPython2024,nakataLargeScaleLinear2020,blumInitioMolecularSimulations2009,tancogne-dejeanOctopusComputationalFramework2020,fleurWeb,koepernikFullpotentialNonorthogonalLocalorbital1999,prenticeONETEPLinearscalingDensity2020,geudtnerDeMon2k2012,willsFullPotentialElectronicStructure2010,bigdft,sundararamanJDFTxSoftwareJoint2017,briddonAccurateKohnSham2011,motamarriDFTFEMassivelyParallel2020,DFTKjcon,onoTOMBOAllelectronMixedbasis2015,SEQUEST,jepsen2000stuttgart,ozakiVariationallyOptimizedAtomic2003} is presented in Fig.~\ref{fig:code_popularity}. 
The immense utility of the method
is self-evident. 

\textcolor{black}{The DFT+$U$ approach has also inspired the development of other advanced electronic structure methodologies such as the Koopmans' Compliant (KC) Functionals~\cite{daboKoopmansConditionDensityfunctional2010,borghiKoopmanscompliantFunctionalsTheir2014,colonnaKoopmansSpectralFunctionals2022,linscottKoopmansOpenSourcePackage2023}. The Koopmans' Integer functional, for example, was explicitly developed to extend a DFT+$U$-type correction to each Kohn Sham (KS) orbital  separately, thus ensuring the retention of the base (i.e., underlying approximate) DFT total energy in the case of insulators. Within the frozen orbital approximation, the unscreened Koopmans' correction $\{\Pi_{i \sigma}^{\rm u}\}$ replaces the non-linearity of the base DFT functional with a term that is linear with respect to orbital occupancy $f_{i \sigma}$. This term takes the form 
\begin{equation}
 \Pi_{i \sigma}^{\rm u}(f_{i \sigma})=f_{i \sigma} \eta_{i \sigma}-\int_0^{f_{i \sigma}}d{s}\braket{\Psi_{i \sigma}|\hat{H}_{\rm KS}(s)|\Psi_{i \sigma}}.
\end{equation}
Here, $\hat{H}_{\rm KS}(s)$ is the approximate KS Hamiltonian where orbital $\Psi_{i \sigma}$ has occupation $s$. A number of methods have been devised for determining the slope of the linear term $\eta_{i \sigma}$, including the Koopmans' Integer functional and the Koopmans' Integer Perdew-Zunger self-interaction corrected functional. In practice, the orbitals relax with respect to occupancy so that the unscreened Koopmans' correction $\{\Pi_{i \sigma}^{\rm u}\}$ will not suffice to enforce linearity. Such orbital relaxation effects are accounted for through orbital dependent screening parameters, which also need to be evaluated. Through rigorous benchmarking, KC functionals have been shown to significantly improve both the predicted bandgaps and ionization potentials compared to the underlying (semi-)local XC approximations~\cite{colonnaKoopmansCompliantFunctionalsPotentials2019,dealmeidaElectronicStructureWater2021,nguyenKoopmansCompliantSpectralFunctionals2018}. }

\textcolor{black}{In contrast,} through the use of a set of fixed projection operators $\hat{P}$, \textcolor{black}{conventional}  DFT$+U$ corrective functionals act only on selected subspaces within the electronic system of interest. The suitability of various choices in projection operator have been investigated including atomic orbitals, ortho-atomic orbitals~\cite{lambertUseMathrmDFT2023,timrovPulayForcesDensityfunctional2020,riccaSelfconsistentDFTStudy2020,timrovElectronicStructurePristine2020}, maximally localized Wannier functions~\cite{shihScreenedCoulombInteractions2012,fabrisTamingMultipleValency2005} and molecular orbitals~\cite{bajajMolecularDFT+UTransferable2021,bajajMolecularOrbitalProjectors2022}. The strength of the DFT$+U$ correction is also dependent on the value of the effective Hubbard parameter $U_{\rm eff}$. Its value is often tuned to ensure agreement with some experimental property of interest~\cite{griffinOriginCorrelatedIsolated2023,wangOxidationEnergiesTransition2006}, but numerous schemes have also been devised for the calculation of the necessary  
 Hubbard $U$ and Hund's $J$ interaction parameters~\cite{cococcioniLinearResponseApproach2005,linscottRoleSpinCalculation2018}.
 More generally, 
 low-energy model embedding or `down-folding'~\cite{PhysRevB.62.R16219},
has a long and successful record in the electronic structure theory
of strongly correlated materials~\cite{PhysRevB.43.7570,springerFrequencydependentScreenedInteraction1998,pickettReformulationMathrmLDA1998,PhysRevB.71.035105,himmetogluFirstprinciplesStudyElectronic2011}. 
 A significant fundamental and practical breakthrough 
 came with the invention of the Green's function
 based constrained random phase approximation (cRPA)~\cite{aryasetiawanCalculationsHubbardFirstprinciples2006, PhysRevB.77.085122, PhysRevB.80.155134, sasiogluEffectiveCoulombInteraction2011,springerFrequencydependentScreenedInteraction1998,kotaniInitioRandomphaseapproximationCalculation2000,aryasetiawanFrequencydependentLocalInteractions2004,aryasetiawanCalculationsHubbardFirstprinciples2006,sasiogluEffectiveCoulombInteraction2011,vaugierHubbardHundExchange2012,amadonScreenedCoulombInteraction2014}.
More recently, the resulting dynamical interaction parameters
 $U\left( \omega \right)$ and $J\left( \omega \right)$ have been 
 physically interpreted~\cite{aryasetiawanFrequencydependentLocalInteractions2004,0295-5075-100-6-67001,PhysRevB.85.035115} 
 and  applied within suitably adapted 
 impurity solvers~\cite{0295-5075-99-6-67003,PhysRevB.94.201106}. Within the SIE-correction interpretation of DFT+$U$~\cite{PhysRevB.71.035105,kulikDensityFunctionalTheory2006}, 
the validity of which is numerically confirmed 
in~\cite{PhysRevB.94.220104}, 
the targeted subspaces are effectively approximated 
as quantum systems in contact with a bath 
that provides and screening and  particle exchange.
The Hubbard $U$ is then  calculated  as 
a measure of the  net self-interaction of the subspace,
which is assumed to be spurious, 
together with the remainder of the system providing a screening 
environment~\cite{linscottRoleSpinCalculation2018,Glenthesis}.

 While DFT+$U$  is routinely applied to 
 calculate diverse electronic structure properties, 
 of both total-energy
 derived and spectroscopic types, 
 its success has been arguably weighted more towards
 the latter; for examples see Refs.~\cite{verma2016does,kirchner-hallExtensiveBenchmarkingDFT2021}. 
 This is due to the extraordinary cost-benefit ratio
 that it provides when the principal drawback of
 the underlying approximate functional is a greatly
 underestimated, or absent, fundamental gap, when
 calculated on the basis of the 
  single-particle eigenvalues.
 \textcolor{black}{Notwithstanding, for just a moment, 
 the validity of such calculation,} DFT+$U$ can often 
 improve such a bandgap
 prediction on an almost cost-free basis, due to the
 implicit derivative discontinuity that it introduces,
 which manifests in a potential that depends on the 
 occupancy of the bands. 
 This is done on an empirical basis 
 or, when the necessary parameters are 
 calculated in situ, an effectively semi-empirical basis.
 \textcolor{black}{Semi-empiricism arises
 because,}
 even when self-consistent projector functions are
used~\cite{eschrig,PhysRevB.82.081102,0953-8984-27-32-325602,fabrisTamingMultipleValency2005,carta2026bridging}, there remains some empiricism related to their means of construction, albeit arguably
 at a level  no worse than that of the 
 typical underlying functional. 

\textcolor{black}{This brings us to the main topic of the present work.} 
Throughout the history of DFT+$U$, it may be said
that there has remained some doubt  
concerning the validity in principle of using the
eigenvalue gap as the basis for calculating the 
fundamental bandgap~\cite{sai2018evaluating,himmetogluHubbardcorrectedDFTEnergy2014}.
DFT+$U$ has, in this regard,  been sometimes viewed as a further layer
of pragmatism on top of an already pragmatic practice,
or something that might be analyzed
as an approximation
to a hybrid functional~\cite{PhysRevB.90.035146}  
 or as an approximation to a many-body
perturbation theory applied on top of the Kohn-Sham
system~\cite{anisimovFirstprinciplesCalculationsElectronic1997}.

The aim of the present work is to rigorously establish
the validity in principle of 
calculating fundamental gaps from the DFT+$U$
(family of corrections) eigenvalue gap, for the case
of pristine crystals,   \textcolor{black}{only where the added electron or
hole delocalizes over the bulk system, and then only when} treated with very well-converged
k-point grids (i.e., in the so-called thermodynamic limit).
This result does not apply to 
crystals simulated 
with insufficiently dense k-point sampling, 
however.
Neither does it apply to molecules, defective
solids, \textcolor{black}{or pristine systems with spontaneous charge carrier localization,   
and so} the fundamental gaps of such systems 
\textcolor{black}{may be} more reliably 
calculated using explicit (spin-polarized, for the
avoidance of doubt)  
electron addition and removal energies.

\textcolor{black}{One needs to proceed with caution   
in this context, as the question whether 
 such a gap  \emph{should} recover the fundamental gap is
 one that depends not only on the functional type, 
 but equally on the domain
 over which it is minimized 
 (e.g., whether the \textcolor{black}{total energy is a functional of the density}),
 and not just the nature of the system under scrutiny~\cite{seidl1996generalized,yangDerivativeDiscontinuityBandgap2012,perdew2017understanding}.}
The result of bandgap validity, \textcolor{black}{as we will discuss,}
holds for the standard
widespread practice of DFT+$U$ \textcolor{black}{for crystals}, 
but not pure \textcolor{black}{Kohn-Sham} DFT settings as in 
optimized effective potential (OEP) 
DFT+$U$~\cite{oepldaplusu,ravindran2025local}.

\textcolor{black}{It is a  well known and  long-standing 
result that only the frontier 
occupied KS eigenvalue generally
corresponds to a total energy difference, 
i.e., minus the ionization potential, in 
\emph{exact}
Kohn-Sham DFT~\cite{almbladhExactResultsCharge1985}.
Why this does not generally hold also for the
unoccupied KS eigenvalue, i.e., 
it is not minus the 
electron affinity, is due  
to the absence of the 
exchange-correlation derivative discontinuity 
in the exact Kohn-Sham 
eigenspectrum~\cite{gould2014kohn}.
In \emph{approximate} KS DFT with
an absent exchange-correlation 
derivative discontinuity (such as conventional local and semi-local functionals), both
levels are valid for bulk 
systems under the aforementioned preconditions~\cite{perdew2017understanding} and in all other cases,
\emph{both} KS eigenvalues
are invalid.}

\textcolor{black}{%
Our work does not revisit KS DFT and has no bearing upon it, 
as it takes place instead within
the different context of  
generalized Kohn-Sham theory (GKS)~\cite{seidl1996generalized},
and moreover an approximated form
with an implicit rather than
explicit derivative discontinuity.} 
\textcolor{black}{%
GKS is, in a nutshell, where the orbitals but not the resulting density provide 
the functional dependence of the total energy, and
it is the practical, commonplace formalism of DFT+$U$.}
\textcolor{black}{%
In GKS theory, unlike KS, any implicit derivative discontinuities 
(i.e., where the energy remains differentiable with respect to the 
density-matrix) are automatically included in the eigenspectrum~\cite{yangDerivativeDiscontinuityBandgap2012},
and it is this type that arises almost
universally in DFT+$U$ (we will later discuss some 
recently-introduced exceptions).}

\textcolor{black}{The excluded case of spontaneous charge carrier localization~\cite{vlvcek2016spontaneous}, i.e., where the added electron or hole localizes in space despite the absence of defects or lattice distortions has, to the best of the authors' knowledge, never been reported to date in DFT+$U$ studies but could conceivably occur for sufficiently large values of the applied Hubbard $U$ parameter. Such symmetry breaking is, however, suppressed in conventional DFT+$U$ calculations where a pristine periodic system is modeled using a primitive unit cell with converged $k$-point sampling.}

Importantly and reassuringly, in connection with
the choice of projection operators $\hat{P}$
that define a DFT$+U$ calculation, we find that the 
validity in principle of the resulting 
eigenvalue gap holds independently of such choice.
The numerical value of the calculated bandgap 
does, of course, generally depend on $\hat{P}$.
\textcolor{black}{Its validity in principle 
holds even for very poor or unphysical choices
of $\hat{P}$ and/or the Hubbard $U$ and
related parameters, reinforcing the point that
we are primarily addressing in the present work a 
different aspect to that of
comparing DFT+$U$ band gaps with experimental ones.}

\textcolor{black}{Notwithstanding, we will return in Sections~\ref{sec:verificationtable} 
and~\ref{sec:projdependence} to shed new light on the
functional and projection dependence, respectively, 
of the DFT+$U$ band gap.}
We prove, in passing, that hybrid functional 
eigenvalue-based bandgaps are perfectly valid for
crystals \textcolor{black}{(under the aforementioned restrictions)}.
We further show, in Appendices~\ref{appendix:ncpp}
and \ref{appendix:ultrasoftpaw}, that this 
GKS eigenvalue-based bandgap validity is generally
preserved by the use of pseudopotentials or the 
projector-augmented wave (PAW) construction.
\textcolor{black}{Validity in the case of
local and semi-local functionals, 
which was addressed
in Ref.~\onlinecite{perdew2017understanding} using a 
somewhat different 
derivation,  
is automatically included in our proof, by setting the Hubbard parameters or 
exact exchange hybrid mixing parameter to zero.}

\section{Numerical test for
band-gap validity}

\begin{figure}
\centering
\includegraphics[scale=0.3]{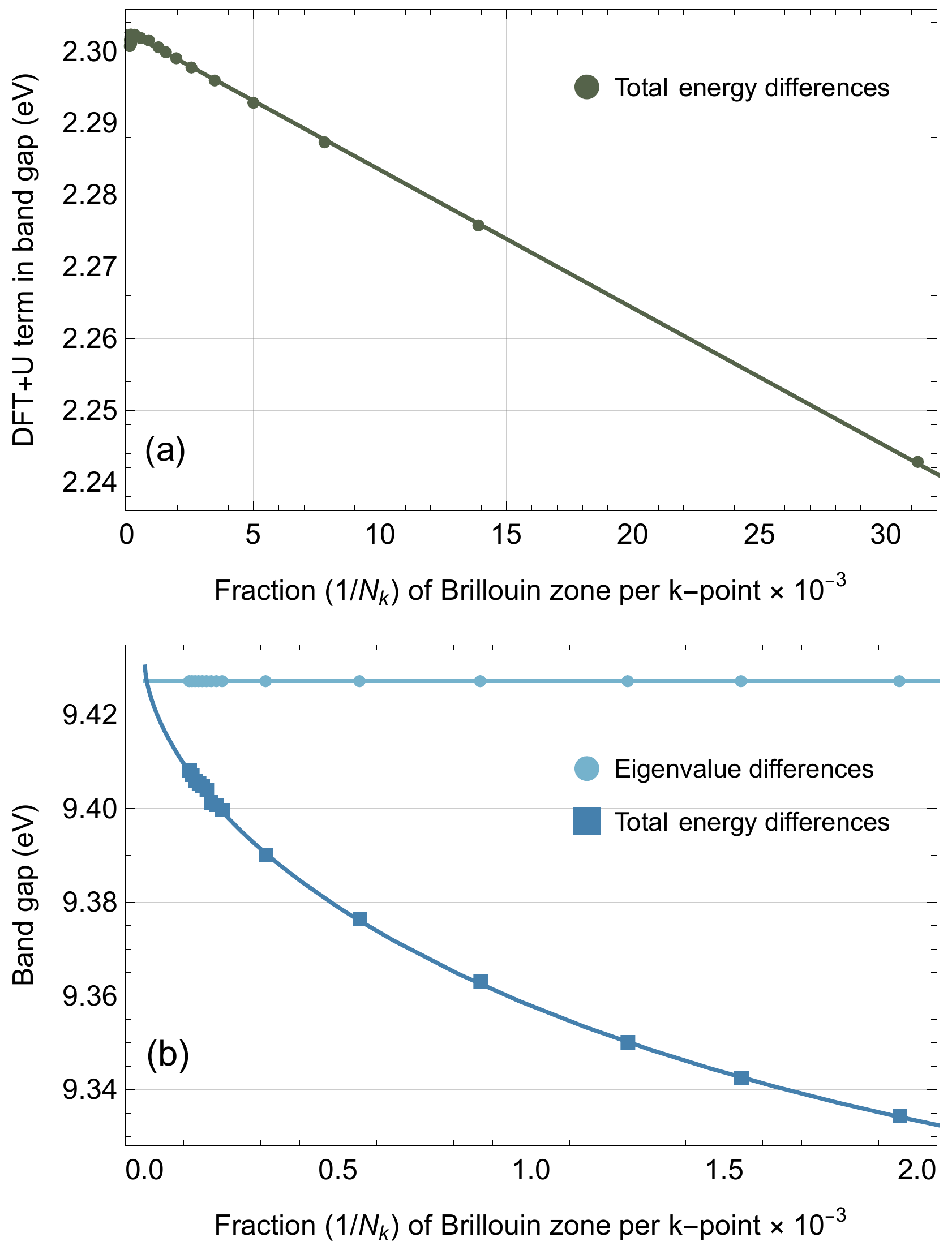}
\caption{\textcolor{black}{Panel (a), above, shows the dependence
of the explicit DFT+$U$ contribution
to the fundamental
band gap of an antiferromagnetic 
two-dimensional square hydrogen lattice, with respect to the inverse of the effective number
of two-atom unit cells that is implemented 
using Brillouin zone sampling.
DFT+$U$ projectors and the
Hubbard $U$ parameter are kept fixed.
The largest calculation, with data points furthest
to the left, corresponds to  
an effective supercell size
$ \gtrsim 1.7 \times
10^4$ atoms. 
In panel (b), below (zooming in 15$\times$ towards the converged limit), total band
gaps based on frontier
eigenvalue differences and total-energy differences are compared. 
Fitted curves, described in the
main text, suggest that  
near-identical values will be reached 
in the thermodynamic limit.}}
\label{fig:numerical}
\end{figure}

\textcolor{black}{In order to 
provide a minimal and motivating
test for  the numerical equivalence of the 
Dudarev et al.~\cite{dudarevElectronenergylossSpectraStructural1998}
functional DFT+$U$ 
fundamental gap and its own single-particle eigenvalue-based gap, 
we performed such calculations on an antiferromagnetic
two-dimensional hydrogen square 
lattice with a large lattice constant.
The Hubbard $U$ parameter and projectors were kept fixed.
Computational details are provided in 
Appendix~\ref{appendix:comp_details}
and main findings are shown in Fig.~\ref{fig:numerical}.
Using Quantum Espresso~\cite{giannozziAdvancedCapabilitiesMaterials2017}, 
we  reached a k-point sampling 
density equivalent to a super-cell of 
$ \gtrsim 1.7 \times 10^4$ atoms, i.e., 
not  at the 
thermodynamic limit required for exact
equality  of the gaps, but sufficiently close to make a reasonable extrapolation.}

\textcolor{black}{In this system, neither the valence nor conduction band
edge are degenerate, and no spontaneous
localization of surplus charges is permitted by 
the two-atom unit cell used.
We found that, with a sufficiently
small cold smearing parameter, an acceptable balance between
convergence and smearing errors could be found in the 
more numerically challenging 
regime of charged systems 
sampled with high k-point density.
Some evidence of remaining $\sim$~meV 
numerical errors are seen in the
top-left corner of Fig.~\ref{fig:numerical} (a), which shows the isolated
explicit DFT+$U$ energy contribution
to the fundamental gap
(i.e., based on DFT+$U$ energy differences).
Some of this noise cancels, 
however, when the other
terms in  the total gap are added.
Panel (a)  shows that the 
finite-size error of the DFT+$U$ 
gap contribution scales 
almost perfectly inversely with respect to the effective volume
of this system, i.e., like $N_k^{-1} $.}

\textcolor{black}{Moving to Fig.~\ref{fig:numerical} (b), 
we see that,
in this system, the bandgap as judged from the (commonplace)
GKS eigenvalue spectrum of the neutral lattice is almost
perfectly constant with respect to k-point density. 
This is because the band edges are included in the sampling
in all cases, and because the density contribution is
near-constant across the Brillouin zone in this
Mott-Hubbard system with very flat bands.
A constant line, with a value $E_{\rm g}^{\rm eigen}$, 
shows the numerically best available estimate of the
eigenvalue-based bandgap.}

\textcolor{black}{The  fundamental gap directly calculated using charged
systems (one electron added or removed per effective
super-cell), as we can see shown
with squares in panel (b), instead 
converges relatively
quite slowly with respect to k-point sampling. 
The power-law behavior of the finite-size error
in the energy-based fundamental gap is known to be rather 
involved~\cite{PhysRevB.78.125106}
when kinetic and XC  contributions
are included. The antiferromagnetic
order may further complicate the
finite-size exponents, and 
their detailed study is beyond the 
scope of the present work.
We include the points shown in a 
fit to this data,  while  excluding those from coarser samplings.
Supposing a $N_k^{-1} $ behavior
of the aforementioned term, and 
permitting a variable
exponent for a second one yields an
(electrostatics dominated) exponent 
very close to $-2/3$, 
echoing the early analysis of Ref.~\onlinecite{PhysRevB.18.3126}, 
noting that the precise value
and that of the intercept depend 
somewhat on the points included. 
As a guide to the eye, therefore,
we show the best fit of the form
$a  + b N_k^{-2/3} + c  N_k^{-1}$. 
This, for example, yields an intercept error of 
$a - E_{\rm g}^{\rm eigen} = 3$~meV.}

\textcolor{black}{We anticipate that the  extrapolated finite-size error
would vanish with yet more k-point data towards the 
thermodynamic limit, 
increasing 
out-of-plane cell dimension, 
and a more sophisticated,
multi-exponent extrapolation formula. 
Overall, this test strongly suggests that
the DFT+$U$ eigenvalue gap agrees with
the corresponding fundamental gap in the
thermodynamic limit.
It further suggests that the eigenvalue
gap converges much faster to this limit,
at least when the band-edge k-points 
are kept included in the sampling.
In a system with more band dispersion
we would expect this convergence to be
slower, of course, but for the eigenvalue
gap to still be by far the more efficient
estimator for the fundamental gap of
periodic systems within the GKS formalism.}

\section{The fundamental bandgap from DFT+U and hybrid functionals}
In this section, we 
provide the main result of this work.
We prove that
the fundamental band gap as
calculated from 
the Generalised Kohn Sham (GKS) eigenvalues~\cite{seidl1996generalized}
(that is to say, 
within commonplace practical DFT), 
in DFT+U and hybrid
functionals, is entirely valid for
pristine periodic systems.
This has previously been shown
for local and semi-local 
functionals and numerically for
conventional hybrid functionals~\cite{perdew2017understanding} 
and our proof includes such cases,
and indeed almost all functionals
in current use, while modifying
the logical argument somewhat.
By \emph{valid} here, we mean that this gap
\emph{exactly} matches what 
would be calculated from
explicit electron removal and addition
energies, for pristine periodic 
systems \textcolor{black}{that remain periodic under charging}.
This is not to say that the result
must agree well with experiment, \textcolor{black}{not by any
means,} but only
that it is a proper and well-founded 
procedure to predict a fundamental 
bandgap based on the gap in the 
conventional single-particle eigenvalues.

\textcolor{black}{As previously mentioned, but still very worth
acknowledging, this may at first seem} contrary to the received, 
often repeated
statement that only the highest occupied eigenvalue has
physical meaning in DFT. 
That statement is indeed true in the case of 
\textcolor{black}{exact DFT applied
generally (and also for 
approximate, yet  
exchange-correlation derivative 
discontinuity exhibiting, functionals 
when they are 
applied to well-converged
pristine bulk systems)  
within a strict 
Kohn-Sham DFT setting.}
\textcolor{black}{There, the density provides the 
functional dependence of the total energy.}

In the in-practice conventional formalism of 
GKS, it is the orbitals \textcolor{black}{but} not the \textcolor{black}{resulting} density, 
that provide the \textcolor{black}{functional dependence
for the total  energy}, 
and this is the key difference.
The KS or GKS distinction only becomes very relevant 
in the presence of a non-multiplicative potential,
\textcolor{black}{e.g., due to
its orbital dependence~\footnote{\textcolor{black}{Orbital-dependent} 
functionals can be applied within strict
Kohn-Sham DFT by means of the optimized 
effective potential (OEP) procedure. 
Studies involving OEP are 
very rare by comparison to those within
the overwhelmingly popular, even ubiquitous 
GKS framework.},}
as when the potential is purely local \textcolor{black}{and
orbital-independent} 
there is no practical distinction and crystalline 
\textcolor{black}{GKS} eigenvalue-based bandgaps
have already been shown to be valid in that  regime~\cite{perdew2017understanding}.

Our result implies that the lowest
unoccupied eigenvalue is as valid as the highest
occupied one, in the standard-practice of DFT+U.
While this is only the case
for defect-free periodic systems
(and only with infinitely
dense $\mathbf{k}$-point sampling, strictly speaking), 
we will also provide expressions
for the leading-order corrections
to the GKS (i.e., usual, familiar eigenspectrum) 
gap needed to recover the fundamental gap, e.g.,
for defective solids or molecules.
However, such corrections will typically be more  demanding 
to compute 
than direct explicit electron (along with their spin)  
removal and addition calculations for the fundamental gap.
In the following, we mirror the 
proof provided in Ref.~\onlinecite{perdew2017understanding}, 
however we instead invoke as our starting point  the 
underpinning total-energy construction
and logical procedure regarding its variation 
\textcolor{black}{following
the Ph.D. thesis of N. Poilvert,} 
Ref.~\onlinecite{poilvert-thesis}.

\subsection{Establishing the variationally optimized
constained energy}

We begin by considering the
total energy constructed using the 
Generalised Kohn Sham system of
$N$ electrons, with orbitals $\{\psi_i\}$
and their complex conjugates as variables.
We allow for possibly non-integer $N$ values, 
representing a statistical ensemble of integer-electron
count  states.
We suppress the spin-index for simplicity, as the extension
to spin-DFT is at least procedurally trivial
(as discussed in Appendix~\ref{appendix:spin}), 
and we  focus on zero temperature.
The real-valued orbital occupancies
$f^{i}$ will also be seen 
as variational variables.
These are defined, for possibly nonorthogonal 
$\{\psi_i\}$, as the matrix elements that
construct the generalized
Kohn-Sham density matrix
according to 
$\hat{\rho} = \sum_{i }  \lvert \psi_i 
\rangle f_{i}
\langle \psi_j  \rvert$,
noting that this only
 represents a valid  ensemble
 state when the orbitals are orthonormal.

The 
total energy, including orthonormality 
and orbital filling constraints, 
in atomic units
and with fixed external potential  $v$, 
reads
\begin{align}
\label{eqn:total_energy_including_orthonormality_and_orbital_filling} 
E_v &{}= - \sum_{i} f_{i}
\frac{ 
 \langle \psi_i \rvert  \nabla^2
\lvert \psi_i \rangle}{2}
+ E_{v\mathrm{Hxc}+U} 
\left[ \{\psi_i\},\{f_{i }\} \right]
\nonumber \\  
&{}\quad\quad+ \sum_{ij}
 \lambda_{i j }
\left( \delta_{j i} -  \langle \psi_j \rvert 
\psi_i \rangle 
 \right) 
  \\  \nonumber
&{}\quad\quad\quad + \epsilon_\mathrm{F} \left( N -  
 \sum_{i } f_{i }  \langle \psi_i \rvert 
\psi_i \rangle   \right).
\end{align}
Here $E_{v\mathrm{Hxc}+U}$ is the
approximated energy due to the Coulomb
interaction, the beyond-independent-particle
kinetic energy, and due to
the external potential. 
The potential contribution
represented as $U$ here could as well 
be an exact exchange term, for the purposes
of the present construction. 
The Lagrange multipliers
$\lambda_{i j}$, as well as their 
complex conjugates and the real-valued
Fermi energy
$\epsilon_\mathrm{F}$, a final Lagrange multiplier, 
are promoted as independent variables.

The energy $E_v$ is simultaneously 
extremized with respect to each
degree of freedom, which includes the Lagrange multipliers. 
First applying $\lvert 0 \rangle = \partial E_v / \partial 
\langle \psi_j \rvert$, noting that the derivatives
are partial, we find the generalized eigenequation 
\begin{align}
 \left(
- \frac{ 
  \nabla^2 }{2}
+  \hat{v}_{v\mathrm{Hxc}+U} - \epsilon_\mathrm{F} 
\right) 
\lvert \psi_j \rangle  f_{j } 
= \sum_{i}
\lvert 
\psi_i \rangle  \lambda_{i j } .
\label{eq:eigenmat}
\end{align}
Here, $\hat{v}_{v\mathrm{Hxc}+U}$  is
the functional 
derivative of  $E_{v\mathrm{Hxc}+U}$ 
with respect to the
density matrix $\hat{\rho}$.
Before going further, we can apply 
 $0 = \partial E_v /  \lambda_{i j}$, 
 from which we deduce that the
 $\delta_{j i} -  \langle \psi_j \rvert 
\psi_i \rangle = 0$, i.e., 
that the orbitals are orthonormalized at the minimum of $E_v$.
Applying next  
$\langle 0 \rvert = \partial E_v / \partial 
\lvert \psi_j \rangle$, we find a very similar equation, 
\begin{align}
 f_j  \langle \psi_j \rvert    \left(
- \frac{ 
  \nabla^2 }{2}
+  \hat{v}_{v\mathrm{Hxc}+U} - \epsilon_\mathrm{F} \right) 
= \sum_{i}
\lambda_{j i  } \langle  \psi_i \rvert  . 
\end{align}
Comparing this  with the complex conjugate of 
Eq.~\ref{eq:eigenmat}, 
we deduce that $ \lambda_{j i  }^\ast  = \lambda_{i j}$.
Here, we have invoked  
 the hermitian
property of $\hat{v}_{v\mathrm{Hxc}+U}$ for
the now assuredly valid 
(due to orthonormality)
GKS density matrix, noting
that while both DFT+$U$ and hybrid functionals introduce
an implicitly orbital dependent Hamiltonian, they do not
introduce an \emph{explicit} orbital dependence, which 
might jeopardize the self-adjoint property of the potential
\textcolor{black}{As an aside, it is preserved for KC functionals by 
their satisfaction of the Pederson 
condition~\cite{linscottKoopmansOpenSourcePackage2023,10.1063/1.446959}.}

Taking  $0 = \partial E_v /  \epsilon_\mathrm{F}$,
together with orbital orthonormality, provides simply that the $f_i$ sum to the correct electron count $N$.
To gain further insight into $\epsilon_\mathrm{F}$, we
must apply  the final condition available, 
$\partial E_v / \partial f_j = 0$. 
Technically, the $f_i$ may be parametrized in
terms of $\sin^2 \theta$ to ensure continuous variability and
hence differentiability.   
This yields, for all states $j$ with $0 < f_j < 1$, the
condition
\begin{align}
H^\textrm{GKS}_{jj} = \langle \psi_j \rvert \left(
- \frac{ 
  \nabla^2 }{2}
+  \hat{v}_{v\mathrm{Hxc}+U} 
\right) 
\lvert \psi_j \rangle  
=  \epsilon_\mathrm{F}.
\label{eq:epsmat}
\end{align}
Thus, $\epsilon_\mathrm{F}$ is indeed the Fermi level
as expected, i.e.,  
the degenerate GKS eigenvalue of $g$ partially filled states.

 Let us explore the generalized eigensystem further.
 We pre-multiply 
Eq.~\ref{eq:eigenmat} with $\langle \psi_k \rvert$
and trace, to find
\begin{equation}
\left( h^\textrm{GKS}_{k j} 
-  \epsilon_\mathrm{F}  \right) f_{j} = 
\sum_{i} \delta_{k i} \lambda_{i j} = \lambda_{k j}. 
\label{eq:lambdamat}
\end{equation}
Consider the \textcolor{black}{unitary
matrix $A$} that diagonalizes the matrix $\lambda$, such
that $\Lambda_{i} = A^\dagger_{i k}
\lambda_{k l } A_{l i}$.
The eigenvectors of $\hat{h}^\textrm{GKS}$
are the canonical GKS orbitals
\begin{equation}
\lvert\Psi_i \rangle = \sum_j 
\lvert\psi_j \rangle A_{j i}.
\end{equation}
Let us verify that unitary 
orbital transformations  
leave
$E_v$ unchanged. 
For the kinetic energy, for example, 
\begin{equation}
\sum_i f_i \frac{ \langle \Psi_i \rvert  \nabla^2
\lvert \Psi_i \rangle }{2} = \sum_i
f_i \frac{ \langle \psi_k \rvert  \nabla^2
\lvert \psi_l \rangle A_{l i } A^\dagger_{i k } }{ 2}, 
\end{equation}
which is as in the original $E_v$ by the property
of unitary matrices. 
Similarly, it is straightforward to show that
all other terms are unitary invariant, 
including the energy
$E_{v\textrm{Hxc+U}}$, 
as this is comprised
of density (and possibly its spatial derivatives)
dependent terms, where the density is
unitary invariant, or density-matrix
dependent terms, as in the DFT+$U$ and/or
exact exchange energies, which can also be readily
shown to be invariant under unitary transformations
of the GKS orbitals. 

The matrix $A$ may be expected to change
as the total electron count $N$ 
(an external parameter) is varied, 
as it shortly will be,
and if
$A$ couples states of different occupancy this would considerably
complicate the analysis.
It is necessary therefore to verify whether the $f_i$
may validly be considered to be independent of $A$,
e.g., for constructing the density matrix $\hat{\rho}$.
Let us again consider, for this, the expression
$\lambda^\ast_{ji} = \lambda_{ij}$.
At the minimum of $E_v$, we  have from Eq.~\ref{eq:lambdamat}, 
therefore 
\begin{equation}
\left( h^{\textrm{GKS}\ast}_{ji} - \epsilon_\textrm{F}\right) f_i = 
\left( h^{\textrm{GKS}}_{ij} - \epsilon_\textrm{F}\right) f_j ,
\end{equation}
but noting that $\hat{h}^\textrm{GKS}$ itself is hermitian
this implies that either $f_i = f_j$ or 
$h^{\textrm{GKS}}_{ij}=0$ for any pair of 
differing states $i$ and $j$.
Thus, somewhat remarkably, the GKS Hamiltonian is block-diagonal, with
no coupling between fully filled, partially filled, or empty states. 
Without loss of generality, then, the unitary matrix $A$ and hence
the canonical eigensystem retains the same block-diagonal structure.
Even if there were multiple different non-integer $f_i$ considered,
it would not be necessary to consider the variation of the degeneracy pattern of
the $f_i$
with external parameters unless the parameters vary sufficiently
to cause crossing of states and a re-arrangement of the block-diagonal
Hamiltonian structure. For sufficiently small variations 
of the external parameters
(as we will see, adding or subtracting
is a sufficiently small change in crystals), 
we conclude that the 
degeneracy grouping of the  $f_i$ remains constant as the
$\epsilon_i$ and $\Psi_i$ change. 

Referring back, as final preliminary observation
with the latter result in hand,  to  Eq.~\ref{eq:lambdamat}, 
and considering the case $f_i = 1$, 
we find that the unitary  transformation
also diagonalizes $h^\textrm{GKS}_{k j}$.
This,  in terms of the  eigenvalues
$\epsilon_i$ 
of the GKS Hamiltonian for fully occupied eigenvalues,
we have 
$ \epsilon_i - \epsilon_\mathrm{F} = \Lambda_{i}$.
For the case $f_i = 0$, i.e., unoccupied
orbitals, we deduce that $\Lambda_{i} = 0$.
For the case of frontier orbitals, with $0 < f_i < 1$, 
let us suppose that there are
$g$ such degenerate GKS eigenstates with the
same $f_i$, so that
$g f_i  = N - \lfloor N \rfloor$.  
This gives
$\left( \epsilon_i - \epsilon_\mathrm{F} \right)
\left( N - \lfloor N \rfloor \right) / g = \Lambda_i$.
However, we already know that for such orbitals 
$H^\textrm{GKS}_{ii} = \epsilon_\mathrm{F}$, 
and so also following diagonalization of this
isolated Hamiltonian block 
(noting that rotation preserves the trace)
we must have
$\epsilon_i = \epsilon_\mathrm{F}$ and hence also in this 
case $\Lambda_i = 0$.

\subsection{Energy variations with respect to
total electron count}

We are now ready to consider the rate of change 
of the minimized total energy $E_v$ with respect to the total electron count $N$, having set the problem up somewhat differently to 
Ref.~\onlinecite{perdew2017understanding}
and instead as per Ref.~\onlinecite{poilvert-thesis}. This derivative is 
necessarily a total one, however for all $N$ we note that 
$E_v$ is minimized with respect to its 
 degrees of freedom
$\left\lbrace \psi_i \right\rbrace$, $\left\lbrace \psi_i^\ast  \right\rbrace$, 
$f_i$, $\lambda_{ij}$, $\lambda^{\ast}_{ij}$,  
and $\epsilon_\mathrm{F}$, and so the partial derivatives of
$E_v$ with respect to these vanishes. 
We therefore have, when evaluated at all but points of 
eigenvalue cross-over or degeneracy change of the frontier orbitals, 
the great simplification 
\begin{equation}
\frac{dE_v}{dN} = \frac{\partial E_v}{\partial N} 
= \epsilon_\mathrm{F}.
    \label{eqn:slope_of_tot_energy}
\end{equation}
This holds, in principle, even for infinite degeneracy $g$ as in a solid.
This result is the generalization of Janak's 
theorem to GKS in the presence
of hermitian, possibly density-matrix dependent non-local potentials
as in DFT+$U$ and hybrid functionals, with possibly degenerate
frontier orbitals. 
From Eq.~\ref{eqn:slope_of_tot_energy}
and under the same conditions, it follows that the 
2$^{\rm nd}$ 
derivative of the total energy with respect to total electron count may be expressed as
\begin{equation}
\label{eqn:2nd-deriv}
\frac{d^2 E_v}{dN^2} = \frac{d  \epsilon_\mathrm{F}}{dN}
= \sum_i \frac{d f_i}{d N}\frac{d  \epsilon_\mathrm{F}}{df_i }
= g \frac{1}{g}\frac{d  \epsilon_\mathrm{F}}{df_i }= 
\frac{d  \epsilon_\mathrm{F}}{df_i },
\end{equation}
where the sum runs over all frontier (i.e., partially filled) orbitals.
The  k$^{\rm th}$ derivative is similarly 
\begin{equation}
\label{eqn:kth-deriv}
\frac{d^k E_v}{dN^k} =  
\frac{d^{k-1}  \epsilon_\mathrm{F}}{df_i^{k-1}}.
\end{equation}

Conventional DFT$+U$ and Hybrid functionals are explicit, 
\textcolor{black}{infinitely} 
differentiable functionals of the first-order reference-system reduced density matrix and as such will vary continuously with respect to 
the $g$ frontier GKS orbital occupancies, 
and hence with respect to $N$. Therefore, one may employ a Taylor expansion of the total energy functional, for example,
around an integer electron count $N_0$ to yield the following expression for the total energy of the $N_0 +1$ electron system,
\begin{equation}
\label{eqn:EA-taylor-expansion}
E(N_0 + 1 )=E(N_0)+\sum_{k=1}^{\infty}\frac{1}{k!}\left.\frac{d^kE_v}{d N^k}\right|_{N = N_0}(1-0)^k.
\end{equation}

Combining Eqs.~\ref{eqn:kth-deriv} \& \ref{eqn:EA-taylor-expansion} with some rearranging yields the following expression for the electron affinity,
\begin{align}
 {\rm EA}&{}\equiv
 E(N_0  ) - E(N_0 + 1) \nonumber \\
&{}= - \epsilon_{\rm F}-\sum_{k=2}^{\infty}\frac{1}{k!}
 \left. \frac{d^{k-1}\epsilon_{\rm F}}{df_{\rm F}^{k-1}} 
 \right|_{f_{\rm F} = 0}
 \nonumber \\
 &{}= - \epsilon_{\rm L}-\sum_{k=1}^{\infty}\frac{1}{k+1!}
 \left. \frac{d^{k}\epsilon_{\rm L}}{df_{\rm L}^{k}} 
 \right|_{f_{\rm L} = 0},
    \label{eqn:general_ea_expression}
\end{align}
where $\epsilon_{\rm L}$ is the 
possibly-degenerate 
lowest unoccupied
GKS eigenvalue of the $N_0$ electron system,
and $f_{\rm L}$ is its occupancy.
Similarly, for the ionization potential, we can deploy a Taylor expansion considering electron removal, of the form 
\begin{equation}
\label{eqn:IP-taylor-expansion}
E(N_0 - 1 )=E(N_0)+\sum_{k=1}^{\infty}\frac{1}{k!}\left.\frac{d^kE_v}{d N^k}\right|_{N = N_0}(0-1)^k.
\end{equation}
Combining Eqs.~\ref{eqn:kth-deriv} \& \ref{eqn:IP-taylor-expansion} with some rearranging yields the following expression for the ionisation potential,
\begin{align}
 {\rm IP}&{}\equiv
\textcolor{black}{ E(N_0 - 1 ) - E(N_0 ) }
 \nonumber \\
 &{}=-\epsilon_{\rm F}-\sum_{k=2}^{\infty}\frac{\left(-1\right)^{k-1}}{k!} \left. \frac{d^{k-1}\epsilon_{\rm F}}{df_{\rm F}^{k-1}}  \right|_{f_{\rm F} = 1}
 \nonumber \\
 &{}=-\epsilon_{\rm H}-\sum_{k=1}^{\infty}\frac{\left(-1\right)^{k}}{k+1!} \left. \frac{d^{k}\epsilon_{\rm H}}{df_{\rm H}^{k}}  \right|_{f_{\rm H} = 1},
    \label{eqn:general_ip_expression}
\end{align}
where $\epsilon_{\rm H}$ is the 
possibly-degenerate 
(with not necessarily the same degeneracy factor as 
$\epsilon_{\rm L}$)
highest occupied
GKS eigenvalue of the $N_0$ electron system,
and $f_{\rm L}$ is its occupancy.  Finally, the fundamental bandgap is given as the difference between the electron affinity and ionisation potential,  and so may be expressed as
\begin{align}
\label{eqn:general_bandgap_expression}
E_{\rm g}\equiv& \;{\rm IP} -  {\rm EA} = \epsilon_{\rm L}-\epsilon_{\rm H}
 \\ \nonumber &+\sum_{k=1}^{\infty}\frac{1}{k+1!}\left(\left. \frac{d^k \epsilon_{\rm L}}{df_{\rm L}^k} \right|_{f_{\rm L} = 0} - (-1)^{k} \left. \frac{d^k \epsilon_{\rm H}}{df_{\rm H}^k} \right|_{f_{\rm H} = 1} \right).
\end{align}

This result accounts for the possibly different
degeneracy of the lowest unoccupied and highest
occupied GKS levels of an $N_0$-electron
insulator, the degeneracy factors having already
canceled out in the expression. 
The result only holds straightforwardly if
the there is no degeneracy 
breaking or \textcolor{black}{discrete} orbital
re-ordering of the lowest unoccupied GKS level
as an electron is continuously added
(or, separately, of the highest occupied
GKS level as an electron is removed), 
however this is assured in an infinite
pristine solid \textcolor{black}{with no spontaneous charge localization}, as we go on to show in
Section~\ref{sec:eigenvalue}.

First, however, we
note that for many commonplace
functionals in many systems,
the quadratic term 
in the total energy 
overwhelmingly 
dominates~\cite{haitDelocalizationErrorsDensity2018}.
In the case we can write
\begin{equation}
E_{\rm g}^{\rm quadratic} =  
\epsilon_{\rm L}-\epsilon_{\rm H}
+ \frac{1}{2}
\left( \frac{d \epsilon_L}{d f_L}
+ \frac{d \epsilon_H}{d f_H}
\right),
\end{equation}
albeit this is no easier to calculate
than the energies of explicit electron
addition and removal. 
For the case instead where 
a functional exhibits
precise piecewise linearity with
respect to electron count,
which the exact 
functional should~\cite{perdewDensityFunctionalTheoryFractional1982,yangDegenerateGroundStates2000,ayersDependenceContinuityEnergy2008},
then both $\epsilon_{\rm L}$ and 
$\epsilon_{\rm H}$ remain constant
over the  electron count intervals
for which they serve, respectively, as
the Fermi level.
Then, all the derivative terms vanish in 
Eq.~\ref{eqn:general_bandgap_expression}.
In such a GKS approximation,
therefore, an analogue of Koopmans'
theorem holds both for 
electron removal and attachment
and we simply have 
for the fundamental gap the result
$E^\textrm{linear}_{\rm g} = 
\epsilon^\textrm{linear}_{\rm L}-
\epsilon^\textrm{linear}_{\rm H}$, 
for any system that it is applied to,
not just pristine solids. 
We caution, however, 
that piecewise linearity
does not imply that the 
exact functional is at hand, 
and in the exact functional there
will be an explicit derivative
discontinuity~\cite{mori-sanchezDiscontinuousNatureExchangeCorrelation2009} 
as unlike conventional DFT+$U$ or hybrid functionals, the exact functional is not an explicit, differentiable functional of the first-order
reference-system reduced density matrix.

The many-body generalization of the Burgess Linscott O'Regan (mBLOR) DFT+$U$
functional~\cite{burgessFlatPlaneBased2024} introduces such an explicit derivative discontinuity, while conventional DFT+$U$ or hybrid functionals exhibit only an implicit derivative discontinuity. 
Unlike implicit derivative discontinuities, such 
explicit contributions
are not incorporated within the
GKS eigenspectrum gap~\cite{yangDerivativeDiscontinuityBandgap2012}, 
i.e., do not appear in Eq.~\ref{eqn:general_bandgap_expression}, and require
explicit addition to $E_{\rm g}$. The need for the exact functional to exhibit such an explicit derivative discontinuity  is discussed in Appendix~\ref{appendix:explicitderiv}).

If DFT+$U$ brings an approximate
functional closer to agreement
with global piecewise linearity, 
we can suppose that DFT+$U$ makes it
a better approximation to use  
$E_{\rm g} \approx 
\epsilon_{\rm L}-\epsilon_{\rm H}$
in defective solids and molecules, 
but it will still be approximate.
Even with the most sophisticated
DFT+$U$ correction or hybrid functional, 
the energy and its
derivatives are not exact, 
and beyond-quadratic, and hence
we can expect non-vanishing 
non-cancelling derivative
contributions in 
Eq.~\ref{eqn:general_bandgap_expression}.
As a result, we might expect
the GKS frontier 
eigenvalue difference 
$\epsilon_{\rm L}-\epsilon_{\rm H}$, 
i.e., the commonplace `gap' extracted
from the DFT+$U$ density of states, 
to be an invalid value for  
the fundamental bandgap. As we will
 see in the next Section, 
 however, its validity is 
perfectly recovered in
\textcolor{black}{extended systems
under the pre-conditions  oft-repeated here.} 

This does not mean
to imply, of course, that the
fundamental bandgap calculated  from
$\epsilon_{\rm L}-\epsilon_{\rm H}$, 
or further corrections to this per
Eq.~\ref{eqn:general_bandgap_expression} 
as may be necessary, 
will bear a good resemblance
to the experimental fundamental gap 
(even leaving aside zero-point phonon
renormalization effects, etc.). 
It only implies that such
a quantity  is valid, 
i.e., gives the same result as explicit
gap calculation by electron addition and
removal.
In particular,
if 
$E\left[ N \right]$ lacks or severely underestimates the derivative
discontinuity at $N_0$, as it does so
typically with  LDA and GGA functionals,
then the predicted fundamental gap will
be zero or severely underestimated at best.
It is part of the success of DFT+$U$
and hybrids that they introduce such
a derivative discontinuity, but 
we find here that their
introduction (because it is an implicit
and not an explicit derivative 
discontinuity) does not affect 
the validity of the GKS 
eigenvalue based bandgap in principle. 

\section{Frontier GKS Eigenvalue Derivatives}
\label{sec:eigenvalue}
To understand why the GKS eigenvalue derivatives in Eq.~\ref{eqn:general_bandgap_expression}  do not contribute to the fundamental bandgap in the case of defect-free
infinitely extended systems, we now derive an explicit expression for  
the first-order derivative of the frontier GKS eigenvalue $\epsilon_\textrm{F}$ with respect to its orbital occupancy $f_\textrm{F}$. Here, F can refer either to the highest occupied (H) or lowest unoccupied (L) GKS orbital,
and it may be degenerate. Due to the orthogonality of 
wavefunctions with their derivatives, 
the derivative of the frontier GKS eigenvalue 
with respect to 
the frontier orbital 
occupancy simplifies to 
\begin{equation}
    \frac{d\epsilon_\textrm{F}}{df_\textrm{F}}=\left\langle\Psi_\textrm{F}\left|\frac{d\hat{H}^{\rm GKS}}{df_\textrm{F}}\right|\Psi_\textrm{F}\right\rangle.
    \label{eqn:Hellmann–Feynman}
\end{equation}
Here, total derivatives imply
that all variational parameters are
permitted to vary as needed
to extremize the total energy.
The GKS Hamiltonian operator in real space is given by 
\begin{equation}
{H}^{\rm GKS}({\bf r},{\bf r'})=-\frac{1}{2}\nabla^2+v_{\rm ext}({\bf r})+v_{\rm Hxc}({\bf r})+v_{ U}({\bf r},{\bf r'}),
\label{eqn:GKS-hamiltonain}
\end{equation}
where $v_{\rm ext}({\bf r})$ is the external potential, which
is kept fixed throughout this work. 
Additionally, $v_{\rm Hxc}({\bf r})$ is the Hartree plus base exchange-correlation potential of local or semi-local character and $v_{U}({\bf r},{\bf r'})$ is the DFT+U(+J) potential. Aside from the latter non-local  potential, the other two terms in the GKS potential include Dirac delta functions $\delta({\bf r}-{\bf r'})$, which have been suppressed in Eq.~\ref{eqn:GKS-hamiltonain} for brevity. The reference system kinetic energy operator and external potential exhibit no dependence on the GKS orbital occupancies. The Hartree plus  exchange-correlation potential will contribute to the derivative via its explicit dependence on the 
ground-state electron-density, $\rho_0$,  via
\begin{align}
    \label{eqn:hartree_plus_bxc}
    \frac{d v_{\rm Hxc} ({\bf r})}{df_\textrm{F}} =  \int d{\bf r'}\left.\frac{\delta  v_{\rm Hxc} ({\bf r})}{\delta\rho({\bf r'})}\right|_{\rho_{0}}\left(\frac{d \rho_{0}({\bf r'})}{d f_\textrm{F}}\right). 
\end{align}
The first term on the right hand side is the Hartree plus base exchange correlation kernel. In the case of the highest occupied or lowest unoccupied GKS orbital, i.e., $f_\textrm{F}=f_{\rm H}$ or $f_{\rm L}$, the final term in Eq.~\ref{eqn:hartree_plus_bxc} is the right or left hand Fukui function $F_{\pm}({\bf r'})$~\cite{parr1984density,ayers2000perspective,parrDensityFunctionalTheoryAtoms,yang2012analytical,ayers2007computing,wang2024extending}, which is a  quantity that incorporates
screening effects.
Therefore, Eq.~\ref{eqn:hartree_plus_bxc} simplifies to 
\begin{align}
\label{eqn:Hartree_plus_bxc_final_result} 
     \frac{d v_{\rm Hxc} ({\bf r})}{df_\textrm{F}}=  \int d{\bf r'}f_{\rm Hxc}({\bf r},{\bf r'})F_{\pm}({\bf r'}), 
\end{align}
where $f_{\rm Hxc}({\bf r},{\bf r'})$ is the Hxc kernel. 
In the frozen orbital approximation, the Fukui function is given by the frontier GKS orbital density, however we do not make such an approximation. The Hxc kernel can be decomposed into a sum of Hartree and exchange-correlation components, where the Hartree kernel is given by the usual expression
\begin{equation}
f_{\rm H}({\bf r},{\bf r'})=\frac{1}{|{\bf r}-{\bf r'}|},
\end{equation}
while the base exchange correlation kernel depends of course on the choice of exchange correlation functional. 
For the Hubbard potential, we invoke here 
as a representative case 
the DFT+$U$ functional of Dudarev et al.~\cite{dudarevElectronenergylossSpectraStructural1998}, although we emphasize that the same proof holds with 
\textcolor{black}{caveats (that are always satisfied or
overcome in practice) for 
all known} DFT+$U$ 
functionals, 
a point which will be elaborated upon further in 
Section~\ref{sec:verificationtable}. 
For concision, we denote the effective
Hubbard parameter simply as $U$ in this
section.
In real space within the spin-restricted formalism, the   Hubbard potential of Dudarev et al. may be expressed as 
\begin{align}
 \label{eqn:hubbard_potential}
    v_{U}^{\sigma}({\bf r},{\bf r'})&=  \sum_I\frac{U^I}{2}\left\{\braket{{\bf r}|\hat{P}^I|{\bf r'}} \right. \\ &  \left.-\int d{\bf r''}d{\bf r'''}\braket{{\bf r}|\hat{P}^I|{\bf r''}}\rho({\bf r''},{\bf r'''})\braket{{\bf r'''}|\hat{P}^I|{\bf r'}}\right\}. \nonumber
\end{align}
The derivative of the Hubbard potential with respect to frontier GKS orbital occupancy is given by 
\begin{align}
    \frac{d v_{U}({\bf r},{\bf r'}) }{df_\textrm{F}}&{}=    \int d{\bf r''}d{\bf r'''}\left.\frac{\delta v_{U}({\bf r},{\bf r'}) }{\delta\rho({\bf r''},{\bf r'''})}\right|_{\rho_{0}}
  \nonumber \\ &{}\quad \times
    \left(
    \frac{d \rho_{0}({\bf r''},{\bf r'''})}{d f_\textrm{F}}\right).
    \label{eqn:hubbard_derivative}
\end{align}
The latter derivative in Eq~\ref{eqn:hubbard_derivative} is the right or left hand,  Fukui matrix $F_{\pm}({\bf r''},{\bf r'''})$~\cite{zhang2015orbital,bultinck2011fukui,peng2013fukui,wang2024slope},
i.e., the screened rate of
change of the ground-state 
(KS or, here, GKS) density
matrix. This quantity is sometimes
termed a `non-interacting' Fukui
matrix since it 
applies to the Kohn-Sham
reference density matrix. This term allows it to be distinguished from its counterpart in first-order reduced density matrix functional theory~\cite{bultinck2012influence,acke2023extending}, 
but given both the GKS and DFT+$U$
context, and the risk that this might
misleadingly 
imply `unscreened', we do not use
this term.
The functional derivative of the Hubbard potential can be explicitly calculated from Eq.~\ref{eqn:hubbard_potential},  to yield 
\begin{align}
  \label{eqn:hubbard_final_result}
    &\frac{d v_{U}({\bf r},{\bf r'}) }{df_\textrm{F}}= 
    -\sum_I \frac{U^I}{2} \int d{\bf r''}d{\bf r'''} \\ & \quad \times \left(\braket{{\bf r}|\hat{P}^I|{\bf r''}}\braket{{\bf r'''}|\hat{P}^I|{\bf r'}}\right)F_{\pm}({\bf r''},{\bf r'''}).
    \nonumber
\end{align}
Combining Eqs.~\ref{eqn:Hartree_plus_bxc_final_result} \& \ref{eqn:hubbard_final_result} with Eq.~\ref{eqn:Hellmann–Feynman} and given that the diagonal element of the  Fukui matrix is equal to the Fukui function~\footnote{The ground-state electron density is equal to the diagonal elements of the ground-state first-order reference system reduced density matrix and as such their left and right hand derivatives with respect to electron count are equal so that $F_{\pm}({\bf r},{\bf r})=F_{\pm}({\bf r})$.}, we have 
\begin{widetext}
\begin{align}
    \frac{d\epsilon_{\rm F}}{df_{\rm F}}=\left\langle\Psi_{\rm F}\right| \int d{\bf r}d{\bf r'} \ket{{\bf r}}\Bigg\{&\int d{\bf r''}f_{\rm Hxc}({\bf r},{\bf r''})F_{\pm}({\bf r''},{\bf r''})\delta({\bf r}-{\bf r'}) \nonumber \\ & 
    -\sum_I\frac{U^I}{2}\int d{\bf r''}d{\bf r'''}\left(\braket{{\bf r}|\hat{P}^I|{\bf r''}}\braket{{\bf r'''}|\hat{P}^I|{\bf r'}}\right)F_{\pm}({\bf r''},{\bf r'''})\Bigg\}\braket{{\bf r'}|\Psi_{\rm F}}.
    \label{eqn:GKS-eigenvalue-deriv-final}
\end{align}
\end{widetext}
\textcolor{black}{The Hxc contribution to the eigenvalue derivative consists of three integrals over real-space, one of which is countered by the Dirac delta function $\delta({\bf r}-{\bf r'})$, which is zero for all $|{\bf r}-{\bf r'}|>0$. Assuming the added electron or hole delocalizes over the bulk system, the frontier orbital density $\rho_{\rm F}({\bf r})=\braket{{\bf r}|\Psi_{\rm F}}\braket{\Psi_{\rm F}|{\bf r}}$ will tend to zero at any given point in space. Likewise the Fukui function $F^{\pm}({\bf r''},{\bf r''})$, which is defined as the derivative of the total electron density at the point ${\bf r''}$ with respect to total electron count, will tend to zero as the addition of a single electron or hole to the system will offer only an infinitesimal change in the total electron density at any point in space. The vanishing frontier orbital density and Fukui function counter the remaining two spatial integrals. This, combined with the Hxc kernel $f_{\rm Hxc}({\bf r},{\bf r''})$ vanishing in the limit of
large $|{\bf r}-{\bf r''}|$, ensures that the Hxc contribution to the eigenvalue derivative vanishes in the bulk limit (excluding cases of spontaneous 
localization of added electrons or holes, 
a caveat that we hereafter assume is satisfied).}

\textcolor{black}{In the case of the Hubbard contribution, there are four spatial integrals and a summation over the Hubbard sites. The $\braket{{\bf r}|\hat{P}^I|{\bf r''}}\braket{{\bf r'''}|\hat{P}^I|{\bf r'}}$ term offers a negligible contribution for large values of $|{\bf R}_I-{\bf r}|$, where ${\bf R}_I$ is the position of the atomic site to which the localized projection operator $\hat{P}^I$ is assigned, same too goes for large values of $|{\bf R}_I-{\bf r'}|$, $|{\bf R}_I-{\bf r''}|$ and $|{\bf R}_I-{\bf r'''}|$.  This counters the four spatial integrals. Analogous to the frontier orbital density and Fukui function, the frontier orbital density matrix $\rho_{\rm F}({\bf r'},{\bf r})=\braket{{\bf r'}|\Psi_{\rm F}}\braket{\Psi_{\rm F}|{\bf r}}$ and  Fukui matrix $F^{\pm}({\bf r''},{\bf r'''})$ will also tend to zero in the bulk limit. This ensures that like the Hxc contribution, the Hubbard contribution to the eigenvalue derivative vanishes in the bulk limit.}

The same argument can also be applied to all higher-order derivatives of the frontier GKS eigenvalue with respect to GKS occupancy as these may also be expressed as integrals over real-space of the Hxc kernel \& $\braket{{\bf r}|\hat{P}^I|{\bf r''}}\braket{{\bf r'''}|\hat{P}^I|{\bf r'}}$ Hubbard term, the frontier orbital density \& frontier orbital density matrix, the Fukui function \&  Fukui matrix and derivatives thereof.  Take for example the second derivative of the frontier GKS eigenvalue with respect to GKS occupancy,
\begin{widetext}
\begin{align}
    \frac{d^2\epsilon_{\rm F}}{df_{\rm F}^2}=&\int d{\bf r}d{\bf r'}\left(\rho'^{\pm}_{\rm F}({\bf r})F^{\pm}({\bf r'})+\rho_{\rm F}({\bf r})F'_{\pm}({\bf r'})\right)f_{\rm Hxc}({\bf r},{\bf r'})+\int d{\bf r}d{\bf r'}d{\bf r''}\rho_{\rm F}({\bf r})F^{\pm}({\bf r'})f_{\rm xc}'({\bf r},{\bf r'},{\bf r''})F^{\pm}({\bf r''}) \nonumber \\ &  -\sum_I\frac{U^I}{2}\int d{\bf r}d{\bf r'}d{\bf r''}d{\bf r'''}\left(\rho'^{\pm}_{\rm F}({\bf r'},{\bf r}) F^{\pm}({\bf r''},{\bf r'''})+\rho_{\rm F}({\bf r'},{\bf r}) F_{\pm}'({\bf r''},{\bf r'''})\right) \braket{{\bf r}|\hat{P}^I|{\bf r''}}\braket{{\bf 
    r'''}|\hat{P}^I|{\bf r'}} ,
    \label{eqn:GKS-eigenvalue-2nd-deriv}
\end{align}
\end{widetext}
where derivatives are denoted by a prime mark. Using the same arguments as before, each term \textcolor{black}{will again vanish in the bulk limit}. It is worth emphasizing that the exact same argument of the frontier GKS eigenvalue derivatives presented here for DFT+$U$-type functionals also applies to hybrid functionals, 
\textcolor{black}{and of course (if $U$=0) the underlying base local or semi-local functionals}. 
By way of example, in the case of global hybrid functionals the first derivative of the GKS eigenvalue with respect to frontier occupancy in the spin-restricted formalism will instead be given by
\begin{widetext}
\begin{align}
    \frac{d\epsilon_{\rm F}}{df_{\rm F}}=\left\langle\Psi_{\rm F}\right| \int d{\bf r}d{\bf r'} \ket{{\bf r}}\Bigg\{\int \Big(f_{\rm Hc}({\bf r},{\bf r''})+(1-a_{\rm HF})f_{\rm x}({\bf r},{\bf r''})\Big)F_{\pm}({\bf r''},{\bf r''})\delta({\bf r}-{\bf r'})  d{\bf r''} 
    -\frac{a_{\rm HF}}{2}\frac{F_{\pm}({\bf r},{\bf r'})}{|{\bf r}-{\bf r'}|}\Bigg\}\braket{{\bf r'}|\Psi_{\rm F}},
    \label{eqn:GKS-hybrid}
\end{align}
\end{widetext}
where $a_{\rm HF}$ is the fraction of exact exchange, $f_{\rm Hc}({\bf r},{\bf r''})$ is the Hartree plus (semi-)local correlation kernel and $f_{\rm x}({\bf r},{\bf r''})$ is the (semi-)local exchange kernel, which is diagonal in spatial coordinates. The exact exchange contribution \textcolor{black}{consists of} two spatial integrals. However, again there will be a negligible contribution for large values of $|{\bf r}-{\bf r'}|$ and the frontier orbital density matrix and  Fukui matrix both \textcolor{black}{tend to zero}. Thus like all other terms, the exact exchange contribution to the GKS eigenvalue derivative \textcolor{black}{vanishes in the bulk limit}. 

Therefore, in the specific case of an infinite pristine solid where the added electron or hole delocalises over the bulk system, the first and all higher-order derivatives of the frontier GKS eigenvalues with respect to orbital occupancy can be neglected and Eqs.~\ref{eqn:general_ea_expression}, \ref{eqn:general_ip_expression} \& \ref{eqn:general_bandgap_expression} reduce to the simple expressions,
\begin{align}
\label{eq:ip_pristine_solid}
  &{\rm IP}=-\epsilon_{\rm H}, \\
 \label{eq:ea_pristine_solid} 
  &{\rm EA}=-\epsilon_{\rm L}, \\
\label{eq:bandgap_pristine_solid}
  &E_{\rm g}=\epsilon_{\rm L}-\epsilon_{\rm H}.
\end{align}
In the case of a conventional DFT$+U$ or hybrid functional, Eq.~\ref{eq:bandgap_pristine_solid} demonstrates that the frontier GKS eigenvalue difference is a rigorous and formally exact approach to evaluating the fundamental bandgap of a pristine solid. Furthermore, from Eqs.~\ref{eq:ip_pristine_solid} \& \ref{eq:ea_pristine_solid}, the negative of the highest occupied and lowest unoccupied GKS eigenvalues are exactly equal to the ionization potential and electron affinity of the bulk system. Eq.~\ref{eq:ip_pristine_solid} is equivalent in form to the ionization potential theorem, also known as the DFT Koopmans' theorem, an exact condition in DFT which the above derivation demonstrates must always be satisfied by conventional DFT+$U$ and hybrid functionals for pristine solids. Perhaps more surprisingly, no derivative discontinuity term appears in Eq.~\ref{eq:ea_pristine_solid}. This is a direct result of conventional DFT$+U$ and hybrid functionals being explicit, differentiable functionals of the first-order reference-system reduced density matrix. Despite Eqs.~\ref{eq:ip_pristine_solid}, \ref{eq:ea_pristine_solid} \& \ref{eq:bandgap_pristine_solid} being formally exact at this level theory, it does not imply that such functionals will yield quantitatively accurate predictions. Rather, for a given functional, equating the negative of the frontier GKS eigenvalues to the functional's predicted ionization potential and electron affinity for the system is a correct and formally exact procedure at this level of theory.

However, we emphasize that this proof only holds in the case of pristine periodic systems \textcolor{black}{where the added electron or hole delocalises over the
bulk system}. For finite electronic systems,  \textcolor{black}{non-pristine solids, or systems with spontaneous charge carrier localization,} one can no longer assume that the added electron or hole delocalises over the bulk system. By extension, one cannot assume that the frontier orbital density $\rho_{\rm F}({\bf r})=\braket{{\bf r}|\Psi_{\rm F}}\braket{\Psi_{\rm F}|{\bf r}}$ will tend to zero at every point in space. Same too goes for the frontier orbital density matrix, the Fukui function and the  Fukui matrix. Thus, the frontier eigenvalue derivatives of Eqs.~\ref{eqn:general_ea_expression}, \ref{eqn:general_ip_expression} \& \ref{eqn:general_bandgap_expression} must be accounted for and the frontier GKS eigenvalue difference is not a valid measure of the fundamental bandgap of such systems.

\textcolor{black}{That said, for defective crystals with  low lowest-energy 
singly-ionized defect binding
energies, i.e., `shallow' defect levels, the localization of the corresponding electron
and hole may be relatively weak, i.e., long-ranged, 
compared to the case of `deep' defect levels.
In such cases, it may be found that such DFT+$U$ GKS eigenvalues 
matches well, in practice, with the corresponding total-energy based
transition Fermi levels.
As a consequence, the
resulting GKS band-edges
eigenvalue gap may be
an acceptable 
\emph{estimate}, in that
case, to the true fundamental
gap of the approximate
DFT method being used.
The magnitude of the corrective 
term describing localization, e.g.,
Eqs.~\ref{eqn:GKS-eigenvalue-deriv-final} 
and~\ref{eqn:GKS-eigenvalue-2nd-deriv} 
in such systems is worthy of future study.}

Eqs.~\ref{eq:ip_pristine_solid}, \ref{eq:ea_pristine_solid} \& \ref{eq:bandgap_pristine_solid} remain valid in the case of a pristine bulk system with frontier GKS eigenvalues evaluated using a conventional DFT$+U$ or hybrid functional. This is also the case with application of norm-conserving pseudopotentials, ultrasoft pseudopotentials, or the projector-augmented wave method, as detailed in Appendices~\ref{appendix:ncpp} and ~\ref{appendix:ultrasoftpaw}. The results can also be readily generalized to spin-unrestricted GKS theory, as we briefly outline in Appendix~\ref{appendix:spin}. In order to elaborate further on how the above proof of the validity of the GKS gap in the bulk limit applies not only to Dudarev's functional but to all other known  DFT+$U$ functionals \textcolor{black}{(subject  only to 
caveats that are  satisfied  or accounted for in practice), }
we next provide a brief overview of the 
mathematical form and original derivation of every \textcolor{black}{collinear-spin} DFT+$U$-type functional known by the present authors to have been published.
\textcolor{black}{We also discuss a 
representative non-collinear  DFT+$U$-type functional
that is more than a generalization of one of the latter.}
We hope that the following section will provide a 
comprehensive, up-to-date compendium of 
\textcolor{black}{collinear-spin} 
DFT+$U$ functionals
within a unified notation.

\section{{The Original DFT+U Functionals}}
\label{sec:allfunctionals1}
The DFT$+U$ method was originally developed by Anisimov et al.~\cite{anisimovBandTheoryMott1991} to alleviate the shortcomings of the local spin density approximation (LSDA)  in describing Mott insulators. In particular, Anisimov et al. noted the LSDA's severe underestimation of the bandgap and local magnetic moments in 3d-transition metal oxides, with the LSDA even predicting CoO and FeO as being metallic~\cite{terakuraBandTheoryInsulating1984}.

However, these systems are well described by the simple multi-band Hubbard model. Anisimov et al. proposed supplementing the local density approximation (LDA) with an additional interaction energy term to treat the localized 3d states, 
\begin{equation}
\label{eqn:anisimmov-interaction-term}
E_{\rm int}=\frac{U}{2}\sum_{\sigma m m'} n_{mm}^{\sigma}n_{m'm'}^{\bar{\sigma}}+\frac{U-J}{2}\sum_{{\sigma m m' \atop m\neq m'}}n_{mm}^{\sigma}n_{m'm'}^{{\sigma}},
\end{equation}
where $U$ and $J$ are the Hubbard $U$ and Hund's rule exchange parameters and $n^{\sigma}_{mm}$ is the spin-$\sigma$ occupancy of orbital $m$, which can be defined as
\begin{equation}
\label{eqn:Anisimov-1991-interaction-energy}
n_{mm'}^{I\sigma}=\braket{\phi^I_m|\hat{\rho}^{\sigma}|\phi^I_{m'}},
\end{equation}
where $\hat{\rho}^{\sigma}$ is the spin-$\sigma$ Kohn-Sham density operator and $\{\phi^I_m\}$ are the set of atomically localized orbitals at atom $I$ (the atomic site index is often suppressed for clarity). This interaction term is simply the expectation value of the Hubbard model Hamiltonian,
\begin{equation}
\hat{H}_{\rm mod}=\frac{U}{2}\sum_{mm'\sigma}\hat{n}_{m\sigma}\hat{n}_{m'\bar{\sigma}}+\frac{U-J}{2}\sum_{{\sigma m m' \atop m\neq m'}}\hat{n}_{m\sigma}\hat{n}_{m'{\sigma}},
\end{equation}
where only the on-site interaction term has been considered, i.e., the hopping term has been excluded. However, the inter-electron interactions described by Eq.~\ref{eqn:Anisimov-1991-interaction-energy} have already been accounted for to a less favourable extent by the LDA functional itself. This necessitates the use of a double counting correction scheme. Anisimov et al. postulated that the on-site inter-electron interactions would be well described by  the LDA in the homogeneous limit, i.e., when each of the $2(2l+1)$ orbitals are equally occupied, where $l$ is the orbtial angular momentum quantum number. The orbital occupancies in Eq.~\ref{eqn:anisimmov-interaction-term} are thus replaced by the deviation from the average occupancy $n_0$. This is referred to as the around mean field double counting correction scheme, where for completeness the average occupancy
is given by 
\begin{equation}
n_0=\frac{\sum_{\sigma m}n^{\sigma}_{mm}}{2(2l+1)}.
\end{equation}
Anisimov et al.'s DFT$+U$-type functional can be written as 
\begin{align}
& E_{\rm u}^{\rm Anisimov-AMF} = \frac{U}{2}\sum_{\sigma m m'}  (n_{mm}^{\sigma}- n_0)(n_{m'm'}^{\bar{\sigma}}-n_0)   \nonumber \\ &  \qquad \quad +\frac{U-J}{2}\sum_{{\sigma m m' \atop m\neq m'}}(n_{mm}^{\sigma}-n_0)(n_{m'm'}^{{\sigma}}-n_0). 
\end{align}
To further develop the DFT$+U$ method, Anisimov et al.~\cite{anisimovDensityfunctionalTheoryNiO1993} devised an alternative double counting correction scheme commonly referred to as the fully localized limit double counting scheme. Anisimov et al. proposed treating the 3d subspace of the transition metal site as an isolated system that exchanges electrons with the bath (the rest of the electronic system). In this case, the total subspace energy, which they assume is solely composed of the 3d inter-electron interaction energy, should obey the piecewise linearity condition with respect to electron count, originally developed by Perdew et al.~\cite{perdewDensityFunctionalTheoryFractional1982}. 

Anisimov et al. argue that the LDA yields relatively accurate total energies but by virtue of it being an explicit and differentiable functional of the electron density, it is a continuous function of the electron count $N$. They therefore approximate that the 3d inter-electron interaction energy within the LDA is given as 
\begin{equation}
\label{eqn:FLL_DC-anisimov}
E_{\rm dc}\approx\frac{U}{2}N(N-1),
\end{equation}
where $N$ is the total subspace occupancy and for now we have ignored the exchange interaction parameter $J$. Using Eq.~\ref{eqn:FLL_DC-anisimov} as our double counting correction term, in combination with a simple Hubbard like interaction term yields
\begin{equation}
E_{\rm u}=\frac{U}{2}\sum_{{ \sigma \sigma' m m' \atop \sigma m\neq \sigma' m'}}n^{\sigma}_{mm}n^{\sigma'}_{m'm'}-\frac{U}{2}N(N-1).
\end{equation}
For systems with integer orbital occupancies ($n^{\sigma}_{mm}=0$ or $1$), whose valence and conduction bands project perfectly onto the 3d subspace, this corrective functional will, to first order in perturbation theory, open a gap of magnitude $U$ between the occupied and unoccupied Generalised Kohn-Sham states, thus alleviating the LDA's significant underestimation of the quasi-particle bandgap. This corrective functional introduces a
discontinuity
in the XC(+$U$) potential within the Kohn Sham scheme but does not do so in the Generalised Kohn Sham scheme 
as it is an explicit functional of the first-order reference
(i.e., non-interacting or partially
interacting, depending on perspective) reduced spin density matrix.

Anisimov et al. accounted for the non-sphericity of the Coulomb interactions by using  orbitally-resolved Coulomb interaction parameters $U_{mm'}$ within the Hubbard like interaction term. Accounting also for the exchange interactions, they arrived at the following final expression for the DFT$+U$-type functional,
\begin{eqnarray}
\label{eqn:anisimov93}
E_{\rm u}^{\rm Anisimov-FLL}=&&\frac{1}{2}\sum_{\sigma m m'}U_{mm'}n_{mm}^{\sigma}n_{m'm'}^{\bar{\sigma}} \nonumber \\&& +\frac{1}{2}\sum_{{\sigma m m' \atop m\neq m'}}(U_{mm'}-J_{mm'})n_{mm}^{\sigma}n_{m'm'}^{{\sigma}}  \nonumber\\ &&- \frac{U}{2}N(N-1)+\frac{J}{4}N(N-2) .
\end{eqnarray}
Motivated by Anisimov et al.'s seminal papers, several groups later developed DFT$+U$-type functionals using either around mean field or fully localized limit double counting schemes. We will explicate some of these functionals in the following three sub-sections.

\smallskip

\subsection{\NoCaseChange{Around Mean Field Double Counting Schemes}}
Czy{\.z}yk et al.~\cite{czyzykLocaldensityFunctionalOnsite1994} proposed using the same orbitally resolved Hubbard like interaction term
\begin{align}
\label{eqn:orbitally-resolved-interaction-energy}
E_{\rm int}=&\frac{1}{2}\sum_{\sigma m m'}U_{mm'}n_{mm}^{\sigma}n_{m'm'}^{\bar{\sigma}} \nonumber \\ &+\frac{1}{2}\sum_{{\sigma m m' \atop m\neq m'}}(U_{mm'}-J_{mm'})n_{mm}^{\sigma}n_{m'm'}^{{\sigma}}
\end{align}
used by Anisimov et al.~\cite{anisimovDensityfunctionalTheoryNiO1993} in their DFT$+U$ functional with a Fully localized Limit Double Counting Scheme, as given by Eq.~\ref{eqn:anisimov93}. Instead of using the average occupancy $n_0$ in their around mean field double counting scheme, Czy{\.z}yk et al. propose using the average spin resolved occupancy $n_0^{\sigma}$,
\begin{eqnarray}
\label{eqn:czyzyk-amf}
&& E_{\rm u}^{\rm {Czy{\dot{z}}yk-AMF}}=\frac{1}{2}\sum_{\sigma m m'}U_{mm'}(n_{mm}^{\sigma}-n_0^{\sigma})(n_{m'm'}^{\bar{\sigma}}-n_0^{\bar{\sigma}})\nonumber \\ &&   +\frac{1}{2}\sum_{{\sigma m m' \atop m\neq m'}}(U_{mm'}-J_{mm'})(n_{mm}^{\sigma}-n_0^{\sigma})(n_{m'm'}^{{\sigma}}-n_0^{\sigma}).
\end{eqnarray}
Pickett et al.~\cite{pickettReformulationMathrmLDA1998} proposed using the spin and orbital resolved LDA occupancies $n_0^{\sigma mm}$ as the reference occupancies as opposed to the average spin resolved occupancies $n_0^{\sigma}$. This ensures that the LDA solution is a stationary solution of the LDA$+U$ functional,
\begin{widetext}
\begin{equation}
\label{eqn:pickett}
E_{\rm u}^{\rm {Pickett}}=\frac{1}{2}\sum_{\sigma m m'}U_{mm'}(n_{mm}^{\sigma}-n_0^{\sigma mm})(n_{m'm'}^{\bar{\sigma} }-n_0^{\bar{\sigma} m'm'}) +\frac{1}{2}\sum_{{\sigma m m' \atop m\neq m'}}(U_{mm'}-J_{mm'})(n_{mm}^{\sigma}-n_0^{\sigma mm})(n_{m'm'}^{{\sigma }}-n_0^{\sigma m'm'}).
\end{equation}
\end{widetext}
Motivated by the failure of Czy{\.{z}yk et al.'s~\cite{czyzykLocaldensityFunctionalOnsite1994} around mean field type functional of Eq.~\ref{eqn:czyzyk-amf} to provide a correction to fully spin polarised systems at half occupancy, Seo developed a new DFT$+U$-type functional with an around mean field double counting scheme. Seo~\cite{seoSelfinteractionCorrectionMathrmLDA2007} rearranged the orbitally resolved Hubbard like interaction term of Eq.~\ref{eqn:orbitally-resolved-interaction-energy} to give
\begin{eqnarray}
E_{\rm int}=&&\frac{1}{2}\sum_{\sigma m m'}U_{mm'}(n_{mm}^{\sigma}n_{m'm'}^{\bar{\sigma}}+n_{mm}^{\sigma}n_{m'm'}^{{\sigma}}) \nonumber \\ && - \frac{1}{2}\sum_{\sigma m m'}U_{mm'}n_{mm}^{\sigma}n_{m'm'}^{{\sigma}}
 \nonumber \\ &&+\frac{1}{2}\sum_{{\sigma m m' \atop m\neq m'}}(U_{mm'}-J_{mm'})n_{mm}^{\sigma}n_{m'm'}^{{\sigma}}.
\end{eqnarray}
Seo identified the first term as the Hartree energy component $E_{\rm H}$ and proposed using average Hubbard $U$ and Hund's $J$ parameters for the second two terms so that
\begin{equation}
\label{eqn:seo-interaction-term}
E_{\rm int}=E_{\rm H} - \frac{U-J}{2}\sum_{\sigma m }n_{mm}^{{\sigma}}n_{mm}^{{\sigma}} -\frac{J}{2}\sum_{\sigma}N^{\sigma}N^{\sigma},
\end{equation}
where $N^{\sigma}$ is the total spin-$\sigma$ subspace occupancy. For the double counting correction, Seo proposed replacing the spin resolved orbital occupancies with the average spin resolved occupancy $n_0^{\sigma}$ and argued that $(U+2lJ)/(2l+1)$ should be replaced simply by the Hund's $J$ parameter. The resulting double counting correction is 
\begin{equation}
\label{eqn:seo-dc-term}
E_{\rm dc}=E_{\rm H}  -\frac{J}{2}\sum_{\sigma}N^{\sigma}N^{\sigma}.
\end{equation}
Seo's DFT$+U$-type functional is given by the combination of the interaction term of Eq.~\ref{eqn:seo-interaction-term} and the double counting term of Eq.~\ref{eqn:seo-dc-term}, yielding 
\begin{equation}
\label{eqn:seo-functional}
E_{\rm u}^{\rm Seo}= - \frac{U-J}{2}\sum_{\sigma m }n_{mm}^{{\sigma}}n_{mm}^{{\sigma}}.
\end{equation}
\subsection{\NoCaseChange{Fully localized Limit Double Counting Schemes}}
Czy{\.{z}yk et al.~\cite{czyzykLocaldensityFunctionalOnsite1994} propose using the same orbitally resolved Hubbard interaction term of Eq.~\ref{eqn:orbitally-resolved-interaction-energy} that Anisimov et al.~\cite{anisimovDensityfunctionalTheoryNiO1993} used for the original DFT$+U$ functional with a Fully localized Limit Double Counting Scheme. Czy{\.{z}yk et al. correctly pointed out that the interaction term as given by Eq.~\ref{eqn:orbitally-resolved-interaction-energy} in the fully localized limit (with $n^{\sigma}_{mm}=0$ or $1$)  and $U_{mm'}$ and $J_{mm'}$ replaced by the average coulomb repulsion and exchange parameters $U$ and $J$, is given by
\begin{equation}
\label{eqn:czyzyk-dc-term}
E_{\rm dc}=\frac{U}{2}N(N-1)-\sum_{\sigma}\frac{J}{2}N^{\sigma}(N^{\sigma}-1),
\end{equation}
which slightly differs from the double counting scheme used by Anisimov et al.~\cite{anisimovDensityfunctionalTheoryNiO1993}. Combining the interaction energy term given by Eq.~\ref{eqn:orbitally-resolved-interaction-energy} with the double counting term given by Eq.~\ref{eqn:czyzyk-dc-term} yields the DFT$+U$ energy functional
\begin{eqnarray}
\label{eqn:czyzyk-fll}
E_{\rm u}^{\rm {Czy{\dot{z}}yk-FLL}}=&&\frac{1}{2}\sum_{\sigma m m'}U_{mm'}n_{mm}^{\sigma}n_{m'm'}^{\bar{\sigma}}  \\  && +\frac{1}{2}\sum_{\sigma m m' \atop m\neq m'}(U_{mm'}-J_{mm'})n_{mm}^{\sigma}n_{m'm'}^{{\sigma}} \nonumber \\  &&-  \frac{U}{2}N(N-1)+\sum_{\sigma}\frac{J}{2}N^{\sigma}(N^{\sigma}-1) .\nonumber
\end{eqnarray}
In Ref.~\cite{solovyevCorrectedAtomicLimit1994}, Solovyev et al. propose a DFT$+U$ functional with a fully localized limit double counting scheme, but the proposed functional is equivalent to Anisimov et al.'s  DFT$+U$ functional~\cite{anisimovDensityfunctionalTheoryNiO1993} as given by Eq.~\ref{eqn:anisimov93}. It is worth emphasising that Anisimov and Solovyev are named authors on both papers, and the later paper should be regarded as a follow up study rather than the introduction of a novel DFT$+U$ functional.

Due to the neglect of off-diagonal terms in the occupancy matrix $n^{\sigma}_{mm'}$, the interaction energy of Eq.~\ref{eqn:orbitally-resolved-interaction-energy} is not rotationally invariant and, by extension, neither are the DFT$+U$ functionals derived from it, i.e., Eqs.~\ref{eqn:anisimov93},\ref{eqn:czyzyk-amf},\ref{eqn:pickett},\ref{eqn:seo-functional}, and Eq.~\ref{eqn:czyzyk-fll}. Liechtenstein et al.~\cite{liechtensteinDensityfunctionalTheoryStrong1995} proposed using the full Hartree-Fock expression for the interaction energy, without neglecting the off diagonal terms,
\begin{align}
\label{eqn:Liechtenstein-interaction}
&E_{\rm int}= \frac{1}{2}\sum_{\{m\}\sigma}U_{mm''m'm'''} n^{\sigma}_{mm'}n^{\bar{\sigma}}_{m''m'''}\nonumber \\ &+\sum_{\{m\}\sigma}(U_{mm''m'm'''}-U_{mm''m'''m'})  n^{\sigma}_{mm'}n^{{\sigma}}_{m''m'''}, 
\end{align}
where $U_{mm''m'''m'}$ is defined as
\begin{equation}
U_{mm''m'm'''}=\braket{m,m''|V_{\rm ee}|m',m'''}
\end{equation}
and $V_{\rm ee}$ is the screened Coulomb potential. Liechtenstein et al.~\cite{liechtensteinDensityfunctionalTheoryStrong1995}  combined this term with the fully localized limit double counting term given by Eq.~\ref{eqn:czyzyk-dc-term} to create a rotationally invariant DFT+$U$ functional,
\begin{align}
\label{eqn:Liechtenstein}
&E_{\rm u}^{\rm Liechtenstein}=\frac{1}{2}\sum_{\{m\}\sigma}U_{mm''m'm'''}n^{\sigma}_{mm'}n^{\bar{\sigma}}_{m''m'''} \nonumber \\ & \quad  +\frac{1}{2}\sum_{\{m\}\sigma} (U_{mm''m'm'''}-U_{mm''m'''m'})n^{\sigma}_{mm'}n^{{\sigma}}_{m''m'''}  \nonumber \\ & \quad - \frac{U}{2}N(N-1)+\sum_{\sigma}\frac{J}{2}N^{\sigma}(N^{\sigma}-1) .
\end{align}
Dudarev et al.~\cite{dudarevElectronenergylossSpectraStructural1998} later developed a rotationally invariant DFT$+U$ functional using only the orbital independent $U$ and $J$ parameters. Combining Anisimov et al.'s interaction term of Eq.~\ref{eqn:anisimmov-interaction-term} with Czy{\.{z}yk et al.'s fully localized limit double counting correction of Eq.~\ref{eqn:czyzyk-dc-term}, Dudarev et al. arrived at the DFT$+U$ functional 
\begin{equation}
\label{eqn:Dudarev-not-rot-invar}
E_{\rm u}=\frac{U-J}{2}\sum_{m \sigma}n^{\sigma}_{mm}-n^{\sigma}_{mm}n^{\sigma}_{mm}.
\end{equation}
However, this functional is not rotationally invariant. Dudarev et al.~\cite{dudarevElectronenergylossSpectraStructural1998}  proposed that this functional should be replaced by the following expression,
\begin{equation}
\label{eqn:Dudarev}
E_{\rm u}^{\rm Dudarev}=\frac{U-J}{2}\sum_{m m'\sigma}n^{\sigma}_{mm'}\delta_{mm'}-n^{\sigma}_{mm'}n^{\sigma}_{m'm}.
\end{equation}
In the case of diagonal occupancy matrices, Eq.~\ref{eqn:Dudarev-not-rot-invar} and Eq.~\ref{eqn:Dudarev} are equal. Crucially Eq.~\ref{eqn:Dudarev} can also be written in terms of the trace over the product of spin resolved occupancy matrices
\begin{equation}
\label{eqn:Dudarev-trace}
E_{\rm u}^{\rm Dudarev}=\frac{U-J}{2}\sum_{\sigma}{\rm Tr}[\mathbf{n}^{\sigma}]-{\rm Tr}[\mathbf{n}^{\sigma}\mathbf{n}^{\sigma}]
\end{equation}
and so, unlike Eq.~\ref{eqn:Dudarev-not-rot-invar}, Eq.~\ref{eqn:Dudarev} is invariant under unitary transformations of the \textcolor{black}{projector orbitals. Several modifications to the DFT+$U$ functional of Dudarev et al.~\cite{dudarevElectronenergylossSpectraStructural1998} have also been explored in the literature. For example, Dabo~\cite{dabothesis} and Moynihan et al.~\cite{PhysRevB.94.220104} have explored the use of separate corrective parameters for the linear and quadratic term of Eq.~\ref{eqn:Dudarev-trace}.
The latter method~\cite{PhysRevB.94.220104}, 
dubbed DFT+$U_1$+$U_2$, was shown to be sufficient to
simultaneously correct the total energy and frontier eigenvalue for 
self-interaction error in the prototypical system stretched H$_2^+$.
More recently,  
Macke et al.~\cite{mackeOrbitalResolvedDFTMolecules2024} 
have explored the use of an orbital-resolved corrective parameter, 
with encouraging results.}

\subsection{\NoCaseChange{A Hybrid Double Counting Scheme}}
Petukhov et al.~\cite{petukhovCorrelatedMetalsMathrm2003} proposed using a linear interpolation between the around mean field and fully localized limit double counting schemes. The authors use a parameter $\alpha$, with $0\leq \alpha \leq 1$, to interpolate between the two double counting schemes. In the fully localized limit, where $\alpha=1$,  Dudarev et al.'s functional is used. This functional as given in Eqs.~\ref{eqn:Dudarev} \& \ref{eqn:Dudarev-trace} can also be written in the form,
\begin{equation}
\label{eqn:dudarev-alt-formulation}
E_{\rm u}^{\rm Dudarev}=-\frac{U-J}{2}\bigg(\sum_{\sigma mm'}n^{\sigma}_{mm'}n^{\sigma}_{m'm}-\sum_{\sigma}n^{\sigma}(2l+1)\bigg).
\end{equation}
For the $\alpha=0$ limit, Petukhov et al. develop a novel DFT$+U$ functional with an around mean field double counting scheme. The authors propose treating the first term of Eq.~\ref{eqn:dudarev-alt-formulation} as the interaction term,
\begin{equation}
\label{eqn:interaction-petukhov}
E_{\rm int}=-\frac{U-J}{2}\sum_{\sigma mm'}n^{\sigma}_{mm'}n^{\sigma}_{m'm}.
\end{equation}
The underlying L(S)DA should yield the correct solution in the limit of uniform spin resolved orbital occupancies so, as before, Petukhov et al.'s around mean field (P-AMF) DFT$+U$ functional can be found by replacing the $n^{\sigma}_{mm'}$ elements with $n^{\sigma}_{mm'}-n^{\sigma}\delta_{mm'}$ so that
\begin{widetext}
\begin{equation}
E_{\rm u}^{\rm P-AMF} =-\frac{U-J}{2}\sum_{\sigma mm'}(n^{\sigma}_{mm'}-n^{\sigma}\delta_{mm'})(n^{\sigma}_{m'm}-n^{\sigma}\delta_{m'm})=-\frac{U-J}{2}\bigg(\sum_{\sigma mm'}n^{\sigma}_{mm'}n^{\sigma}_{m'm}-\sum_{\sigma}(n^{\sigma})^2(2l+1)\bigg).
\end{equation}
\end{widetext}
To interpolate between these two limits Petukhov et al.~\cite{petukhovCorrelatedMetalsMathrm2003} proposed the double counting correction term,
\begin{align}
\label{eqn:dc-petukhov}
E_{\rm dc}=&-\frac{U-J}{2}\sum_{\sigma}(1-\alpha)(2l+1)(n^{\sigma})^2 \nonumber \\ &-\frac{U-J}{2}\sum_{\sigma}\alpha(2l+1)n^{\sigma}.
\end{align}
Combining the interaction and double counting terms of Eqs.~\ref{eqn:interaction-petukhov} \& ~\ref{eqn:dc-petukhov}, Petukhov et al.~\cite{petukhovCorrelatedMetalsMathrm2003} proposed the following functional with a hybrid double counting scheme,
\begin{align}
\label{eqn:petukhov-hybrid}
E_{\rm u}^{\rm P-Hybrid}=&-\frac{U-J}{2}\sum_{\sigma mm'}n^{\sigma}_{mm'}n^{\sigma}_{m'm} \nonumber \\ & +\frac{U-J}{2}\sum_{\sigma}(1-\alpha)(2l+1)(n^{\sigma})^2 \nonumber \\ & +\frac{U-J}{2}\sum_{\sigma}\alpha(2l+1)n^{\sigma}.
\end{align}
They proposed setting the value of the parameter $\alpha$ such that the DFT$+U$ functional of Eq.~\ref{eqn:petukhov-hybrid} corrects only the potential, i.e., offers no correction to the total energy. To satisfy this constraint, $\alpha$ is defined as 
\begin{equation}
\alpha=\frac{\sum_{\sigma mm'}n^{\sigma}_{mm'}n^{\sigma}_{m'm}-\sum_{\sigma}(n^{\sigma})^2(2l+1)}{(2l+1)\sum_{\sigma}n^{\sigma}(1-n^{\sigma})}.
\end{equation}

\section{\NoCaseChange{Extended DFT+U Functionals}}
\label{sec:allfunctionals2}
\subsection{DFT+U+J}
Himmetoglu et al.~\cite{himmetogluFirstprinciplesStudyElectronic2011} developed an extended DFT$+U$ functional, nowadays often referred to as the DFT+$U$+$J$ functional, or sometimes DFT+$U$+$J0$. They proposed that the full Hartree-Fock interaction term of Eq.~\ref{eqn:Liechtenstein-interaction} is dominated by the two-orbital Coulomb and exchange terms with interaction parameters of the form $U_{mm'mm'}$ and $U_{mm'm'm}$, respectively. Replacing these interaction parameters with their atomic averages,
\begin{equation}
U=\frac{1}{(2l+1)^2}\sum_{mm'}\braket{m,m'|V_{\rm ee}|m,m'}
\end{equation}  
and
\begin{equation}
J=\frac{1}{(2l+1)^2}\sum_{mm'}\braket{m,m'|V_{\rm ee}|m',m},
\end{equation}
Himmetoglu et al.~\cite{himmetogluFirstprinciplesStudyElectronic2011} arrived at the interaction term
\begin{align}
\label{eqn:Himmetoglu-interaction-term}
&E_{\rm int}=\frac{U}{2}\sum_{mm'\sigma}n^{\sigma}_{mm}n^{\sigma}_{m'm'}+n^{\sigma}_{mm}n^{\bar{\sigma}}_{m'm'}-n^{\sigma}_{mm'}n^{\sigma}_{m'm} \nonumber \\ & +\frac{J}{2}\sum_{mm'\sigma}n^{\sigma}_{mm'}n^{\sigma}_{m'm}+n^{\sigma}_{mm'}n^{\bar{\sigma}}_{m'm}-n^{\sigma}_{mm}n^{\sigma}_{m'm'}.  
\end{align}
This interaction term can also be written in the form
\begin{align}
E_{\rm int}=&\frac{U}{2}\sum_{\sigma}{\rm Tr}[\mathbf{n}^{\sigma}]^2+{\rm Tr}[\mathbf{n}^{\sigma}]{\rm Tr}[\mathbf{n}^{\bar{\sigma}}]-{\rm Tr}[\mathbf{n}^{\sigma}\mathbf{n}^{{\sigma}}]\nonumber \\&+\frac{J}{2}\sum_{\sigma}{\rm Tr}[\mathbf{n}^{\sigma}\mathbf{n}^{{\sigma}}]+{\rm Tr}[\mathbf{n}^{\sigma}\mathbf{n}^{\bar{\sigma}}]-{\rm Tr}[\mathbf{n}^{\sigma}]^2.
\end{align}
Analogous to Dudarev's functional of Eq.~\ref{eqn:Dudarev-trace}, this interaction term is invariant under unitary transformations of the 
\textcolor{black}{projector} orbitals. Himmetoglu et al. choose to apply a fully localized limit double counting scheme. Using the orbital representation that diagonalizes the occupation matrix and recalling that in the fully localized limit the orbital occupations (i.e., the diagonal elements of the occupation matrix) are equal to zero or one, one can readily show that in the fully localized limit the interaction term of Eq.~\ref{eqn:Himmetoglu-interaction-term} becomes
\begin{equation}
\label{eqn:Himmetoglu-dc}
E_{\rm dc}=\frac{U}{2}N(N-1) -\frac{J}{2}\sum_{\sigma}N^{\sigma}(N^{\sigma}-1)+JN^{\sigma_{\rm min}}.
\end{equation}
Combining the interaction and double counting terms of Eq.~\ref{eqn:Himmetoglu-interaction-term} and Eq.~\ref{eqn:Himmetoglu-dc}, Himmetoglu et al. proposed the following DFT+$U$+$J$ functional,
\begin{align}
\label{eqn:Himmetoglu-functional}
E_{\rm u}^{\rm Himmetoglu}=& \frac{U-J}{2}\sum_{\sigma mm'}n^{\sigma}_{mm'}\delta_{mm'}-n^{\sigma}_{mm'}n^{\sigma}_{m'm} \nonumber \\ & +\frac{J}{2}\sum_{\sigma mm'}n^{\sigma}_{mm'}n^{\bar{\sigma}}_{m'm}-JN^{\sigma_{\rm min}}.
\end{align}
The final term on the right-hand-side of Eq.~\ref{eqn:Himmetoglu-functional} is referred to as the minority spin term and is almost always omitted in practical calculations to aid numerical convergence. Despite being derived by very distinct techniques, Eq.~\ref{eqn:Himmetoglu-functional} bears a remarkable resemblance to Dudarev's functional of Eq.~\ref{eqn:Dudarev} but also contains an unlike spin interaction term.

\subsection{DFT+U+U$_{\uparrow \downarrow}$}
Instead of using the Hartree-Fock interaction term of Eq.~\ref{eqn:Himmetoglu-interaction-term}, Shishkin et al. used the Hubbard-like interaction term of Eq.~\ref{eqn:anisimmov-interaction-term} and  applied the same arguments as Himmetoglu et al. in deriving their double counting correction term to arrive at the following expression,
\begin{align}
E_{\rm u}=&\frac{U-J}{2}\sum_{\sigma mm'}n^{\sigma}_{mm'}\delta_{mm'}-n^{\sigma}_{mm'}n^{\sigma}_{m'm} \nonumber \\ & +\frac{U}{2}\sum_{\sigma mm'}n^{\sigma}_{mm'}n^{\bar{\sigma}}_{m'm}-UN^{\sigma_{\rm min}}.
\end{align}
Finally, Shishkin et al.~\cite{shishkinDFTDudarevFormulation2019} chose to denote the like-spin interaction $(U-J)$ simply by $U$ and the unlike-spin interaction $U$ by $U_{\uparrow \downarrow}$, 
\begin{align}
E_{\rm u}^{\rm Shishkin}=&\frac{U}{2}\sum_{\sigma mm'}n^{\sigma}_{mm'}\delta_{mm'}-n^{\sigma}_{mm'}n^{\sigma}_{m'm} \nonumber \\ &+\frac{U_{\uparrow \downarrow}}{2}\sum_{\sigma mm'}n^{\sigma}_{mm'}n^{\bar{\sigma}}_{m'm}-U_{\uparrow \downarrow}N^{\sigma_{\rm min}}.
\end{align}
Shishkin et al.'s functional is equiavlent to Himmetoglu et al.'s DFT+$U$+$J$ except that the unlike-spin interaction is of Coulomb type rather than exchange. Shishkin et al. proposed evaluating~\cite{shishkinEvaluationRedoxPotentials2021} the unlike-spin Hubbard $U$ parameter per a formula 
\begin{equation}
U_{\uparrow \downarrow}=-U \frac{\sum_{\sigma}{\rm Tr}\left[\bf{n}_{\sigma}\bf{n}_{\bar{\sigma}} \right]}{\sum_{\sigma}{\rm Tr}\left[\bf{n}_{\sigma}\bf{n}_{{\sigma}} \right]},
\end{equation}
yielding $0$ and $-U$ at the fully spin polarized and non-spin polarized 
limits, \textcolor{black}{respectively.}

\subsection{DFT+U+V}
Campo et al.~\cite{campoExtendedDFTMethod2010} proposed using an extended DFT$+U$ functional typically referred to as the DFT$+U+V$ method, to treat both on-site and inter-site interactions. Accounting for both on-site and inter-site interactions, Campo et al.~\cite{campoExtendedDFTMethod2010} started with the Hartree-Fock interaction energy,
\begin{align}
\label{eqn:hf-interaction}
&E_{\rm int}=\frac{1}{2}\sum_{ \{I\} \{m\}\sigma}U_{mm''m'm'''}^{I\hspace{0.25em} I'' \hspace{0.25em} I' \hspace{0.25em} I'''}n^{\sigma}_{mm'}n^{\bar{\sigma}}_{m''m'''} \\ & +\frac{1}{2}\sum_{\{I\} \{m\}\sigma}(U_{mm''m'm'''}^{I\hspace{0.25em} I'' \hspace{0.25em} I' \hspace{0.25em} I'''}-U_{mm''m'''m'}^{I\hspace{0.25em}I''\hspace{0.25em}I'''\hspace{0.25em}I'})n^{\sigma}_{mm'}n^{{\sigma}}_{m''m'''},  \nonumber
\end{align}
where the interaction matrix elements
\begin{equation}
U_{mm''m'm'''}^{I\hspace{0.25em}I''\hspace{0.25em}I'\hspace{0.25em}I'''}=\braket{\phi^I_m,\phi^{I''}_{m''}|V_{\rm ee}|\phi^{I'}_{m'},\phi^{I'''}_{m'''}},
\end{equation}
have both site and orbital indices and the occupancy matrix elements 
\textcolor{black}{are given by}
\begin{equation}
\label{eqn:occ-matrix}
n_{mm'}^{\sigma II'}=\braket{\phi^I_m|\hat{\rho}^{\sigma}|\phi^{I'}_{m'}}.
\end{equation}
Campo et al. neglected all but the two-orbital Coulomb terms with interaction parameters of the form $U_{mm'mm'}^{I\hspace{0.25em}I'\hspace{0.25em}I\hspace{0.25em}I'}$ and replaced these parameters with orbital averages

\begin{equation}
V^{II'}=\frac{\sum_{ij}\braket{\phi^I_i,\phi^{I'}_{i'}|V_{\rm ee}|\phi^{I}_{i},\phi^{I'}_{i'}}}{(2l_I+1)(2l_{I'}+1)}.
\end{equation}
They arrived at the interaction energy expression
\begin{equation}
\label{eqn:campo-interaction}
E_{\rm int}=\sum_{ I I'}\frac{V^{II'}}{2}n^In^{I'}-\sum_{II'\sigma mm'}\frac{V^{II'}}{2}n^{\sigma II'}_{mm'}n^{\sigma I'I}_{m'm}.
\end{equation}
The interaction energy expression of Eq.~\ref{eqn:campo-interaction} can also be written in terms of the trace over the product of occupancy matrices
\begin{align}
\label{eqn:campo-interaction-trace}
E_{\rm int}=&\sum_{ I I'\sigma \sigma'}\frac{V^{II'}}{2}{\rm Tr}\big[{\mathbf n}^{I\sigma}\big]{\rm Tr}\left[{\mathbf n}^{I'\sigma'}\right] \nonumber \\ &-\sum_{II'\sigma }\frac{V^{II'}}{2}{\rm Tr}\left[{\mathbf n}^{\sigma II'}{\mathbf n}^{\sigma I'I}\right],
\end{align}
and so is invariant under unitary transformations of the 
\textcolor{black}{projector} 
orbitals. Campo et al. proposed that in the fully localized limit, the off-diagonal terms ${\rm Tr}\left[{\mathbf n}^{\sigma II'}{\mathbf n}^{\sigma I'I}\right]$, where $I\neq I'$, are equal to zero. The on-diagonal terms ${\rm Tr}\left[{\mathbf n}^{\sigma II}{\mathbf n}^{\sigma II}\right]$ are approximated by ${\rm Tr}\left[{\mathbf n}^{\sigma II}\right]$, as discussed in the derivation of Himmetoglu et al.'s DFT+U+J functional. The fully localized limit double counting correction is
\begin{equation}
\label{eqn:campo-dc}
E_{\rm dc}=\sum_{ I I'}\frac{V^{II'}}{2}n^In^{I'}-\sum_{I}\frac{V^{II}}{2}n^I.
\end{equation}
Subtracting the double counting term of Eq.~\ref{eqn:campo-dc} from the interaction term of Eq.~\ref{eqn:campo-interaction}, Campo et al. arrived at the DFT$+U$ functional 
\begin{align}
\label{eqn:campo}
E_{\rm u}^{\rm Campo}=&\sum_{II'\sigma mm'}\frac{U^{II'}}{2}n^{\sigma II'}_{mm'}\delta_{mm'}\delta_{II'} \nonumber \\&-\sum_{II'\sigma mm'}\frac{U^{II'}}{2}n^{\sigma II'}_{mm'}n^{\sigma I'I}_{m'm},
\end{align}
where all but the same site $U^I = U^{II}$ is usually denoted by $V^{I I'}$.

\subsection{DFT+U+F}
Soler-Polo et al.~\cite{soler-poloLocalorbitalDensityFunctional2021} proposed using a Kanamori-like Hamiltonian for the on-site interaction,
\begin{align}
\label{eqn:kanamori}
&\hat{H}_{\rm int}=  \nonumber \sum_{m \sigma}\epsilon_{m }^{\sigma}\hat{n}_{mm}^{ \sigma} +U\sum_m \hat{n}_{mm}^{ \upharpoonright}\hat{n}_{mm }^{\downharpoonright} \\& \nonumber +\frac{U-J}{2}\sum_{m \neq m' ,\sigma} \hat{n}_{mm }^{\sigma}\hat{n}_{m'm'}^{ \bar{\sigma}}+\frac{U-2J}{2}\sum_{m \neq m', \sigma}\hat{n}_{mm}^{ \sigma}\hat{n}_{m'm'}^{ {\sigma}}  \nonumber \\ & \qquad-\frac{J}{2}\sum_{m \neq m' ,\sigma} \hat{c}^{\dag}_{m' \sigma} \hat{c}_{m' \bar{\sigma}}\hat{c}^{\dag}_{m \bar{\sigma}}\hat{c}_{m \sigma} ,
\end{align}
where the last term in Eq.~\ref{eqn:kanamori} is the spin-flip term, which is expressed in terms of creation and annihilation operators. At integer occupancy, the expectation value of Eq.~\ref{eqn:kanamori} can be expressed as 
\begin{numcases}{E_{\rm int}=} 
\frac{U-2J}{2}N(N-1),
& 
$ N \leq {\rm Tr}[{\hat{P}}]$,
\nonumber \\ \label{eqn:interaction_simple} \\ 
\frac{U-2J}{2}N(N-1)\nonumber \\ +J({\rm Tr}[{\hat{P}}]+1)(N-{\rm Tr}[{\hat{P}}]), \nonumber
&  
$N>{\rm Tr}[{\hat{P}}]$,
\end{numcases}
\vspace{0.1cm}

\noindent
where the one-electron contribution has been omitted and it is assumed that Hund's First rule is satisfied. At non-integer occupancy, Soler-Polo et al. argued that the total energy should be expressed as a linear combination of the $N_0$ and $N_0+1$  energies, where $N_0=\lfloor N \rfloor$, which yields the interaction energy,
\begin{numcases}{E_{\rm int}=} 
\frac{U-2J}{2}N_0(2N-N_0-1),
& 
$ N \leq {\rm Tr}[{\hat{P}}]$,
\nonumber \\ \label{eqn:piecewise-interaction} \\ 
\frac{U-2J}{2}N_0(2N-N_0-1) \nonumber \\ +J({\rm Tr}[{\hat{P}}]+1)(N-{\rm Tr}[{\hat{P}}]), \nonumber
&  
$N>{\rm Tr}[{\hat{P}}]$. 
\end{numcases}
\vspace{0.1cm}

\noindent
Like many other DFT$+U$ functionals, the above expression for the total energy was derived by explicitly ignoring the hopping term of the model Hamiltonian, i.e., the expressions were derived in the $U/T\rightarrow \infty$ limit. By explicitly solving the Kanamoi-Hamiltonian-like model system, for a range of values of $U/T$ and analyzing the corresponding potential, Soler-Polo et al. argue that the interaction term of Eq.~\ref{eqn:piecewise-interaction} should be modified using a set of four fitting parameters $\{F_i\}$ to better model systems beyond the $U/T\rightarrow \infty$ limit. Soler-Polo et al. split the interaction energy of Eq.~\ref{eqn:piecewise-interaction} into a mean-field Hartree contribution and a correlation contribution. The fitting parameters $\{F_i\}$ were then set as pre-factors for each of the four terms in the correlation energy expression so that
\begin{widetext}
\begin{align}
\label{eqn:soler-polo-final-interaction-energy}
E_{\rm int}= & \frac{U-2J}{2}(N-N_0)(1+N_0-N)F_2-\frac{U-2J}{2}\sum_{m \sigma}n_{mm}^{ \sigma}(1-n_{mm}^{ \sigma})F_1  -JF_3\left(N^{\upharpoonright}N^{\downharpoonright}+\sum_m n_{mm}^{ \upharpoonright}n_{mm}^{ \downharpoonright}\right)  \\ &+JF_4({\rm Tr}[{\hat{P}}]+1)(N-{\rm Tr}[{\hat{P}}])  +\frac{U-2J}{2}N(N-1)+\frac{U-2J}{2}\sum_{i \sigma}n_{mm}^{ \sigma}(1-n_{mm}^{ \sigma})+JN^{\upharpoonright}N^{\downharpoonright}+J\sum_m n_{mm}^{ \upharpoonright}n_{mm}^{ \downharpoonright}, \nonumber
\end{align}
\end{widetext}
where the final line in Eq.~\ref{eqn:soler-polo-final-interaction-energy} is the Hartree contribution to the interaction energy. Finally, Soler-Polo et al. employed the non-spin polarized, fully localized limit double counting scheme of Anisimov et al. (with the $U$ parameter replaced by $U-J$) to arrive at the following expression for their DFT$+U$-type functional,
\begin{widetext}
\begin{align}
E_{\rm u}^{\rm Soler-Polo}= & (1-F_1)\frac{U-2J}{2}\sum_{m \sigma}n_{mm}^{ \sigma}(1-n_{mm}^{ \sigma})+\frac{U-2J}{2}(N-N_0)(1+N_0-N)F_2 \nonumber \\ & +(1-F_3)J\left(N^{\upharpoonright}N^{\downharpoonright}+\sum_m n_{mm}^{ \upharpoonright}n_{mm}^{ \downharpoonright}\right)+JF_4({\rm Tr}[{\hat{P}}]+1)(N-{\rm Tr}[{\hat{P}}]) -\frac{J}{4}N^2.
\end{align}
\end{widetext}
\section{DFT+U Functionals from the Flat Plane Condition}
\label{sec:allfunctionals3}
It has long been argued in the literature that the success of the widely used DFT$+U$ functional of Dudarev et al., can be attributed to its ability to correct for many electron self interaction error. Ignoring off-diagonal occupancy matrix elements $n^{\sigma}_{mm'}$ where $m\neq m'$, Dudarev's functional,
\begin{equation}
    \frac{U_{\rm eff}}{2}\sum_{\sigma m}n^{\sigma}_{mm}-n^{\sigma}_{mm}n^{\sigma}_{mm},
\end{equation}
will enforce a linearity condition in the total energy with respect to orbital occupancy, assuming that the total energy at the bare DFT level is quadratic with respect to orbital occupancy with a curvature of $U_{\rm eff}$. There are is of course a serious caveat to this, many electron self interaction error is defined as the energetic deviation from the piecewise linearity condition of Perdew et al., which pertains to the total electron count
\textcolor{black}{rather than to} 
the subspace orbital occupancy. Hence enforcing a linearity condition on the total energy with respect to orbital occupancy is not rigorously justified. 

However, in the case of highly localized subspaces such as that of the 3d transition metal atomic sites where DFT$+U$ functionals are traditionally applied, the neglect of subspace-bath interactions for the purpose of deriving improved DFT$+U$-type functionals may be justified. By neglecting such interactions, the subspace can be treated as an isolated electronic system and many of the exact conditions which are often used to motivate approximate XC functional expressions can then be leveraged. Once a DFT$+U$-type functional has been derived, subspace-bath interaction effects can then be re-accounted for during corrective parameter evaluation, by ensuring that they are appropriately screened (we emphasize, however, that this is very much a topic of active research).  

Several modern DFT$+U$-type functionals have been developed to enforce exact conditions on localized subspaces, most notably the flat plane condition~\cite{yangDegenerateGroundStates2000,chanFreshLookEnsembles1999,mori-sanchezDiscontinuousNatureExchangeCorrelation2009,perdewExactExchangecorrelationPotentials2009,goshenEnsembleGroundState2024} and its generalization, the tilted plane condition~\cite{burgessTiltedPlaneStructureEnergy2024}. The flat plane condition defines the shape of the total energy surface of a finite electron system with respect to electron count $N$ and spin magnetization $M$. In the case of an electronic system with $N \leq 2$, the $E_v[N,M]$ surface will typically be composed of two triangular shaped planes with a derivative discontinuity along the $N=1$ line,
\begin{widetext}
\begin{numcases}{E_{v}[N,M]=} 
\frac{1}{2}(N+M) E_{v}[1,1]+\frac{1}{2}(N-M) E_{v}[1,-1],
& 
$N \leq 1$, 
\nonumber \\ \label{eqn:he_flat_plane_condition1} \\ 
\frac{1}{2}(2-N+M) E_{v}[1,1] +\frac{1}{2}(2-N-M) E_{v}[1,-1]+(N-1) E_{v}[2,0], \nonumber
&  
$N>1$,
\end{numcases}
\end{widetext}
where for simplicity we have ignored the highly magnetized $M=\pm 2$ triplet states. Standard local and semi-local XC approximations fail to reproduce this planar energy surface and instead exhibit a spurious curvature in the total energy with respect to electron count, resulting in many electron self interaction error, and a spurious curvature with respect to magnetization, resulting in static correlation error. 

\subsection{Judiciously Modified DFT}
Bajaj et al.~\cite{bajajCommunicationRecoveringFlatplane2017,bajajNonempiricalLowcostRecovery2019} were the first to develop a DFT$+U$-type functional from the flat plane condition. They proposed that the like-spin interaction term of the DFT+$U$+$J$ functional can help mitigate static correlation error, in the same spirit that Dudarev's functional acts as a corrector for many-electron self-interaction error. They asserted that the standard DFT+$U$+$J$ functional of Himmetoglu et al.~\cite{himmetogluFirstprinciplesStudyElectronic2011} can be used to enforce the flat plane condition in s-valence species when the s-subshell is less than half filled $(N < 1)$. Bajaj et al. argued that the like-spin interaction term should be modified for cases when $N> 1$ as the term increases with increased filling of the subshell, while the subspace analogue of SCE vanishes at full occupancy. Upon assessing several functional forms, Bajaj et al.~\cite{bajajCommunicationRecoveringFlatplane2017,bajajNonempiricalLowcostRecovery2019} arrived at the following modified DFT+$U$+$J$ functional which they refer to as judiciously modified DFT (jmDFT),
\begin{widetext}
\begin{numcases}{E_{\rm u}^{\rm Bajaj}=} 
{{\frac{U-J}{2}}}{{\sum}}_{\sigma}{\rm Tr}[\mathbf{n}^{\sigma}(1-\mathbf{n}^{\sigma})]
+
{{\frac{J}{2}}}{{\sum}}_{\sigma}{\rm Tr}[\mathbf{n}^{\sigma}\mathbf{n}^{\bar{\sigma}}],
& 
$N \leq 1$, 
\nonumber \\ \nonumber  \\ \label{eqn:BLOR_trace_version}
{{\frac{U-J}{2}}}{{\sum}}_{\sigma}{\rm Tr}[\mathbf{n}^{\sigma}(1-\mathbf{n}^{\sigma})]
+
{{\frac{J}{2}}}{{\sum}}_{\sigma}{\rm Tr}[(\mathbf{n}^{\sigma}-1)(\mathbf{n}^{\bar{\sigma}}-1)],
&  
$N>1$.
\end{numcases}
\end{widetext}
Bajaj et al. successfully employed their functional to correct for energetic errors in finite atomic and molecular test systems at fractional values of electron count and magnetization.

\subsection{BLOR}
In deriving the Burgess Linscott O'Regan (BLOR) DFT$+U$-type functional from the flat plane condition~\cite{burgessMathrmDFTTexttypeFunctional2023}, the authors departed from the DFT+$U$+$J$ functional of Himmetoglu et al. and 
even dispensed with the Hubbard model.
Instead, they specifically derived their DFT$+U$-type functional 
based only the principle of imposing the 
flat-plane (or, more generally, tilted-plane~\cite{burgessTiltedPlaneStructureEnergy2024})
condition to DFT+$U$ subspaces. 
Specifically, they asserted that the corrective functional should 
satisfy four key conditions for a single orbital subspace, namely that it should: 
(1) be a continuous function of the subspace occupancy $N$ and spin-magnetization $M$. 
(2) Yield no correction at integer values of $N$ and $M$.
(3) Have a constant curvature of $-U^{\sigma}$ in order to mitigate the subspace analogue of many electron self interaction error.
(4) Have a constant curvature of $J$ with respect to $M$ in order to mitigate the subspace analogue of static correlation error. BLOR is the unique simplified rotationally-invariant (i.e., Dudarev-type) DFT+$U$  functional which satisfies these four conditions for a single orbital subspace. In the case of a multi-orbital subspace, it can be readily extended to enforce a helium like flat plane condition on each orbital of the localized subspace, which gives the following general expression for the BLOR functional,
\begin{widetext}
\begin{numcases}{E_{\rm BLOR}=} 
\frac{U^{\upharpoonright}+U^{\downharpoonright}}{4}{\rm Tr}[\mathbf{N}-\mathbf{N}^2]+\frac{J}{2}{\rm Tr}[\mathbf{M}^2-\mathbf{N}^2]+\frac{U^{\upharpoonright}-U^{\downharpoonright}}{4}{\rm Tr}[\mathbf{M}-\mathbf{N}\mathbf{M}],
&
${\rm Tr}[\mathbf{N}] \leq {\rm Tr}[\mathbf{P}]$, 
\nonumber \\ \label{eqn:he_flat_plane_condition2} \\ 
\frac{U^{\upharpoonright}+U^{\downharpoonright}}{4}{\rm Tr}[(\mathbf{N}-\mathbf{P})-(\mathbf{N}-\mathbf{P})^2]+\frac{J}{2}{\rm Tr}[\mathbf{M}^2-(\mathbf{N}-2\mathbf{P})^2]  +\frac{U^{\upharpoonright}-U^{\downharpoonright}}{4}{\rm Tr}[\mathbf{M}-\mathbf{N}\mathbf{M}], \nonumber 
& ${\rm Tr}[\mathbf{N}] > {\rm Tr}[\mathbf{P}]$,
\label{eqn:blor_trace_version}
\end{numcases}
\end{widetext}
where $\mathbf{N} = \mathbf{n}^{\upharpoonright}+\mathbf{n}^{\downharpoonright}$ and $\mathbf{M} = \mathbf{n}^{\upharpoonright}-\mathbf{n}^{\downharpoonright}$ are the total subspace occupancy and magnetization matrices, respectively,  and $\mathbf{P}$ is the 
subspace projection matrix (the identity matrix, in practice). In the case of multi-orbital subspaces, Eq.~\ref{eqn:blor_trace_version} corrects for the orbital analogue of self interaction error and static correlation error but, by construction, does not correct explicitly for inter-orbital errors.

As an alternative, multi-orbital generalization that includes inter-orbital many electron self interaction error and static correlation error corrections, Burgess et al.~\cite{burgessFlatPlaneBased2024} proposed a DFT$+U$-type functional that instead enforces a flat plane condition once on the subspace as a whole, i.e., a neon like flat plane condition in the case of p-orbitals, or a zinc like flat plane condition in the case of d orbitals. This alternative  is known as the many-body generalization of the BLOR functional or mBLOR for concision,
and is given by 
\begin{widetext}
\begin{numcases}{E_{\rm mBLOR}=} 
\frac{U^{\upharpoonright}+U^{\downharpoonright}}{4}\left[{(N-N_0)}-{(N-N_0)}^2\right]+\frac{J}{2}\left[{M}^2-{N}^2\right]  +{\frac{U^{\upharpoonright}-U^{\downharpoonright}}{4} F_{\rm AMSIE}^{\rm early}},
& 
${\rm Tr}[\mathbf{N}] \leq {\rm Tr}[\mathbf{P}]$, 
\nonumber \\ \label{eqn:he_flat_plane_condition3} \\ 
\frac{U^{\upharpoonright}+U^{\downharpoonright}}{4}\left[{(N-N_0)}-{(N-N_0)}^2\right]+\frac{J}{2}\left[{M}^2-{(N-2{\rm Tr}[\hat{P}])}^2\right]  +\frac{U^{\upharpoonright}-U^{\downharpoonright}}{4} F_{\rm AMSIE}^{\rm late}, \nonumber
&  
${\rm Tr}[\mathbf{N}] > {\rm Tr}[\mathbf{P}]$,
\label{eqn:mblor}
\end{numcases}
\end{widetext}
where $F_{\rm AMSIE}$ is known as the asymmetric many electron self-interaction error function. The latter is a simple function of the subspace occupancy $N$ and magnetization $M$, but which has eight different forms depending on occupancy conditions, 
and is irrelevant when $U^{\upharpoonright}=U^{\downharpoonright}$
or in practice when these are not separately calculated.

Due to their piecewise definition, the BLOR and mBLOR functionals exhibit an explicit derivative discontinuity with respect to subspace occupancy, which is a feature of the exact functional
(see Appendix~\ref{appendix:explicitderiv}).
Explicit deriative discontinuities can contribute to the fundamental bandgap but, unlike (and additional to) 
the usual implicit derivative discontinuity of DFT+$U$,
do not contributed to the Generalized Kohn Sham eigenvalue gap. 
Burgess et al. proposed, however a 
computationally convenient scheme to `potentialize' this
derivative discontinuity, 
by adding 
a derivative discontinuity correction term $\hat{v}^{\sigma}_{\Delta \rm{xc}}$ to the potential affecting only the
unoccupied bands, to incorporate this contribution to the fundamental gap within the Generalized Kohn Sham gap. 
This correction is given by
\begin{equation}
\label{eqn:additional_potential}
\hat{v}^{\sigma}_{\Delta \rm{xc}}
=\widetilde{\Delta}_{\rm xc}\left[N\right](1-\hat{\rho})\hat{P}(1-\hat{\rho}),
\end{equation}
where, for non-spin polarized systems $\widetilde{\Delta}_{\rm xc} = U^{\sigma} = U-J$ except at half filling. The corresponding energy term $E_{\Delta \rm{xc}}$, will yield a vanishing correction to the total energy in the case of insulators. 

\subsection{DFT+UCI}
Janesko ~\cite{janesko2024multiconfigurational} recently reported the development of an extended DFT$+U$ functional referred to as the DFT plus Hubbard $U$ Configuration Interaction method (DFT+UCI). Although its derivation is not explicitly motivated by the flat plane condition, the functional nevertheless been shown to reduce both many electron self interaction error and static correlation error in challenging molecular test systems and as such has been included within this section of the review. In deriving the DFT+UCI functional, Janesko defines a partially interacting reference system through the use of the subspace projected electron interaction operator,
\begin{align}
\label{eqn:projected-vee}
&V_{\rm ee}^{\hat{P}}=\sum_{i>j}\sum_{m \sigma \sigma'}\ket{\phi_{m \sigma}(i)\phi_{m \sigma'}(j)}U_m\bra{\phi_{m \sigma}(i)\phi_{m \sigma'}(j)} \quad \& \nonumber \\ &  U_m=\braket{\phi_{m}\phi_{m}|\hat{V}_{\rm ee}|\phi_{m}\phi_{m}}.
\end{align}}
The exact total energy of the fully interacting system is given by 
\begin{equation}
E[\rho]=\min_{\Psi \rightarrow \rho}\braket{\Psi|\hat{T}+\hat{V}_{\rm ext}+V_{\rm ee}^{\hat{P}}|\Psi}+E_{\rm H}^{\hat{P}}[\rho]+E_{\rm xc}^{\hat{P}}[\rho],
\label{eqn:fully_interacintg_total_energy}
\end{equation}
where here $\Psi$ is a multi-determinant wave function. The projected Hartree $E_{\rm H}^{\hat{P}}[\rho]$ and Exchange Correlation $E_{\rm xc}^{\hat{P}}[\rho]$ functionals account for the remaining Hartree and Exchange Correlation energies not incorporated within the partially interacting reference system. Janesko argues that these two functionals can be approximated as 
\begin{align}
&E_{\rm H}^{\hat{P}}[\rho]= E_{\rm H}[\rho]-\frac{1}{2}\sum_{m \sigma \sigma '}U_m n^{\sigma}_m n^{\sigma'}_m \qquad\& \nonumber \\ & E_{\rm xc}^{\hat{P}}[\rho^{\upharpoonright},\rho^{\downharpoonright}]=E_{\rm xc}[\rho^{\upharpoonright},\rho^{\downharpoonright}]-E_{\rm xc}[\rho_m^{\upharpoonright},\rho_m^{\downharpoonright}],
\end{align}
where $\rho^{\sigma}$ is the total spin-$\sigma$ density and $\rho_m^{\sigma}$ is the projected spin-$\sigma$ density,
\begin{equation}
    \rho_m^{\sigma}({\bf r})=n^{\sigma}_m|\phi_m({\bf r})|^2.
\end{equation}
Eq.~\ref{eqn:fully_interacintg_total_energy} can be solved within the Generalized Kohn Sham scheme, to find the optimum ground-state single-determinant wave function $\Phi_0$. The total energy of which can be expressed as a standard DFT total energy plus a Hubbard-type corrective term,
\begin{equation}
E_{\rm u}= \sum_{m } -\frac{U_m}{2}\left(n^{\upharpoonright}_m n^{\upharpoonright}_m+n^{\downharpoonright}_m n^{\downharpoonright}_m\right)-E_{\rm xc}[\rho_m^{\upharpoonright},\rho_m^{\downharpoonright}].
\end{equation}
The occupied and unoccupied GKS spin resolved orbitals from the evaluation of $\Phi_0$ can then be used to construct a multi-determinant wavefunction, which Janesko uses as an approximation for the ground state wavefunction $\Psi$ of the partially interacting system. Applying separate unitary transformations to the occupied and unoccupied GKS spin resolved orbitals Janesko showed that the Configuration Interaction Hamiltonian has an exceedingly simple structure and can be diagonalized exactly to yield a total energy which is the sum of a DFT total energy and a Hubbard type corrective term and a non-self consistent multi-configurational correlation energy term,
\begin{align}
    &E_{\rm DFT}+\sum_{m }\bigg\{ -\frac{U_m}{2}\left(n^{\upharpoonright}_m n^{\upharpoonright}_m+n^{\downharpoonright}_m n^{\downharpoonright}_m\right)-E_{\rm xc}[\rho_m^{\upharpoonright},\rho_m^{\downharpoonright}] \nonumber \\& +\Delta_m-\left(\Delta_m^2-4U_m^2n^{\upharpoonright}_m n^{\downharpoonright}_m n_{m(v) }^{\upharpoonright } n_{m(v) }^{\downharpoonright }\right)^{1/2} \bigg\},
\end{align}
where $n_{m(v) }^{\sigma }$ is the occupancy of unoccupied GKS orbital with spin-$\sigma$ and orbital index $m$. \textcolor{black}{Here,} $2\Delta_m$ is the promotion energy associated with the doubly excited state configuration, which Janesko approximates using the GKS eigenvalues, Hubbard $U_m$ parameters and spin resolved orbital occupancies. 

Finally, Janesko used a weighting scheme in the multi-configurational correlation energy term to account for the non-orthogonality between the projector states. Full exact exchange was introduced into the projector states by projecting $\hat{V}_{\rm ee}$ onto the entire set of non-orthogonal projector states as opposed to projecting $\hat{V}_{\rm ee}$ onto one state at a time, as specified by Eq.~\ref{eqn:projected-vee}. The final expression for the DFT+UCI functional then becomes,
\begin{align}
    &E_{\rm u}^{\rm Janesko}= -\frac{1}{2} \sum_{\sigma{\{I\}\{m\}}}   U_{mm'm''m'''}^{I\hspace{0.3em}I'\hspace{0.3em}I''\hspace{0.3em}I'''} n^{II'''\sigma}_{mm'''} n^{I'I''\sigma}_{m'm''} \nonumber \\ &\quad -\sum_{Im}w_{Im}\left(\Delta_{Im}^2-4(U^I_m)^2n^{I \upharpoonright}_{m} n^{I \downharpoonright}_m n_{m(v) }^{I \upharpoonright } n_{m(v) }^{I \downharpoonright }\right)^{1/2} \nonumber \\ & \quad +\sum_{I m}w_{Im}\Delta_{Im}-E_{\rm xc}[\rho_{\hat{P}}^{\upharpoonright},\rho_{\hat{P}}^{\downharpoonright}],
\label{eqn:janesko}
\end{align}
where $\rho_{\hat{P}}^{\sigma}$ is the total spin-$\sigma$ subspace projected density and the summation on the first line runs over spins $\sigma$, orbitals $\{m\}$ and the combined shell and site indices $\{I\}$. Furthermore, we emphasize again that the multi-configurational correlation energy contribution to Eq.~\ref{eqn:janesko}, i.e., the terms with site and orbital indexed weights $\{w_{Im}\}$, is applied non-self consistently. 

\section{DFT+U Functionals for Non-collinear Systems}
\label{sec:allfunctionals4}
Dudarev et al. developed a DFT$+U$ functional to treat non-collinear magnetic systems~\cite{dudarevParametrizationMathrmLSDANoncollinear2019}. They started with the on-site interaction operator used for cubic harmonic p-orbitals, namely 
\begin{align}
\label{eqn:dudarev-non-collinear-hamiltonian}
\hat{H}=&\frac{1}{2}\left( U -\frac{J}{2}\right) \left(\hat{N}^2-\hat{N}\right)-\frac{J}{4}\left(\hat{{\mathbf M}}^2-3\hat{N}\right) \nonumber \\ &+\frac{J}{2}\sum_{mm'\sigma \sigma'}\hat{c}_{m\sigma}^{\dagger}\hat{c}_{m\sigma'}^{\dagger}\hat{c}_{m'\sigma'}\hat{c}_{m'\sigma}.
\end{align}
The operators $\hat{c}_{m\sigma}^{\dagger}$ and $\hat{c}_{m\sigma}$ are the orbital $m$, spin-$\sigma$ creation and annihilation operators while the total subspace electron count and magnetic moment vector operators,
\begin{equation}
\hat{N}=\sum_{m\sigma}\hat{c}_{m\sigma}^{\dagger}\hat{c}_{m\sigma},
\quad \& \quad
\hat{{\mathbf M}}=\sum_{m\sigma \sigma'}\hat{c}_{m\sigma}^{\dagger}{\boldsymbol\sigma}_{\sigma \sigma'}\hat{c}_{m\sigma'},
\end{equation}
and ${\boldsymbol\sigma}_{\sigma \sigma'}$ are the Pauli matrices. Focusing solely on terms in the Hamiltonian composed of two creation and two annihilation operators acting on the same electronic state, i.e., terms of the form $\hat{n}_{m \sigma}\hat{n}_{m \sigma}$, Dudarev et al. argued that with approximate DFT functionals such terms contribute an energy proportional to $n_{m \sigma}^2$, when they should contribute an energy proportional to $n_{m \sigma}$. In this case, the terms with a $J$ prefactor cancel to yield the DFT$+U$ functional 
\begin{equation}
E_{\rm u}=\frac{U}{2}\sum_{\sigma m}n_{mm}^{\sigma}-n_{mm}^{\sigma}n_{mm}^{\sigma}.
\end{equation}
Dudarev et al.~\cite{dudarevParametrizationMathrmLSDANoncollinear2019} propose modifying this functional, using the same technique applied to Eq.~\ref{eqn:Dudarev-not-rot-invar}, so that the functional is invariant with the respect to the choice of orbitals and spin quantization axis, specifically 
\begin{equation}
E_{\rm u}^{\rm Dudarev-NC}=\frac{U}{2}\sum_{\sigma \sigma'  m m'}n_{mm'}^{\sigma \sigma'}\delta_{mm'}\delta_{\sigma \sigma'}-n_{mm'}^{\sigma \sigma'}n_{m'm}^{\sigma' \sigma},
\end{equation}
where the occupancy matrix elements
\begin{equation}
n_{mm'}^{\sigma \sigma'}=\braket{\phi_m^{\sigma}|\hat{\rho}|\phi_{m'}^{\sigma'}}.
\end{equation}
Interestingly, in the case of collinear spin systems, this functional reduces to Dudarev's standard functional of Eq.~\ref{eqn:Dudarev} except with an effective parameter $U_{\rm eff}=U$ as opposed to $U-J$. 

\textcolor{black}{Several other DFT+$U$ functionals have also been generalized to non-collinear spin systems, e.g., those discussed in Refs.~\onlinecite{bultmark2009multipole,bousquet2010j,ryee2018comparative,gomez2023compatibility,binci2023noncollinear}. 
A detailed exposition on these falls beyond the scope of the present study. However, while these may differ in their
invariance with respect to the projection axes, no non-collinear
DFT+$U$ functional known to the authors 
exhibits an explicit derivative discontinuity
or other additional cause for band-gap invalidity
within the GKS bulk-limit context.}

\section{Checklist for eigenvalue-based bandgap
validity}
\label{sec:verificationtable}
\begin{table*}
\caption{\label{tab:table4}%
The DFT+$U$ functional checklist specifies which physical quantities must be kept fixed across the evaluation of the bulk total energies of the $N-1$, $N$ and $N+1$ electron systems to ensure agreement with the respective GKS gap.}
\begin{ruledtabular}
\label{hubbard_checklist}
\begin{tabular}{ccccc }
{DFT+$U$  Functional} & {Hubbard Parameters} 
& {Projection Axes}& {Auxiliary Parameters} 
& Functional Form \\ \hline
Anismiov et al. (AMF)~\cite{anisimovBandTheoryMott1991}& \checkmark & \checkmark & \checkmark & \\
Anismiov et al. (FLL)~\cite{anisimovDensityfunctionalTheoryNiO1993} & \checkmark & \checkmark  &  &  \\
Czy{\.z}yk et al. (AMF)~\cite{czyzykLocaldensityFunctionalOnsite1994}   & \checkmark & \checkmark & \checkmark & \\
Pickett et al. (AMF)~\cite{pickettReformulationMathrmLDA1998}  & \checkmark & \checkmark & \checkmark & \\
Seo~\cite{seoSelfinteractionCorrectionMathrmLDA2007}  & \checkmark & \checkmark &  & \\
Czy{\.z}yk et al. (FLL)~\cite{czyzykLocaldensityFunctionalOnsite1994}   & \checkmark & \checkmark &  & \\
Liechtenstein et al.~\cite{liechtensteinDensityfunctionalTheoryStrong1995} & \checkmark &  & & \\
Dudarev  et al.~\cite{dudarevElectronenergylossSpectraStructural1998}  & \checkmark & &  & \\
Petukhov  et al.~\cite{petukhovCorrelatedMetalsMathrm2003} & \checkmark &  & \checkmark & \\
DFT+$U$+$J$~\cite{himmetogluFirstprinciplesStudyElectronic2011}  & \checkmark & &  & \\
DFT+$U$+$J$ (with min. spin)~\cite{himmetogluFirstprinciplesStudyElectronic2011}  & \checkmark &  &  & \\
DFT+$U$+$U_{\uparrow\downarrow}$~\cite{shishkinDFTDudarevFormulation2019}   & \checkmark &  & & \\
DFT+$U$+$V$~\cite{campoExtendedDFTMethod2010}  & \checkmark &  &  & \\
DFT+$U$+$F$~\cite{soler-poloLocalorbitalDensityFunctional2021}  & \checkmark & \checkmark &  \checkmark  &  \checkmark\\
jmDFT~\cite{bajajCommunicationRecoveringFlatplane2017}  & \checkmark & &  &  \checkmark\\
BLOR~\cite{burgessMathrmDFTTexttypeFunctional2023}  & \checkmark & &  &  \checkmark\\
mBLOR~\cite{burgessFlatPlaneBased2024} & \checkmark & &  &  \checkmark\\
DFT+UCI~\cite{janesko2024multiconfigurational} & \checkmark   & 
&  \checkmark &  \\
Dudarev et al. (Non-collinear)~\cite{dudarevParametrizationMathrmLSDANoncollinear2019}  & \checkmark & &  & \\
\end{tabular}
\end{ruledtabular}
\end{table*}

The proof of the validity of the DFT+$U$ \textcolor{black}{bulk} bandgap has been 
established above using the conventional DFT+$U$ functional of Dudarev et al.~\cite{dudarevElectronenergylossSpectraStructural1998}, however, the same proof holds for \textcolor{black}{all known} DFT+$U$ functionals, discussed in the preceding section, subject 
only to \textcolor{black}{caveats
that are either satisfied or accounted for in practice.} In the case of all DFT+$U$ functionals it \textcolor{black}{is, for example,
necessary but also safely} assumed that the bulk total energy calculations of the $N-1$, $N$ and $N+1$ electron systems are evaluated using a fixed set of Hubbard corrective parameters. In the case of the DFT+$U$ functional of Liechtenstein et al.~\cite{liechtensteinDensityfunctionalTheoryStrong1995} we refer here to fixing the orbital averaged $U$ and $J$ parameters as opposed to fixing the full four orbital indexed corrective parameters $\{U_{mm'm''m''}\}$, similarly only the single orbital and site indexed parameter $U_m^I$ is fixed in the DFT+UCI functional. 

For some corrective functionals, additional auxiliary parameters must also be fixed across all three bulk total energy calculations. 
\textcolor{black}{Again, in practice, for the bulk limit}
\textcolor{black}{it is natural to treat such auxiliary parameters on the same footing as the Hubbard corrective parameters.}
In the case of DFT+$U$ functionals with around mean field double counting schemes, the auxiliary parameters refers to the reference occupancies $\{n_0\}$ which, depending on the DFT+$U$ functional, may include spin and orbital indices. In the case of the DFT+$U$ functional of Petukhov  et al.~\cite{petukhovCorrelatedMetalsMathrm2003} and the DFT+$U$+$F$~\cite{soler-poloLocalorbitalDensityFunctional2021} functional, the auxiliary parameters refers to the $\alpha$ parameter and the four fitting parameters $F_1$-$F_4$ respectively, while in the case of the DFT+UCI functional it refers to the full multi-configurational correlation energy term which is evaluated non-self consistently. 

The \textcolor{black}{on-site} atomic orbital basis set axes through which the subspace projection operator is defined must also be kept fixed for all three bulk total energy calculations in the case of DFT+$U$ functionals which are not invariant with respect to 
\textcolor{black}{an on-site 
unitary transformation of 
spatial projector orbitals. 
This corresponds to ordinary practice, and there is no physical cause to do otherwise in the bulk limit.}
It goes without saying that spatial form of the orbitals must be kept fixed, \textcolor{black}{as well 
as any 
unitary orbital mixing between  
different sites.}
Finally, several modern DFT+$U$ functionals are defined piecewise with respect to subspace occupancy, namely the jmDFT, BLOR, mBLOR and DFT+$U$+$F$ functionals,  due to \textcolor{black}{their} explicit dependence on $N_0=\lfloor N \rfloor$. In such cases, the same functional form must be applied in the three bulk total energy calculations to ensure agreement with the GKS gap. However, in the unusual scenario that application of two or more distinct functional forms was deemed necessary \textcolor{black}{when simulating the bulk limit}, the deviation between the GKS gap and the fundamental gap amounts to an explicit derivative discontinuity. \textcolor{black}{This} may be easily evaluated using the derivative discontinuity \textcolor{black}{`potentialization'} scheme  mentioned in the BLOR subsection. These various caveats, \textcolor{black}{which 
are of theoretical interest more so than practical consequence}, 
are summarized in Table~\ref{hubbard_checklist}.

\subsection{DFT+U Energetic Correction for the Idealized Hydrogen Lattice}
The DFT$+U$ functionals examined 
in Sections~\ref{sec:allfunctionals1}-\ref{sec:allfunctionals4}
will each offer a distinct energetic correction to the total electronic energy. In Fig.~\ref{fig:hubbard_corrective_energy}, these energetic corrections within the spin-unrestricted formalism are compared and contrasted. As arguably the chemical system that most closely resembles the Hubbard model, the hydrogen lattice at the infinite separation limit was selected for this comparative study. In the infinite separation limit, a single electron will localize on each of the hydrogen atomic sites. In the case of the non-spin polarized hydrogen lattice $n^{\sigma}=0.5$, while in the fully spin-polarized case $n^{\upharpoonright}=1.0$ and $n^{\downharpoonright}=0.0$. 

For this highly idealized system, almost all DFT$+U$ functionals offer no energetic correction in the spin-polarized case. The notable exceptions to this are the original DFT$+U$ functionals of Anisimov et al. The unusual performance of these  two DFT$+U$ functionals can be attributed to the spin-agnostic character of their double counting correction schemes. Seo's DFT$+U$ functional also offers an energetic correction in the spin polarized case. Unlike the well known functional of Dudarev et al.,  Seo's functional lacks a linear term in $n^{\sigma}_{mm}$. Therefore, in the case of single orbital subspaces its energetic correction will increase quadratically with respect to spin resolved orbital occupancy.  

By contrast, in the non-spin polarized case, almost all DFT$+U$ functionals offer a non-zero energetic correction. The spin polarized and non-spin polarized hydrogen lattices should be degenerate in the infinite separation limit. However, standard local and semi-local XC functionals are known to energetically favor the spin-polarized hydrogen lattice, an error which Fig.~\ref{fig:hubbard_corrective_energy} shows will be exacerbated by application of a DFT$+U$ functional, despite such corrective functionals being explicitly designed for Mott-Hubbard insulators. With the exception of BLOR and Seo'07, the spin-polarized energy correction $E_{\rm u}\leq 0$ while the non-spin-polarized energy correction $E_{\rm u} \geq 0$ and hence the energy difference will only increase upon application of a DFT$+U$ functional. Seo '07 offers a negative energetic correction for both systems, however the spin polarized lattice will receive a larger energetic correction and this functional too will exacerbate this error. By contrast, the BLOR DFT$+U$ functional offers a negative energetic correction to the non-spin polarized hydrogen lattice and no energetic correction for the spin-polarized system. It is thus the sole DFT$+U$ functional considered here that can, in principle, successfully correct this energy error given 
reasonable values of the parameters, i.e., whenever $U> 0$, $J > 0$ and $U> J$. Therefore, almost all DFT$+U$ functionals will exacerbate the above mentioned hydrogen lattice energy error for any reasonable choice in value of the corrective parameters. The only exception to this result is the DFT$+U$+$J$ functional with the minority spin term, which like the BLOR functional will offer a negative energetic correction to the non-spin polarized hydrogen lattice whenever $U<4J$. 

\begin{figure}
\centering
\includegraphics[scale=0.5]{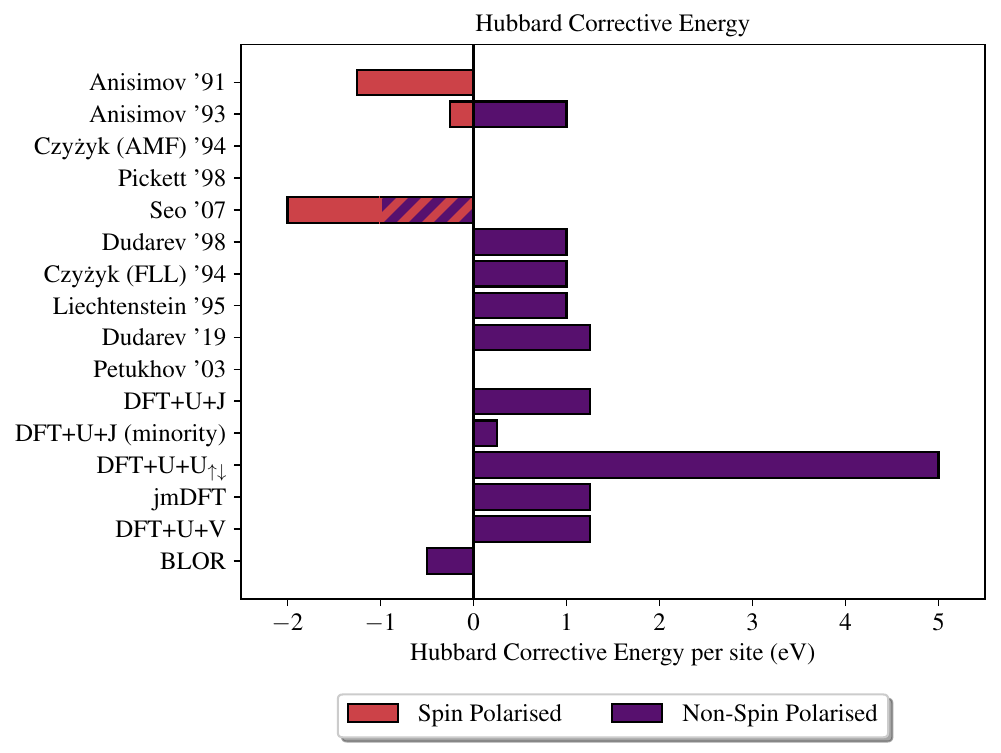}
\caption{DFT$+U$ correction to the total energy of the spin polarized and non-spin polarised hydrogen lattice at the infinite separation limit upon non-self consistent application of each DFT+$U$-type functional. The Hubbard $U$, Hund's $J$ and intersite $V$ parameters were set at 5 eV, 1 eV  and 0 eV, respectively. Seo's DFT$+U$ functional yields a negative energetic correction for both the spin polarized and non-spin polarized hydrogen lattices of -2 eV and -1 eV, respectively.}
\label{fig:hubbard_corrective_energy}
\end{figure}

\subsection{DFT+U Bandgap Correction for the Idealized Hydrogen Lattice}
In practice, DFT$+U$ functionals are often employed to ameliorate the bandgap problem~\cite{perdewDensityFunctionalTheory1985,kirchner-hallExtensiveBenchmarkingDFT2021}, i.e., standard (semi-)local XC functionals' tendency to vastly underestimate the fundamental bandgap, frequently even predicting insulators as being metallic. 

In order to elaborate further on the use of DFT$+U$-type functionals to alleviate the bandgap problem, in
Figs.~\ref{fig:spin_polarised_h_lattice} \& \ref{fig:non_spin_polarised_h_lattice} the bandgap correction offered by a variety of DFT$+U$-type functionals within the spin-unrestricted formalism is presented. To ascertain the reported corrections to the bandgap, the main result of this work is used, i.e., that the fundamental bandgap of a pristine bulk system is given by the frontier GKS eigenvalue difference as specified by Eq.~\ref{eq:bandgap_pristine_solid}. Furthermore, it is assumed that in the infinite separation limit, the GKS orbitals at the bare DFT level can be approximated as a linear combination of the hydrogen atomic orbitals, the same atomic orbitals which are used in the definition of the projection operator $\hat{P}$ for non-self consistent application of the DFT$+U$-type corrective functionals. 
\begin{figure}
\centering
\includegraphics[scale=0.5]{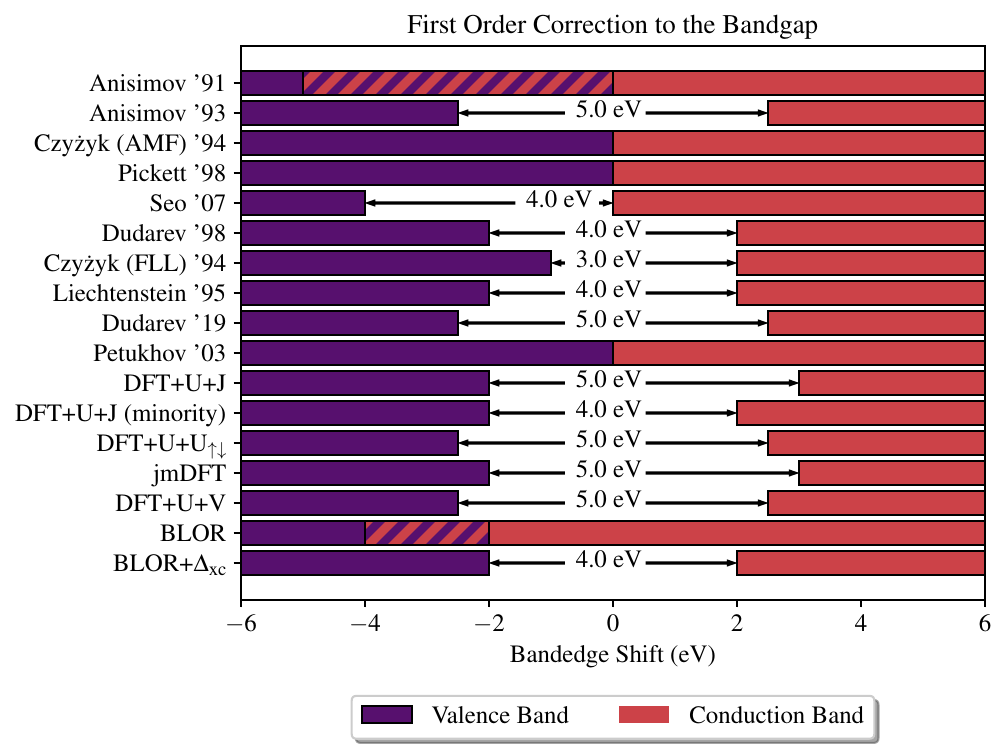}
\caption{Correction to the bandgap of the spin polarized hydrogen lattice at the infinite separation limit upon non-self consistent application of each DFT+$U$-type functional. The Hubbard $U$, Hund's $J$ and intersite $V$ parameters were set at 5 eV, 1 eV  and 0 eV, respectively. Two DFT$+U$ functionals, namely Anisimov '91 and BLOR \textcolor{black}{(if neglecting its
unusual explicit derivative discontinuity contribution $\Delta_{\rm xc}$)},  reduce the Kohn Sham gap by 5 eV and 2 eV, respectively, while most other functionals increase the Kohn Sham gap of this system.}
\label{fig:spin_polarised_h_lattice}
\end{figure}

\begin{figure}[t]
\centering
\includegraphics[scale=0.5]{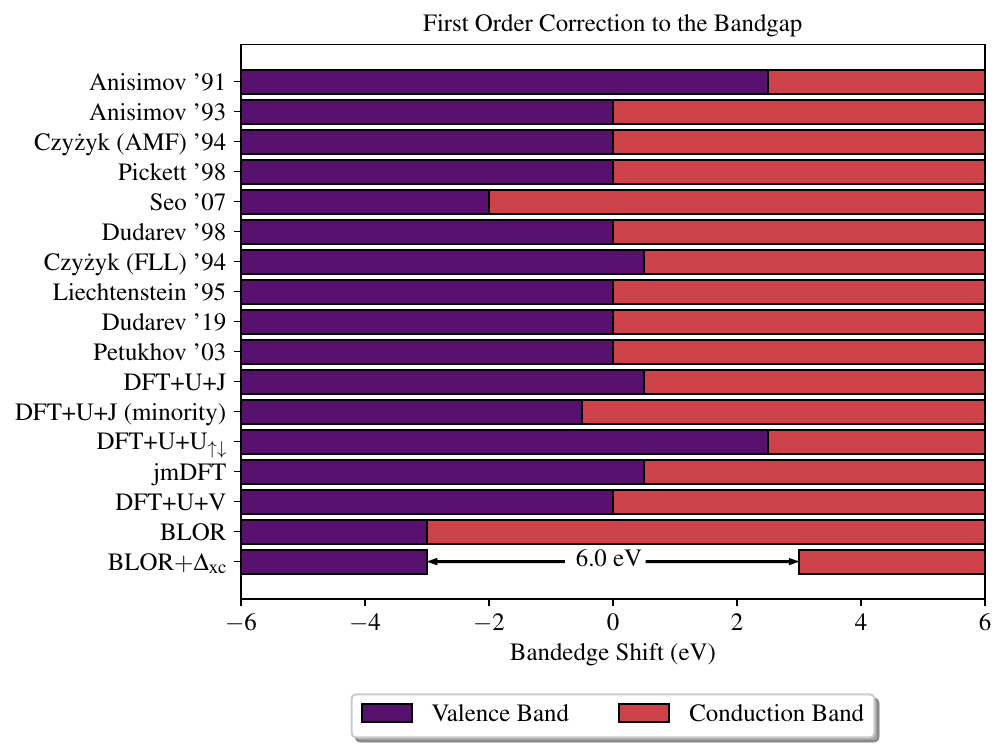}
\caption{Correction to the bandgap of the non-spin polarized hydrogen lattice at the infinite separation limit upon non-self consistent application of each DFT+$U$-type functional. The Hubbard $U$, Hund's $J$ and intersite $V$ parameters were set at 5 eV, 1 eV  and 0 eV, respectively. Most DFT+$U$ functionals fail to modify the Kohn Sham gap of this system.}
\label{fig:non_spin_polarised_h_lattice}
\end{figure}

Specifically, in Fig.~\ref{fig:spin_polarised_h_lattice}, the shift to the valence and conduction band edges of the spin-polarized hydrogen lattice upon non-self consistent application of a variety of DFT$+U$-type functionals is shown. The Hubbard $U$, Hund's $J$ and intersite $V$ parameters were set at 5 eV, 1 eV  and 0 eV, respectively. We emphasize that setting $V=0$ eV is, of course, a sensible choice in the infinite separation limit. It is worth noting for single orbital subspaces, multiple DFT$+U$ functionals take the same functional form, such as Dudarev '98 and Liechtenstein '95,  which both simplify to 
\begin{equation}
E_{\rm u}=\frac{U-J}{2}\sum_{I \sigma}n^{I \sigma}-n^{I \sigma}n^{I \sigma}.
\end{equation}
It is thus unsurprising that, for the spin-polarized hydrogen lattice, many DFT$+U$ functionals yield similar corrections to bandgap, increasing its value by 4 or 5 eV. Interestingly, with the exception of Seo '07, the DFT$+U$ functionals with around mean field double counting schemes fail to increase the bandgap of the spin-polarized hydrogen lattice, with Czy{\.z}yk (AMF) '94 and Pickett '98 yielding no shift to the valence and conduction band edges, while Anisimov '91 reduces the magnitude of the Kohn Sham gap by 5 eV. The BLOR functional also reduces the Kohn Sham gap, in this case by 2 eV. However, the derivative discontinuity corrected version of the BLOR functional (BLOR$+\Delta_{\rm xc}$), which is the recommended version for achieving suitable bandgap predictions, opens the gap of the spin polarized hydrogen lattice by 4 eV, in line with most other DFT$+U$-type functionals. 

In Fig.~\ref{fig:non_spin_polarised_h_lattice}, the analogous results are presented for the non-spin polarized hydrogen lattice. Interestingly, in this case most DFT$+U$ functionals fail to offer any correction to the valence and conduction band edges, while the functionals which do offer a correction all shift both band edges by the same amount, resulting in no first order correction to the bandgap. The noteable exception to this is the BLOR$+\Delta_{\rm xc}$ functional which opens the gap by 6 eV.

\begin{figure*}
  \includegraphics[width=\textwidth]{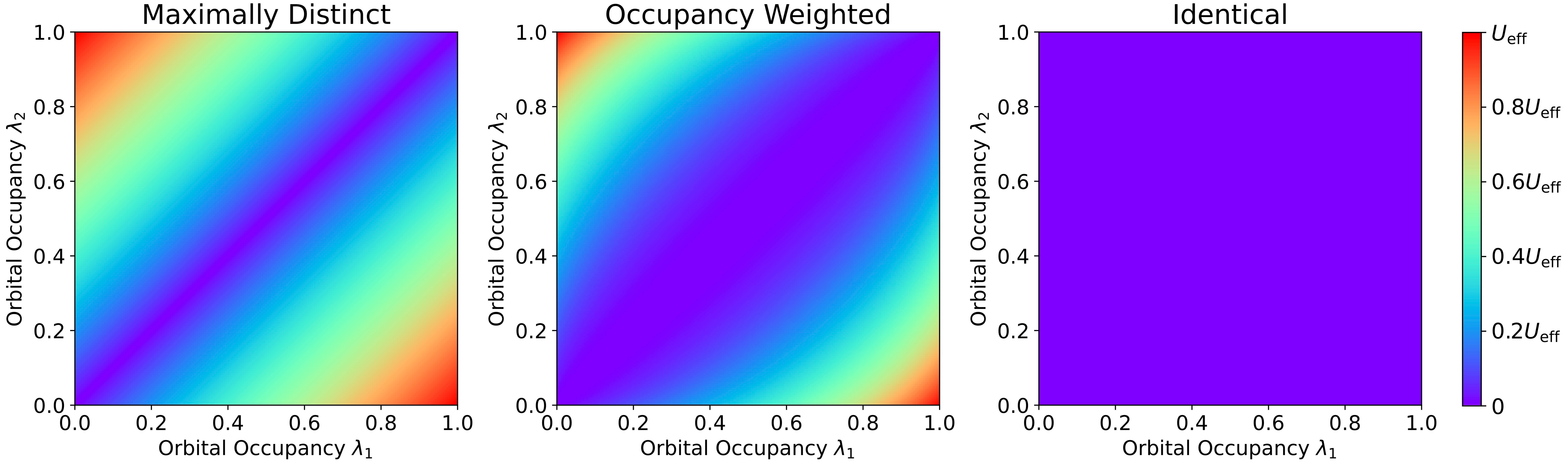}
\caption{Dudarev's DFT+$U$ correction to the bandgap $\Delta_{\rm GKS}^{\sigma}$ as a function of the spin-resolved subspace occupancy eigenvalues of a d-orbital subspace under octahedral or tetrahedral crystal field splitting.  Under octahedral or tetrahedral crystal field splitting, in the absence of further symmetry breaking, there will be two eigenspaces (of dimensions 2 and 3) of the d-orbital spin-resolved subspace occupancy matrix, the eigenvalues of which we label $\lambda^{\sigma}_{\rm e_{\rm g}}$ and $\lambda^{\sigma}_{\rm t_{\rm 2g}}$. The existence of only two distinct spin-resolved subspace occupancy eigenvalues enables the DFT+$U$ bandgap correction of Eq.~\ref{eqn:hubbard_potential_working_equation} to be readily plotted as a function of $\lambda^{\sigma}_{\rm e_{\rm g}}$ and $\lambda^{\sigma}_{\rm t_{\rm 2g}}$ once a suitable model for the orbital resolved subspace projection, i.e., the inner summation of Eq.~\ref{eqn:hubbard_potential_working_equation} has been selected. In subplot (a) the highest occupied GKS orbital projects onto whichever eigenspace, be it the e$_{\rm g}$ or t$_{\rm 2g}$ orbitals, that has a higher subspace occupancy eigenvalue and vice-versa for the lowest unoccupied GKS orbital. Hence the total projection of the e$_{\rm g}$ orbitals onto the highest occupied and lowest unoccupied GKS orbitals is equal to 1 or 0 , and the difference in the total projection onto the e$_{\rm g}$ orbitals, $\sum_{I,\;j\,\in \,{\rm e}_{\rm g}}\left(|\braket{\Psi_{\rm L}^{\sigma}|\varphi^{I\sigma}_j}|^2-|\braket{\Psi_{\rm H}^{\sigma}|\varphi^{I\sigma}_j}|^2\right)=\pm 1$ and similarly the difference in the total projection onto the ${\rm t}_{\rm 2g}$ orbitals, $\sum_{I,\;j\,\in \,{\rm t}_{\rm 2g}}\left(|\braket{\Psi_{\rm L}^{\sigma}|\varphi^{I\sigma}_j}|^2-|\braket{\Psi_{\rm H}^{\sigma}|\varphi^{I\sigma}_j}|^2\right)=\mp 1$. In subplot (b), the projection of the  highest occupied and lowest unoccupied GKS orbitals  onto the subspace orbitals is instead weighted by their respective subspace occupancy
eigenvalues as detailed in Eq.~\ref{eqn:subspace-occupancy-weighted-projection}. 
In subplot (c),  the highest occupied and lowest unoccupied GKS orbitals have the same orbital resolved subspace occupancies so that $\sum_{I}\left(|\braket{\Psi_{\rm L}^{\sigma}|\varphi^{I\sigma}_j}|^2-|\braket{\Psi_{\rm H}^{\sigma}|\varphi^{I\sigma}_j}|^2\right)=0$, in which case \textcolor{black}{there} can be no first-order correction to the bandgap.}
\label{figure:hubbard_bandgap_correction} 
\end{figure*}

\section{Projection dependence of DFT+U Bandgap Correction 
for Crystalline Materials}
\label{sec:projdependence}
We have shown that the frontier GKS eigenvalue difference of a 
\textcolor{black}{well-converged periodic} system, evaluated using a
conventional DFT+U functional, is rigorously equal to the fundamental bandgap at the same level of theory, 
and that this equality holds independently of the nature of the subspace projection.
This does not necessarily imply, however, that application of a DFT+$U$ functional will open a bandgap of magnitude $U_{\rm eff}$. To elaborate further on this, we again consider the widely used Hubbard corrective functional of Dudarev et al.~\cite{dudarevElectronenergylossSpectraStructural1998}, whose corrective potential in the  spin-unrestricted formalism may be expressed in terms of the effective Hubbard parameter $U_{\rm eff}^I$ by 
\begin{equation}
    \hat{v}^{\sigma}=\sum_{I}\frac{U^I_{\rm eff}}{2}\left(\hat{P}_I-2\hat{P}_I\hat{\rho}^{\sigma}\hat{P}_I\right),
\end{equation}
where $\hat{P}_I$ is the subspace projection operator at site $I$ and $\hat{\rho}^{\sigma}$ is the spin-$\sigma$ Generalized Kohn Sham (GKS) density operator. This expression for the corrective potential can be readily recast in terms of the subspace occupancy eigenvalues $\{\lambda_j^{I \sigma}\}$ and corresponding eigenstates $\{\varphi_j^{I \sigma}\}$,
\begin{equation}
    \hat{v}^{\sigma}=\sum_{I,j}\frac{U^I_{\rm eff}}{2}\left(\ket{\varphi^{I \sigma}_j}\bra{\varphi^{I\sigma}_j}-2\ket{\varphi^{I\sigma}_j}\lambda_j^{I\sigma}\bra{\varphi^{I\sigma}_j}\right).
\end{equation}
To first-order in perturbation theory and assuming no GKS orbital reordering or geometry re-optimization on application of the Hubbard $U$ correction, Dudarev's functional will offer a correction to every spin-resolved GKS state $\{\Psi_{i}^{\sigma}\}$, the strength of which may be expressed as
\begin{equation}
   \braket{ \Psi_{i}^{\sigma}|\hat{v}^{\sigma}|\Psi_{i}^{\sigma}}=\sum_{I,j}\frac{U^I_{\rm eff}}{2}(1-2\lambda_j^{I\sigma})|\braket{\Psi_{i}^{\sigma}|\varphi^I_j}|^2.
\end{equation}
To first-order in perturbation theory, the spin-resolved correction to the GKS bandgap is given by 
\begin{align}
   &\Delta^{\sigma}_{\rm GKS}=   \\ & \quad \sum_{I,j}\frac{U^I_{\rm eff}}{2}(1-2\lambda_j^{I\sigma})\left(|\braket{\Psi_{\rm L}^{\sigma}|\varphi^{I\sigma}_j}|^2-|\braket{\Psi_{\rm H}^{\sigma}|\varphi^{I\sigma}_j}|^2\right), \nonumber
\end{align}
where $\Psi_{\rm H}^{\sigma}$ and $\Psi_{\rm L}^{\sigma}$ are the highest occupied and lowest unoccupied spin-$\sigma$ GKS states. If we now assume that we are treating only one equivalent set of sites at the DFT+$U$ level, the site index dependence of both the Hubbard parameter and the subspace occupancy eigenvalue, $\lambda_j^{I\sigma}$, can be removed to yield the  expression
\begin{align}
\label{eqn:hubbard_potential_working_equation}
   \Delta_{\rm GKS}^{\sigma}=&\frac{U_{\rm eff}}{2}\sum_{j}\bigg\{(1-2\lambda_j^{\sigma}) \\
  &\sum_{I}\left(|\braket{\Psi_{\rm L}^{\sigma}|\varphi^{I\sigma}_j}|^2-|\braket{\Psi_{\rm H}^{\sigma}|\varphi^{I\sigma}_j}|^2\right)\bigg\}.
  \nonumber 
\end{align}
The inner summation of Eq.~\ref{eqn:hubbard_potential_working_equation} evaluates the difference in the $\Psi_{\rm L}^{\sigma}$ and $\Psi_{\rm H}^{\sigma}$ occupancy of orbital $\varphi_j^{I\sigma}$ over all sites with index label $I$. From Eq.~\ref{eqn:hubbard_potential_working_equation}, it is immediately apparent  that Dudarev's functional will offer no first-order correction to the bandgap if $\Psi_{\rm H}^{\sigma}$ and $\Psi_{\rm L}^{\sigma}$ have the same orbital resolved subspace occupancies, i.e., identical projection. In contrast, Dudarev's functional will offer the maximum bandgap correction when $\Psi_{\rm L}^{\sigma}$ and $\Psi_{\rm H}^{\sigma}$ project onto different orbitals within the subspace, i.e., maximally distinct projection. This is a subtle but important point.

In at least some solid-state materials that are gapless at the (semi-)local DFT level, one can expect both $\Psi_{\rm H}^{\sigma}$ and $\Psi_{\rm L}^{\sigma}$ to have the same orbital resolved subspace occupancies as they come from the same partially filled band at the Fermi level. In such cases Dudarev's functional cannot, to first-order in perturbation theory, offer a correction to the bandgap. 

In contrast, in the case of materials which exhibit an erroneously small but finite gap at the (semi-)local DFT level, $\Psi_{\rm H}^{\sigma}$ and $\Psi_{\rm L}^{\sigma}$ are in distinct valence and conduction bands and as such, in most cases one can expect $\Psi_{\rm H}^{\sigma}$ and $\Psi_{\rm L}^{\sigma}$ to have different orbital resolved subspace occupancies. Such cases are thus amenable to bandgap correction. However, even in the case of such maximally distinct projection, Dudarev's functional offers no first-order correction to the bandgap if all the spin-resolved subspace occupancy eigenvalues are equal, this is due to the $(1-2\lambda_j^{\sigma})$ factor in Eq.~\ref{eqn:hubbard_potential_working_equation}. 

These findings are summarized in the subplots (a) and (c) of Fig.~\ref{figure:hubbard_bandgap_correction}  in the case of a d-orbital subspace with octahedral or tetrahedral crystal field splitting and both $\Psi_{\rm L}^{\sigma}$ and $\Psi_{\rm H}^{\sigma}$ projecting perfectly onto the Hubbard subspace. In subplot (b), assuming no symmetry breaking we interpolate between these two limits, allowing a subspace occupancy weighted projection taking the form,
\begin{align}
&\sum_I|\braket{\Psi_{\rm H}^{\sigma}|\varphi^{I\sigma}_j}|^2=\frac{\lambda_j^{\sigma}}{N^{\sigma}} \qquad \& \nonumber \\  & \sum_I|\braket{\Psi_{\rm L}^{\sigma}|\varphi^{I\sigma}_j}|^2=\frac{1-\lambda_j^{\sigma}}{(2l+1)-N^{\sigma}},
\label{eqn:subspace-occupancy-weighted-projection}
\end{align}
where like in Eq.~\ref{eqn:hubbard_potential_working_equation}, we assume only one equivalent set of sites and $N^{\sigma}$ \& $l$ are the spin-$\sigma$ subspace occupancy and subspace orbital angular momentum quantum number respectively. This subspace occupancy weighted projection ensures that $\Psi_{\rm H}^{\sigma}$ projects more strongly on the more occupied subspace orbitals and vice-versa for  $\Psi_{\rm L}^{\sigma}$. 

In Fig.~\ref{figure:hubbard_bandgap_correction} we label the two distinct spin-resolved subspace eigenvalues as $\lambda_{{\rm e_g}}^{\sigma}$ and $\lambda_{{\rm t_{2g}}}^{\sigma}$, although strictly speaking this labeling should be reserved for the atomic d-orbitals through which the subspace projection operator is defined and not for labeling the subspace occupancy eigenstates. Nevertheless, under octahedral or tetrahedral crystal field splitting in the absence of further symmetry breaking there will be two eigenspaces (of dimensions 2 and 3) of the d-orbital spin-resolved subspace occupancy matrix. The decision of what labels to ascribe to these two eigenspaces is arbitrary and does not affect the results. Here, we choose the e$_{\rm g}$ and t$_{\rm 2g}$ labels as in many materials of interest with octahedral and tetrahedral crystal field splittings, the subspace occupancy eigenvectors will closely resemble the e$_{\rm g}$ and t$_{\rm 2g}$ atomic orbitals, respectively. Furthermore, we also note that, strictly speaking, the gerade (g) symmetry subscript should be omitted in the case of tetrahedral crystal field splitting. 

In summary, for this Section, the band edge states must project onto DFT+$U$
subspaces in such a way that their corresponding (GKS)-orbital resolved subspace occupancies differ, if DFT+$U$ is to be able to change the bandgap (to first order in perturbation theory). The efficacy of even the Dudarev DFT+$U$ functional, in this regard, has a subtle dependency on GKS-orbital resolved, DFT+$U$ projector orbital resolved occupancies. This is a different matter entirely, however, to the question on bandgap validity from DFT+$U$ in principle, which hold for well-converged crystals even with very poor DFT+$U$ subspace choices.

\section{Future Outlook}
Strongly correlated materials, for which the DFT$+U$ method was originally designed, critically underpin numerous current and nascent 
technologies, including high-performance ion-insertion battery cathodes~\cite{C1EE01598B}, multiferroics~\cite{Fiebig2016}, heterogeneous catalysts for renewable energy~\cite{B800489G}, and high-temperature superconductivity~\cite{SHIOHARA20131,Si2016}. The challenge of modeling these materials is a 
compelling and rewarding one due to the very rich variety of emergent  
many-body effects that these materials exhibit, such as exotic 
magnetism~\cite{Eerenstein2006,ADMA:ADMA201202018}, multiple exciton generation~\cite{Sambur63}, quasiparticle satellites~\cite{PhysRevB.87.155147,PhysRevLett.99.266402}, and heavy-fermion behaviour~\cite{PhysRevB.75.024503}. 

A particularly urgent emphasis 
is falling upon new materials that may facilitate a prompt
transition to renewable energy technologies, such as catalysts and
electrodes for batteries and photovoltaic cells.
Moreover, the state-of-the-art in the applied simulation of 
  energy differences of 
 relevance to heterogeneous catalysis is also focused on 
correlated-electron metals and metal oxides~\cite{B800489G,C1EE01598B},
due to their multiple-valence character 
(i.e., changeable oxidation state) and adsorption energies
ranging from low (for metals) to high and even  
tuneable (for oxides)~\cite{ANIE:ANIE200705739}. In all of the areas mentioned, the energies central to phase stability 
and functionality are 
now  almost
ubiquitously  calculated using 
DFT+U~\cite{PhysRevB.56.4900,anisimovFirstprinciplesCalculationsElectronic1997,doi:10.1063/1.4922693,
Cococcioni147,himmetogluHubbardcorrectedDFTEnergy2014} 
with empirically-fitted parameters~\cite{loschenFirstprinciples$mathrmLDA+mathrmU$$mathrmGGA+mathrmU$2007}.
Although hybrid DFT~\cite{doi:10.1021/jp203112p,QUA:QUA24548,PSSB:PSSB201046384}
is very widely used in solid-state chemistry,
it has become evident that DFT+$U$ currently provides the
 \emph{only} empirically accurate methodological framework  that
remains  practicable in high-throughput contexts~\cite{JAIN20112295,SETYAWAN2010299,Curtarolo2013} 
and suitable for the large simulation cell sizes  needed 
for complex physics~\cite{Norskov2009,0953-8984-14-11-303}.
However, it has been shown that DFT+$U$ does \emph{not}  yield  reliable total 
energies, in practice~\cite{doi:10.1063/1.3660353,PhysRevB.84.045115, doi:10.1021/ct200202g,scherlisSimulationHemeUsing2007,0953-8984-29-21-215701,
PhysRevB.95.045141,PhysRevB.93.085135}. 
It is so potent and highly efficient, however, being an excellently scalable  and
almost a cost-free correction to approximate
DFT, that it is very worthwhile to try to understand the reasons \emph{why} this 
current failure  comes about~\cite{burgessMathrmDFTTexttypeFunctional2023}.
The DFT+$U$ family of functionals, over their three decade
history, have  been applied in thousands of 
published research articles as detailed in Fig.~\ref{fig:code_popularity}.
It is  an important challenge to see if it 
can be made reliable.

On one hand, there is no reliable numerical 
comparability of total energies across 
stoichiometries and structures when 
using a fixed Hubbard $U$ parameter, since the latter 
is an inherently system-dependent quantity~\cite{PhysRevB.85.155208,wangOxidationEnergiesTransition2006,zhouFirstprinciplesPredictionRedox2004}.
On the other hand, when one advances to using
 system-dependent first-principles $U$ parameters, 
within state-of-the-art linear-response theory~\cite{PhysRevB.71.035105,himmetogluFirstprinciplesStudyElectronic2011}, 
the direct comparability of total energies  
can sometimes break down~\cite{doi:10.1063/1.3660353}.
Various pragmatic approaches have been 
devised to deal with the necessary stoichiometry dependence of the
Hubbard $U$.
One pioneering empirical  technique, 
developed in the context of high-throughput 
battery-material simulation, 
defined a Hubbard $U$ that interpolated between
first-principles values 
based on the stoichiometric ratio
of the compound of interest~\cite{doi:10.1021/cm702327g}.
Another proposed solution is to
mix the DFT and DFT+$U$ total
energies~\cite{PhysRevB.84.045115}, which has been   
found to be a successful  work-around
for calculating first-row transition metal oxide 
formation energies and phase 
diagrams, resulting in  a good agreement with experiment.
Indeed, the accurate calculation of phase diagrams are of 
central importance in this area~\cite{doi:10.1021/cm0620943,PhysRevB.85.155208,C1EE01782A}.
Pragmatic measures such as the fitting of $U$ values 
to experimental results such as binary formation 
energies~\cite{wangOxidationEnergiesTransition2006}, or the 
 mixing of energies from different theories, 
albeit  insightful and successful~\cite{doi:10.1063/1.4812323,C1EE01782A,Hautier2012,lambertRSC},  
 highlight the need for a fundamental breakthrough in this area of condensed matter theory.
Without it, progress in strongly-correlated materials prediction is hampered.

Recent advances made by incorporating findings from quantum chemistry~\cite{flat_plane_and_deriv_discon,quantifying_flat_plane_error,bajajNonempiricalLowcostRecovery2019,functional_to_correct_flat_plane}  
in the development of DFT$+U$-type functionals have shown much promise. %
In particular, developing  
 corrective functionals   
 to enforce the most relevant  exact conditions~\cite{perdewDensityFunctionalTheoryFractional1982,doi:10.1063/1.2987202,PhysRev.123.420,mori-sanchezDiscontinuousNatureExchangeCorrelation2009,flat_plane_and_deriv_discon},  
 rather than following the traditional approach of requiring  derivability from the  
 Hubbard model~\cite{Hubbard_original_paper,anisimovBandTheoryMott1991}. Underpinning this approach is the working hypothesis
that significant accuracy gains are inexpensively within reach, 
within generalised Kohn-Sham 
DFT~\cite{seidl1996generalized},   
 by efficiently imposing known exact conditions~\cite{perdewDensityFunctionalTheoryFractional1982,doi:10.1063/1.2987202,daboKoopmansConditionDensityfunctional2010}
 on selected error-prone subspaces.   %
Indeed, there exists a large body of evidence,
found in the literature over the past decade and more~\cite{PhysRevB.71.035105,doi:10.1063/1.4922693,Cococcioni147}, 
that has strongly indicated that DFT+$U$ may be 
interpreted as an efficient 
corrector for electronic self-interaction error (SIE)
or more generally delocalisation error~\cite{doi:10.1063/1.2403848,perdew_bond_dissocation1,perdew_bond_dissocation2,delocalisation_error_is_quadratic}, 
in approximate DFT~\cite{cohen2008insights}, 
given a differential definition of that  error 
based on spurious energy-occupancy 
curvature.

As well as affecting the total energy, SIE is associated  with
a  characteristic underestimation of the insulating gap~\cite{MSIE_bandgap_error,bandgap_benchmarking}. This definition is technically distinct but complementary to the SIE
corrected by Perdew-Zunger type DFT-SIC~\cite{perdewSelfinteractionCorrectionDensityfunctional1981,PhysRevB.37.9919,PhysRevB.47.4029}.
Further progress in this area will  require 
suitably-constructed subspaces, 
together with the necessary form of
corrective functional,
which requires parameters appropriately calculated from first-principles. 
Ideally, these features together 
might be packaged in a self-contained
and automated way, from the user's
perspective, so that reliable, minimal-empiricism 
total energy differences may be extracted
from DFT+$U$ with the same convenience
(we can now be assured, for crystals) 
that band gaps may be.

\section{Conclusions}
We have rigorously proven that the GKS eigenspectrum gap of a pristine periodic system as computed by a DFT+$U$ or hybrid functional is a valid measure of the fundamental gap of a material, in the sense that it matches its own fundamental gap calculated using total-energy differences. 
Furthermore, under the same conditions, the negative of the highest occupied and lowest unoccupied GKS eigenvalues are rigorously equal to the ionisation potential and electron affinity of the material, respectively. This proof was \textcolor{black}{theoretically verified} to hold for all known  DFT+$U$-type functionals, subject only to caveats that are met or accounted for in practice.
Our proof recovers the previously demonstrated 
same result for local and semi-local functionals, with a somewhat
different argument.}

The diverse set of  DFT$+U$-type functionals will continue to grow in the coming decades. To date, the vast majority of DFT+$U$ functionals have been constructed by starting with one of a  range of interaction terms, typically inspired either by Hartree Fock theory or the Hubbard model, and subtracting off a double counting correction, the form of which often strongly depends on the choice of interaction term. 

Future DFT+$U$ functionals will almost certainly be constructed within the `extended DFT+$U$' category, via the retention of additional subspace averaged parameters from the Hartree Fock inspired interaction term, and within the `flat plane condition' category, via the imposition of additional exact conditions on the selected subspaces or the identification of more advanced interaction terms inspired by quantum chemistry techniques. 

\textcolor{black}{The proof of the DFT+$U$ eigenspectrum gap being a valid measure of the fundamental gap of a material does not, however, ensure close agreement with experiment.
In this sense, DFT+$U$ is no different to conventional hybrid functionals.}
If progress is to be made, novel functional forms must retain the desirable features of conventional DFT+$U$ functionals, such as improved bandgaps and magnetic moments of transition metal oxides, while also offering improved total energies and bandgaps in more challenging electronic systems. 
\textcolor{black}{Model yet
chemical 
systems, exemplified by the 
non-spin polarized hydrogen lattice, can helpfully simplify and reveal
the differences in behavior between different functionals and methods, and guide
progress.}

\section{ACKNOWLEDGMENTS}
This research was supported by Taighde-Éireann 
-- \textcolor{black}{Research} Ireland 
under Grant No. GOIPG/2020/1454 and under Prime Award No. 12/RC/2278$\_$P2, the latter of  which is co-funded by the European Regional Development Fund.
\textcolor{black}{The authors further acknowledge Enterprise Ireland, project CS20212086.} 
The authors wish to thank 
Tim Gould, Stefano de Gironcoli,
\textcolor{black}{Nathalie Vast,
and Nicola Marzari} for 
stimulating discussions.

\appendix
\section{Computational Details}
\label{appendix:comp_details}
\textcolor{black}{Electronic structure calculations to investigate the numerical difference between the DFT+$U$ eigenspectrum gap and the fundamental gap (as calculated via total-energy differences) were executed using  Quantum Espresso code version 7.4~\cite{giannozziAdvancedCapabilitiesMaterials2017}. The conventional DFT+$U$ functional of Dudarev et al.~\cite{dudarevElectronenergylossSpectraStructural1998} with an effective Hubbard $U$ parameter of 5 eV was applied with the Perdew-Burke-Ernzerhof XC functional~\cite{perdewGeneralizedGradientApproximation1996}. 
DFT+$U$ projectors were of standard pseudo-atomic type, without
orthonormalization. 
A 2-dimensional square hydrogen lattice with an inter-nuclear separation length of 7$a_0$ was selected for this study. A rectangular primitive unit cell composed of two antiferromagnetically ordered hydrogen atoms was used, giving
ordering in the x-direction. 
A large out of plane unit cell length of 50$a_0$ was used, in conjunction with the Coulomb interaction truncation technique for 2-dimensional systems devised by Sohier et al.~\cite{sohier2017density}.  A SG15 v1.0~\cite{SCHLIPF201536}
scalar-relativistic optimized norm-conserving Vanderbilt pseudopotential~\cite{hamann2013optimized} was applied and the plane wave cutoff energy was set at 400 Ry (2000 Ry for the density).
The default mixing scheme proved adequate.} 

\textcolor{black}{For the charged calculations, the total charge of the primitive unit cell was set as $\pm 1/N_k$, where $N_k$ is the total number of k-points~\cite{perdew2017understanding}. This is equivalent to a collection of $N_k$ primitive unit cells in periodic boundary conditions with a single electron added or removed. In the bulk limit, i.e., where $N_k \rightarrow \infty$, the total energy per unit cell is equal to that of the $\pm 1$ charged system, assuming the added electron or hole delocalizes over the lattice. The total magnetization of the primitive unit cell was set in accordance with the total charge.
The cold-smearing method of Marzari et al.~\cite{marzari1999thermal} was applied with a low smearing temperature of $1\times 10^{-9}$ Ry to ensure that integer GKS orbital occupancies were retained under such conditions.}

\section{Explicit Derivative Discontinuities}
\label{appendix:explicitderiv}
The need for the exact functional  to exhibit an explicit derivative discontinuity in the case of the hydrogen atom has already been discussed in detail by Mori-S\'{a}nchez et al.~\cite{mori-sanchezDiscontinuousNatureExchangeCorrelation2009}.  Here we review this hitherto under-appreciated point in the case of exact GKS theory. By exact GKS theory here, we refer to all possible forms of exact GKS theory as any functional of the occupied GKS orbitals can be made exact through the use of a complimentary remainder functional~\cite{seidl1996generalized,garrick2020exact}. The absence of a minority spin population in the hydrogen atom complicates matters, so we consider here the neutral lithium atom with two spin up electrons and one spin down electron and a ground state electron configuration 1s$^2$2s$^1$. By comparison,  the lithium anion has a spin singlet ground state of the form  1s$^2$2s$^2$ composed of two spin up and two spin down electrons, while the lithium cation has a ground state electron configuration 1s$^2$, with one spin up and one spin down electron. \textcolor{black}{This simple atomic system has also been analyzed extensively within the context of exact KS theory~\cite{gritsenkoSpinunrestrictedMolecularKohn2004} and ensemble DFT~\cite{gould2026ensemblization}. }

\textcolor{black}{In the case of GKS theory,} in order to reproduce the correct fundamental gap in the absence of an explicit derivative discontinuity, the difference between the highest occupied spin up and lowest unoccupied spin down GKS eigenvalues of the neutral lithium atom must be equal \textcolor{black}{to} 4.77 eV, which is the difference between lithium's ionization potential (5.39 eV) and electron affinity (0.62 eV). However, in order to reproduce the correct spin-flip energy~\cite{burgessTiltedPlaneStructureEnergy2024,capelleSpinGapsSpinflip2010,galEnergySurfaceChemical2010,hayman2025spin,chanFreshLookEnsembles1999,yangDerivativeDiscontinuityBandgap2012,gritsenkoAnalogKoopmansTheorem2002,gritsenkoSpinunrestrictedMolecularKohn2004,galNonuniquenessMagneticFields2009,galvan1992spin} in the absence of an explicit derivative discontinuity, defined in this case as the energy difference between the neutral lithium atom with an unpaired spin up electron versus an unpaired spin down electron, the same frontier GKS eigenvalue difference must, due to spin degeneracy, be equal to 0 eV. Both of these conditions cannot be satisfied simultaneously, which thus warrants the introduction of an explicit derivative discontinuity~\cite{yangDerivativeDiscontinuityBandgap2012}.

\section{Bandgaps in the presence of  Norm Conserving Pseudopotentials}
\label{appendix:ncpp}
In most practical calculations the bare Coulomb potential of the static nuclei is replaced by a pseudopotential to avoid explicit and costly treatment of the core electrons. Norm-Conserving Pseudopotentials (NCPP) as originally devised by Hamann et al.~\cite{hamann1979norm} are local in the radial coordinate but non-local in angular coordinates $\Omega$ and may be expressed in terms of the set of spherical harmonics $\{\ket{Y_{lm}}\}$ and element-specific local and non-local contributions, $v_I^L$ and $\Delta v_{Il}^{NL}$, located at atom site $I$ as given by ${v}_{\rm NCPP}({\bf r},{\bf r'})$ in Eq.~\ref{eqn:NCCP}. In practice, in order to improve computational efficiency, most modern NCPPs are converted into a fully non-local form via the Kleinman-Bylander scheme~\cite{kleinman1982efficacious}. The resulting non-local potentials may be expressed as
\begin{widetext}
\begin{align}
&{v}_{\rm NCPP}({\bf r},{\bf r'})=\sum_Iv_I^L(|{\bf r}-{\bf R}_I|)\delta({\bf r}-{\bf r'})+\sum_I\sum_{lm}Y_{lm}\left(\Omega_{{\bf r}-{\bf R}_I}\right)\Delta v_{Il}^{NL}\left(|{\bf r}-{\bf R}_I|\right)\delta\left(|{\bf r}-{\bf R}_I|-|{\bf r'}-{\bf R}_I|\right)Y^*_{lm}\left(\Omega_{{\bf r'}-{\bf R}_I}\right), \nonumber \\ 
&{v}_{\rm KB}({\bf r},{\bf r'})=\sum_I\braket{{\bf r}-{\bf R}_I|v_I^L|{\bf r'}-{\bf R}_I}\delta({\bf r}-{\bf r'}) +\sum_I\sum_{lm}\frac{\braket{{\bf r}-{\bf R}_I|\Delta v_{Il}^{NL}\phi_{Ilm}} \braket{\Delta v_{Il}^{NL}\phi_{Ilm}|{\bf r'}-{\bf R}_I}}{\braket{\phi_{Ilm}|\Delta v_{Il}^{NL}|\phi_{Ilm}}},
\label{eqn:NCCP}
\end{align}
\end{widetext}
where the radial coordinate dependence of $v_I^L$ and $\Delta v_{Il}^{NL}$ has been suppressed in ${v}_{\rm KB}({\bf r},{\bf r'})$ for clarity and $\{\phi_{Ilm}\}$ are the reference pseudo-atomic wavefunctions. 
Unlike the Hubbard, Hartree and Base Exchange-Correlation potentials the NCPPs of Eq.~\ref{eqn:NCCP} are independent of the first-order reference-system reduced density matrix and, by extension, are independent of the frontier orbital occupancy. As such, the NCPPs do not offer an explicit contribution to the GKS eigenvalue derivative of Eq.~\ref{eqn:GKS-eigenvalue-deriv-final} or any higher-order derivatives. Thus, the simple expressions for the ionisation potential, electron affinity and fundamental bandgap of defect-free infinitely extended systems, given by Eqs.~\ref{eq:ip_pristine_solid}, \ref{eq:ea_pristine_solid} \& \ref{eq:bandgap_pristine_solid}, remain valid in the case of frontier GKS eigenvalues evaluated using NCPPs. To be clear, these values will match the ionisation potential, electron affinity and fundamental bandgap, respectively,  evaluated via explicit total-energy differences employing the same set of NCPPs.

\section{Bandgaps within the    Ultrasoft and Projector-Augmented Wave Formulations}
\label{appendix:ultrasoftpaw}
In the case of Ultrasoft pseudopotential (USPP)~\cite{vanderbilt1990soft,laasonen1993car} and Projector-Augmented Wave (PAW)~\cite{blochl1994projector,kresse1999ultrasoft} formulations, the orthonormality condition of the pseudo-GKS states $\{\tilde{\psi}_i\}$ is relaxed to ensure a relatively `soft', i.e., smooth profile in the atomic core region. The pseudo-GKS states instead satisfy a generalized orthonormality condition of the form,
\begin{equation}
\braket{\tilde{\psi}_i|\hat{S}|\tilde{\psi}_j}=\delta_{ij},
\end{equation}
where $\hat{S}$ is an overlap (a.k.a., augmentation)
operator. In both the USPP and PAW formalisms, the overlap operator may be defined using a set of smooth projector functions $\{\tilde{p}_{\mu}\}$ and takes the following form,
\begin{align}
&\hat{S}_{\rm USPP}=1+\sum_{\mu\nu} q_{\mu \nu }\ket{\tilde{p}_{\mu}}
\bra{\tilde{p}_{\nu}} \qquad \& \nonumber \\ & \hat{S}_{\rm PAW}=1+\sum_{\mu\nu}\left[\braket{\phi_{\mu}|\phi_{\nu}}-\braket{\tilde{\phi}_{\mu}|\tilde{\phi}_{\nu}}\right]\ket{\tilde{p}_{\mu}}\bra{\tilde{p}_{\nu}} ,
\label{eqn:overlap_operator}
\end{align}
where $\{\phi_{\mu}\}$ and $\{\tilde{\phi}_{\mu}\}$ are the all-electron atomic and pseudo atomic states and $\{q_{\nu \mu}\}$ are the USPP augmentation charges. Here, $\mu$ and $\nu$ are combined atomic state and site indices and the summations in both Eq.~\ref{eqn:overlap_operator} and below are restricted to on-site terms only.  This necessitates a modification of the total energy expression of Eq.~\ref{eqn:total_energy_including_orthonormality_and_orbital_filling} which may be expressed as 
\begin{align}
E_v &{}= - \sum_{i} \tilde{f}_{i}
\frac{ 
 \langle \tilde{\psi}_i \rvert  \nabla^2
\lvert \tilde{\psi}_i \rangle}{2}
+ E_{\mathrm{Hxc}+U+\Delta E_\mathrm{a}} 
\left[ \{\tilde{\psi}_i\},\{\tilde{f}_{i }\} \right]
\nonumber \\  
&{}\quad\quad+ \sum_{ij}
 \lambda_{i j }
\left( \delta_{j i} -  \langle \tilde{\psi}_j \rvert \hat{S}\rvert 
\tilde{\psi}_i \rangle 
 \right) 
  \\  \nonumber
&{}\quad\quad\quad + \epsilon_\mathrm{F} \left( N -  
 \sum_{i } \tilde{f}_{i }  \langle \tilde{\psi}_i \rvert \hat{S} \rvert 
\tilde{\psi}_i \rangle   \right),
\end{align}
where the external energy due to the  bare Coulomb potential of the static nuclei has been replaced by a USPP/PAW specific atomic energy correction $\Delta E_a$.  This total energy expression is similarly minimized with respect to its 
 degrees of freedom
$\left\lbrace \tilde{\psi}_i \right\rbrace$, $\left\lbrace \tilde{\psi}_i^\ast  \right\rbrace$, 
$\tilde{f}^i$, $\lambda_{ij}$, 
and $\epsilon_\mathrm{F}$, and so again the partial derivatives of
$E_v$ with respect to these vanish. The total derivative of $E_v$ with respect to electron count is thus, again, given by 
\begin{equation}
\frac{dE_v}{dN} = \frac{\partial E_v}{\partial N} 
= \epsilon_\mathrm{F},
    \label{eqn:slope_of_tot_energy_paw}
\end{equation}
in all but points of 
eigenvalue cross-over or degeneracy change of the frontier orbitals. Therefore, the general expressions for the electron affinity, ionization potential and bandgap as given by Eqs.~\ref{eqn:general_ea_expression}, \ref{eqn:general_ip_expression} \& \ref{eqn:general_bandgap_expression} are also valid within the USPP and PAW formalisms once the GKS eigenvalues and occupancies are replaced by their soft-pseudo analogues.

Unlike NCPPs, the USPP/PAW formulations offer non-negligible contributions to the frontier GKS eigenvalue derivatives. Therefore, we must again examine the 
\textcolor{black}{bulk limit} of the frontier GKS eigenvalue derivatives to assess if the electron affinity, ionization potential and fundamental bandgap expressions of Eqs.~\ref{eq:ip_pristine_solid}, \ref{eq:ea_pristine_solid} \& \ref{eq:bandgap_pristine_solid} for a pristine \textcolor{black}{periodic system} remain valid.  In the case of the USPP formalism, the first derivative of the frontier pseudo-GKS eigenvalue with respect to its  occupancy consists of a Hxc, Hubbard and USPP contribution, and may be expressed as 
\begin{widetext}
\begin{align}
\label{eqn:uspp-eigenvalue-deriv}
   &\frac{d\tilde{\epsilon}_{\rm F}}{d\tilde{f}_{\rm F}}= \bra{\tilde{\Psi}_{\rm F}}\Bigg\{ \int d{\bf r} d{\bf r'} \ket{{\bf r}}f_{\rm Hxc}({\bf r},{\bf r'})\left(\tilde{F}_{\pm}({\bf r'},{\bf r'})+\sum_{\mu \nu}\int d{\bf r''}d{\bf r'''}\braket{{\bf r''}|\tilde{p}_{\mu}}\braket{\tilde{p}_{\nu}|{\bf r'''}}\tilde{F}_{\pm}({\bf r'''},{\bf r''})Q_{\mu \nu}({\bf r'})\right)\bra{{\bf r}} \nonumber \\ &  -\sum_I \int d{\bf r}d{\bf r'} d{\bf r''}d{\bf r'''}\ket{{\bf r}}\left(U^I\braket{{\bf r}|\hat{S}\hat{P}^I\hat{S}|{\bf r''}}\braket{{\bf r'''}|\hat{S}\hat{P}^I\hat{S}|{\bf r'}}\right)\tilde{F}_{\pm}({\bf r''},{\bf r'''})\bra{\bf r'} +\sum_{\mu \nu}\int d{\bf r}d{\bf r'} d{\bf r''''}d{\bf r'''''} \ket{{\bf r}}  \braket{{{\bf r}}|\tilde{p}_{\mu}} 
    \\ &  \times \bigg(\tilde{F}_{\pm}({\bf r''''},{\bf r''''}) +\sum_{\mu' \nu'}\int d{\bf r''}d{\bf r'''}\braket{{\bf r''}|\tilde{p}_{\mu'}}\braket{\tilde{p}_{\nu'}|{\bf r'''}}\tilde{F}_{\pm}({\bf r'''},{\bf r''})Q_{\mu' \nu'}({\bf r''''})  \bigg)  f_{\rm Hxc}({\bf r'''''},{\bf r''''})Q_{\mu \nu}({\bf r'''''}) \braket{\tilde{p}_{\nu}|{\bf r'}}\bra{{\bf r'}} \Bigg\}\ket{\tilde{\Psi}_{\rm F}}, \nonumber
\end{align}
\end{widetext}
where $\{Q_{\mu \nu}({\bf r})\}$ are the augmentation charge densities, the integrals of which yield the augmentation charges $\{q_{\nu \mu}\}$ of Eq.~\ref{eqn:overlap_operator}, $\tilde{F}_{\pm}({\bf r},{\bf r'})$  is the  soft-pseudo  Fukui matrix and the Hxc kernel is evaluated on the total electron density, i.e., including both soft and hard contributions. Focusing on the Hxc contribution, the term in parentheses consists of a Fukui matrix and a second term with two spatial integrals and a summation over the combined atomic state and site indices $\mu$ and $\nu$.   \textcolor{black}{For a given site $I$,  the $\braket{{\bf r''}|\tilde{p}_{\mu}}\braket{\tilde{p}_{\nu}|{\bf r'''}}Q_{\mu \nu}({\bf r'})$ term will offer a negligible contribution for large values of  $|{\bf R}_I-{\bf r'}|$, $|{\bf R}_I-{\bf r''}|$ and $|{\bf R}_I-{\bf r'''}|$. This counters the two spatial integrals and summation over $\mu$ and $\nu$, noting that the augmentation charge density $Q_{\mu \nu}({\bf r'})$ is diagonal in site index. Finally, the two Fukui matrices are left to ensure that both terms in parentheses  vanish in the bulk limit.} Outside the parentheses, there is a further two spatial integrals but these are countered by the Hxc kernel and the frontier orbital density $\braket{\tilde{\Psi}_{\rm F}|{\bf r}}\braket{{\bf r}|\tilde{\Psi}_{\rm F}}$. Therefore, the Hxc contribution to the GKS eigenvalue derivative evaluated with USPPs \textcolor{black}{vanishes in the bulk limit}. Same too goes for the Hubbard contribution, whose only modification is the subspace projection operator $\hat{P}^I$ which has been replaced by $\hat{S}\hat{P}^I\hat{S}$~\cite{timrovSelfconsistentHubbardParameters2021}. 

In the case of the explicit USPP contribution to the GKS eigenvalue derivative, the term in parentheses is the same as that discussed for the Hxc term and thus vanishes in the bulk limit. Outside of the parentheses, \textcolor{black}{ the  $\braket{{\bf r}|\tilde{p}_{\mu}}\braket{\tilde{p}_{\nu}|{\bf r'}}Q_{\mu \nu}({\bf r''''})$ term offers a negligible contribution for large values of  $|{\bf R}_I-{\bf r}|$, $|{\bf R}_I-{\bf r''}|$ and $|{\bf R}_I-{\bf r''''}|$. This term along with the Hxc kernel and frontier orbital density matrix $\braket{\tilde{\Psi}_{\rm F}|{\bf r'}}\braket{{\bf r}|\tilde{\Psi}_{\rm F}}$ counter the summation and four spatial integrals.} Therefore, the explicit USPP contribution to the GKS eigenvalue derivative also vanishes in the bulk limit.


For PAW, we follow the formalism \textcolor{black}{proposed} by Kresse and Joubert~\cite{kresse1999ultrasoft} to arrive at an expression for the derivative of the soft-pseudo GKS eigenvalue 
with respect to frontier pseudo-GKS orbital occupancy, namely 
\onecolumngrid
\begin{align}
\label{eqn:PAW_eigenvalue_deriv}
\frac{d\tilde{\epsilon}_{\rm F}}{d\tilde{f}_{\rm F}}=& \bra{\tilde{\Psi}_{\rm F}}\Bigg\{ -\sum_I\int d{\bf r}d{\bf r'} d{\bf r''}d{\bf r'''} \ket{{\bf r}}\left(U^I\braket{{\bf r}|\hat{S}\hat{P}^I\hat{S}|{\bf r''}}\braket{{\bf r'''}|\hat{S}\hat{P}^I\hat{S}|{\bf r'}}\right)\tilde{F}_{\pm}({\bf r''},{\bf r'''})\bra{\bf r'} \nonumber  \\& \qquad +\int d{\bf r}d{\bf r'}\ket{{\bf r}}f_{\rm Hxc}[\tilde{n}+\hat{n}+\tilde{n}_{\rm c}]({\bf r},{\bf r'})\bigg(\tilde{F}_{\pm}({\bf r'},{\bf r'})+ \sum_{\mu \nu lm}\int d{\bf r''} d{\bf r'''}\braket{{\bf r''}|\tilde{p}_{\mu}}\braket{\tilde{p}_{\nu}|{\bf r'''}} \tilde{F}_{\pm}({\bf r'''},{\bf r''}){Q}_{\mu \nu}^{lm}({\bf r'})\bigg)\bra{{\bf r}}  \nonumber \\ &   \qquad  +\sum_{\mu \nu}\int d{\bf r}d{\bf r'} \ket{\bf r} \braket{{\bf r}|\tilde{p}_{\mu}}\frac{d{D}_{\mu \nu}}{d\tilde{f}_{\rm F}}\braket{\tilde{p}_{\nu}|\bf r'}\bra{\bf r'}\Bigg\}\ket{\tilde{\Psi}_{\rm F}},
\end{align}
\twocolumngrid
\noindent
where $D_{\mu \nu}=\hat{D}_{\mu \nu}+{D}^1_{\mu \nu}-\tilde{D}_{\mu \nu}$. The latter three quantities are defined in Eqs. 44, 45 $\&$ 46 of Kresse and Joubert~\cite{kresse1999ultrasoft} and their derivatives with respect to frontier pseudo GKS occupancy $\tilde{f}_{\rm F}$ may be expressed as 
\begin{widetext}
\begin{align}
\frac{d \hat{D}_{\mu \nu}}{d\tilde{f}_{\rm F}}=&\sum_{lm}\int d{{\bf r}Q_{\mu \nu}^{lm}({\bf r})} \int d{{\bf r'}}f_{\rm Hxc}[\tilde{n}+\hat{n}+\tilde{n}_{\rm c}]({\bf r},{\bf r'})\Bigg(\tilde{F}_{\pm}({\bf r'},{\bf r'})+\nonumber \\ & \sum_{\mu'\nu'l'm'}\int d{\bf r''} d{\bf r'''}\braket{{\bf r''}|\tilde{p}_{\mu'}}\braket{\tilde{p}_{\nu'}|{\bf r'''}} \tilde{F}_{\pm}({\bf r'''},{\bf r''})Q_{\mu'\nu'}^{l'm'}({\bf r'})\Bigg),
\nonumber \\  
\nonumber \\ 
\frac{d {D}^1_{\mu \nu}}{d\tilde{f}_{\rm F}}=&\bra{\phi_{\mu}}\Bigg\{\int d{\bf r}\ket{{\bf r}}\Bigg(\int d{{\bf r'}}f_{\rm Hxc}[n^1+n_{\rm c}]({\bf r},{\bf r'})\nonumber \\ &\sum_{\mu' \nu'}\int d{\bf r''}d{\bf r'''}\braket{{\bf r''}|\tilde{p}_{\mu'}}\braket{\tilde{p}_{\nu'}|{\bf r'''}} \tilde{F}_{\pm}({\bf r'''},{\bf r''})\braket{\phi_{\mu'}|{\bf r'}}\braket{{\bf r'}|\phi_{\nu'}}\Bigg)\bra{{\bf r}}\Bigg\}\ket{\phi_{\nu}} \qquad  \& 
\nonumber \\  
\nonumber \\ \nonumber
\end{align}
\end{widetext}
\begin{widetext}
\begin{align}
\frac{d \tilde{D}^1_{\mu \nu}}{d\tilde{f}_{\rm F}}=&\bra{\tilde{\phi}_{\mu}}\Bigg\{\int d{\bf r}\ket{{\bf r}}\Bigg(\int d{{\bf r'}}f_{\rm Hxc}[\tilde{n}^1+\hat{n}+\tilde{n}_{\rm c}]({\bf r},{\bf r'})\nonumber \\ &\sum_{\mu'\nu'}\int d{\bf r''}d{\bf r'''}\braket{{\bf r''}|\tilde{p}_{\mu'}}\braket{\tilde{p}_{\nu'}|{\bf r'''}} \tilde{F}_{\pm}({\bf r'''},{\bf r''})\left(\braket{\tilde{\phi}_{\mu'}|{\bf r'}}\braket{{\bf r'}|\tilde{\phi}_{\nu'}}+\sum_{l'm'}Q^{l'm'}_{\mu'\nu'}({\bf r'})\right)\Bigg) \bra{{\bf r}}\Bigg\}\ket{\tilde{\phi}_{\nu}} \nonumber \\ &
+\sum_{lm}\int_{\Omega_{r}} d{\bf r}Q_{\mu \nu}^{lm}({\bf r})\Bigg(\int d{{\bf r'}}f_{\rm Hxc}[\tilde{n}^1+\hat{n}+\tilde{n}_{\rm c}]({\bf r},{\bf r'})\nonumber \\ &\sum_{\mu'\nu'}\int d{\bf r''}d{\bf r'''}\braket{{\bf r''}|\tilde{p}_{\mu'}}\braket{\tilde{p}_{\nu'}|{\bf r'''}} \tilde{F}_{\pm}({\bf r'''},{\bf r''})\left(\braket{\tilde{\phi}_{\mu'}|{\bf r'}}\braket{{\bf r'}|\tilde{\phi}_{\nu'}}+\sum_{l'm'}Q^{l'm'}_{\mu'\nu'}({\bf r'})\right)\Bigg) .
\end{align}
\end{widetext}
The charge densities $\tilde{n}$,  $\tilde{n}_c$ and $\hat{n}$ are the soft pseudo-charge density, the frozen-core all-electron charge density, and the total augmentation charge density, the latter of which can be written as a one-electron multipole expansion of the form
\begin{equation}
    \hat{n}({\bf r})=\sum_{\mu \nu lm}{Q}^{lm}_{\mu \nu}({\bf r})\sum_i\tilde{f}_i\braket{\tilde{\Psi}_i|\tilde{p}_{\mu}}\braket{\tilde{p}_{\nu}|\tilde{\Psi}_i}.
\end{equation}
The GKS eigenvalue derivative in the PAW formulation as exposed by Eq.~\ref{eqn:PAW_eigenvalue_deriv} takes an almost analogous form to that of the USPP formulation with the Hxc and Hubbard terms \textcolor{black}{vanish in the bulk limit}. As for the explicit PAW contribution to the GKS eigenvalue derivative, using the same arguments as before it can be readily shown that the $ d{D}_{\mu \nu} / d\tilde{f}_{\rm F}$ term \textcolor{black}{vanishes in the bulk limit}. The two spatial integrals and summation \textcolor{black}{are} countered by the frontier orbital density matrix and the $\braket{{\bf r}|\tilde{p}_{\mu}}\braket{\tilde{p}_{\nu}|{\bf r'}}$ term, which offers a negligible contribution for large values of  $|{\bf R}_I-{\bf r}|$ and $|{\bf R}_I-{\bf r'}|$.  Thus, the PAW contribution to the GKS eigenvalue derivative also \textcolor{black}{vanishes in the bulk limit} and the simple expressions for the ionisation potential, electron affinity and fundamental bandgap of \textcolor{black}{pristine periodic systems as} given by Eqs.~\ref{eq:ip_pristine_solid}, \ref{eq:ea_pristine_solid} \& \ref{eq:bandgap_pristine_solid}, remain valid in the case of frontier GKS eigenvalues evaluated using the USPP/PAW formulation.

\section{Generalization to Spin-Unrestricted GKS Theory}
\label{appendix:spin}
In the case of the spin-unrestricted GKS formalism, we must derive an explicit expression for the first-order
derivative of the frontier spin-$\sigma$ GKS eigenvalue with respect to its own occupancy, which for collinear spin-systems may be expressed as 
\begin{equation}
    \frac{d\epsilon^{\sigma}_\textrm{F}}{df_\textrm{F}^{\sigma}}=\left\langle\Psi_\textrm{F}^{\sigma}\left|\frac{d\hat{H}^{{\rm GKS}}_{\sigma}}{df_\textrm{F}^{\sigma}}\right|\Psi_\textrm{F}^{\sigma}\right\rangle.
    \label{eqn:Hellmann–Feynman-spin-resolved}
\end{equation}
The derivative of the spin-$\sigma$ GKS Hamiltonian will again consist of two terms, namely the derivative of the spin-$\sigma$ Hxc potential $\hat{v}_{\rm Hxc}^{\sigma}$ and the derivative of the Hubbard potential $\hat{v}_{\rm u}^{\sigma}$. In the case of the spin-$\sigma$ Hxc potential its dependence on both the spin-up and spin-down ground-state densities must be accounted for and, as such, its derivative with respect to frontier spin$-\sigma$ GKS orbital occupancy may be expressed as,
\begin{align}
\label{eqn:Hartree_plus_bxc_final_result_spin_resolved} 
     \frac{d v_{\rm Hxc}^{\sigma} ({\bf r})}{df_\textrm{F}^{\sigma}}=  \sum_{\sigma'}\int d{\bf r'}f_{\rm Hxc}^{\sigma \sigma'}({\bf r},{\bf r'})F^{\sigma' \sigma}_{\pm}({\bf r'}), 
\end{align}
where $F^{\sigma' \sigma}_{\pm}({\bf r'})$ is the spin-resolved analogue of the right or left hand Fukui function, or to to be explicit, the derivative of the ground-state spin-$\sigma'$ density with respect to spin-$\sigma$ electron count. In contrast, the spin-$\sigma$ Hubbard potential of Dudarev et al.~\cite{dudarevElectronenergylossSpectraStructural1998} depends only on the ground-state spin-$\sigma$ GKS density matrix $\rho^{\sigma}_0({\bf r''},{\bf r'''})$. Its derivative with respect to frontier spin$-\sigma$ GKS orbital occupancy may be expressed as 
\begin{align}
  \label{eqn:hubbard_final_result_spin_resolved}
    &\frac{d v_{U}^{\sigma}({\bf r},{\bf r'}) }{df_\textrm{F}^{\sigma}}= \\ & -\sum_I U^I \int d{\bf r''}d{\bf r'''}\left(\braket{{\bf r}|\hat{P}^I|{\bf r''}}\braket{{\bf r'''}|\hat{P}^I|{\bf r'}}\right)F^{\sigma\sigma}_{\pm}({\bf r''},{\bf r'''}),
    \nonumber
\end{align}
where $F^{\sigma\sigma}_{\pm}({\bf r''},{\bf r'''})$ is the spin-resolved analogue of the the right or left hand Fukui matrix. More advanced spin resolved Hubbard potentials such as that of the DFT+$U$+$J$ or BLOR functionals~\cite{himmetogluFirstprinciplesStudyElectronic2011,burgessMathrmDFTTexttypeFunctional2023} will warrant a summation over spin-index $\sigma'$, akin to that of the Hxc term, due to their dependence on both the spin up and spin down subspace occupancies. From Eqs.~\ref{eqn:Hellmann–Feynman-spin-resolved}, \ref{eqn:Hartree_plus_bxc_final_result_spin_resolved}  \& \ref{eqn:hubbard_final_result_spin_resolved}, it follows that the derivative of the spin-resolved GKS eigenvalue may be expressed as,
\begin{widetext}
\begin{align}
    \frac{d\epsilon_{\rm F}^{\sigma}}{df_{\rm F}^{\sigma}}=\left\langle\Psi_{\rm F}^{\sigma}\right| \int d{\bf r}d{\bf r'} \ket{{\bf r}}\Bigg\{&\sum_{\sigma'}\int d{\bf r''}f_{\rm Hxc}^{\sigma \sigma'}({\bf r},{\bf r''})F^{\sigma' \sigma}_{\pm}({\bf r''},{\bf r''})\delta({\bf r}-{\bf r'}) \nonumber \\ & 
    -\sum_I\int d{\bf r''}d{\bf r'''}\left(U^I\braket{{\bf r}|\hat{P}^I|{\bf r''}}\braket{{\bf r'''}|\hat{P}^I|{\bf r'}}\right)F^{\sigma\sigma}_{\pm}({\bf r''},{\bf r'''})\Bigg\}\braket{{\bf r'}|\Psi_{\rm F}^{\sigma}}.
    \label{eqn:GKS-eigenvalue-deriv-final2}
\end{align}
\end{widetext}
Applying the same scaling arguments as before, one can readily show that each term in the spin resolved GKS eigenvalue derivative \textcolor{black}{vanishes in the bulk limit}. The spin resolved analogues of Eqs.~\ref{eq:ip_pristine_solid}, \ref{eq:ea_pristine_solid} \& \ref{eq:bandgap_pristine_solid} thus immediately follow in the pristine bulk limit. The same argument of course also holds for hybrid functionals, in which case the derivative of the frontier spin-resolved GKS eigenvalue may instead be expressed as,
\onecolumngrid
\begin{align}
    \frac{d\epsilon_{\rm F}^{\sigma}}{df_{\rm F}^{\sigma}}=\left\langle\Psi_{\rm F}^{\sigma} \right| \int d{\bf r}d{\bf r'} \ket{{\bf r}}\Bigg\{ &
     -a_{\rm HF}\frac{F_{\pm}^{\sigma \sigma}({\bf r},{\bf r'})}{|{\bf r}-{\bf r'}| } \nonumber \\ & +\sum_{\sigma'}\int d{\bf r''}\left(f_{\rm Hc}^{\sigma \sigma'}({\bf r},{\bf r''})+(1-a_{\rm HF})f_{\rm x}^{\sigma \sigma'}({\bf r},{\bf r''})\right)F_{\pm}^{\sigma' \sigma}({\bf r''},{\bf r''})\delta({\bf r}-{\bf r'}) \Bigg\}\braket{{\bf r'}|\Psi_{\rm F}^{\sigma}}.
    \label{eqn:GKS-hybrid2}
\end{align}
\twocolumngrid


%

\end{document}